\documentclass[11pt,a4paper]{article}
\DeclareMathAlphabet{\scr}{U}{rsfs}{m}{n}

\usepackage{etoolbox}

\usepackage{latexsym}
\usepackage{epsfig}
\usepackage[mathscr]{eucal}
\usepackage{amsfonts}
\usepackage{amscd}
\usepackage{array}   
\usepackage{bbold}
\usepackage{amsmath}
\usepackage[colorlinks=true,citecolor=blue,linkcolor=blue,urlcolor=blue]{hyperref}
\usepackage{cite}
\usepackage{amssymb}
\usepackage{colordvi}
\usepackage{enumerate}
\usepackage{graphicx}
\usepackage[dvipsnames]{xcolor}
\usepackage{booktabs}
\usepackage{theorem}
\usepackage{multirow}
\usepackage{hhline}
\usepackage{xspace}
\usepackage[format=plain,labelfont={bf},font={small}]{caption}
\usepackage{bbm}
\usepackage{units}
\usepackage{comment}
 \usepackage{slashed}
\usepackage{color}
\usepackage[normalem]{ulem}

\let\newabs=\abs
\let\newnorm=\norm
\let\newbraket=\braket
\let\newpb=\pb
\usepackage{physics}
\usepackage{float}

\let\abs=\newabs
\let\norm=\newnorm
\let\braket=\newbraket
\let\pb=\newpb

\usepackage{mathtools}

\usepackage[T1]{fontenc}

\usepackage{romanbar}
\newcommand{\MHpmm}[1][h]{\ensuremath{M_{#1}^{\text{\tiny \Romanbar{2}}}}\xspace}
\newcommand{\MHqcm}[1][h]{\ensuremath{M_{#1}^{\text{\tiny \Romanbar{4}}}}\xspace}
\newcommand{\pmii}{{\text{\tiny \Romanbar{2}}}}

\newcommand{\BP}[1]{\ensuremath{\texttt{BP#1}}}

\DeclarePairedDelimiter\abs{\lvert}{\rvert}%
\DeclarePairedDelimiter\norm{\lVert}{\rVert}%

\usepackage[a4paper,left=2.5cm,right=2.5cm,top=2.5cm,bottom=2.5cm]{geometry}
\usepackage{setspace}
\usepackage[capitalize]{cleveref}

\newcommand{\newc}{\newcommand}
\newc{\be}{\begin{equation}}
\newc{\ee}{\end{equation}}
\newc{\bea}{\begin{eqnarray}}
\newc{\eea}{\end{eqnarray}}
\newc{\ol}{\overline}
\newc{\wt}{\widetilde}
\newc{\bs}{\boldsymbol}
\newc{\m}{\mathcal}
\newc{\lan}{\langle}
\newc{\ra}{\rangle}
\newc{\pa}{\partial}

\newcommand{\cb}{c_\beta}   
\newcommand{\sbb}{s_\beta}

\newcommand{\ccdot}{\!\cdot\!}

\newcommand{\non}{\nonumber} 

\newcommand{\beq}{\begin{eqnarray}} 
\newcommand{\eeq}{\end{eqnarray}} 
\newcommand{\bpmatrix}{\begin{pmatrix}}
\newcommand{\epmatrix}{\end{pmatrix}}
\newcommand{\ba}{\begin{array}}
\newcommand{\ea}{\end{array}}
\newcommand{\braket}[1]{\left(#1\right)}

\newcommand{\fr}{\frac}

\newcommand{\al}{\alpha}

\newcommand{\doublet}[2]{\begin{pmatrix} #1 \\ #2 \end{pmatrix}}
\newcommand{\figref}[1]{Fig.~\ref{#1}}
\renewcommand{\eqref}[1]{Eq.~(\ref{#1})}
\newcommand{\bib}[1]{Ref.~\cite{#1}}

\newcommand{\DRb}{\overline{\text{DR}}}

\newcommand{\OS}{\text{OS}}
\newcommand{\MSb}{\overline{\text{MS}}}

\renewcommand{\Re}{\text{Re}\,}
\renewcommand{\Im}{\text{Im}\,}

\newcommand{\ie}{{\it i.e.\;}}
\newcommand{\eg}{{\it e.g.\;}}
\newcommand{\cf}{{\textit{cf.}\;}}
\newcommand{\bc}{\begin{center}}
\newcommand{\ec}{\end{center}}

\newcommand{\gev}{~\text{GeV}}

\newcommand{\pb}{{~\text{pb}}}

\newcommand{\MSbar}{\ensuremath{\overline{\text{MS}}}\xspace}
\newcommand{\DRbar}{\ensuremath{\overline{\text{DR}}}\xspace}
\newcommand{\ti}{\tilde}

\usepackage{bm}
\allowdisplaybreaks

\newcommand{\NMSSMCALC}{\ensuremath{\texttt{NMSSMCALC}}\xspace}
\newcommand{\mr}{\ensuremath{\texttt{mr}}\xspace}
\newcommand{\SMDR}{\ensuremath{\texttt{SMDR}}\xspace}
\newcommand{\SARAH}{\ensuremath{\texttt{SARAH}}\xspace}

\newcommand{\sm}{{\text{SM}}}
\newcommand{\SM}{\sm}
\newcommand{\nmssm}{{\text{NMSSM}}}
\newcommand{\NMSSM}{\nmssm}

\newcommand{\abskap}{|\kappa|}
\newcommand{\abslam}{|\lambda|}

\newcommand{\phis}{\varphi_s}
\newcommand{\phiw}{\varphi_w}
\newcommand{\phiy}{\varphi_y}

\newcommand{\msusy}{\ensuremath{M_{\text{\tiny SUSY}}}\xspace}
\newcommand{\Msusy}{\msusy}
\newcommand{\msm}{m_{\text{\tiny SM}}}
\newcommand{\vs}{v_s}
\renewcommand{\Re}{\text{Re}}
\renewcommand{\Im}{\text{Im}}

\newcommand{\ReAkap}{\Re A_{\kappa}}

\newcommand{\ReAlam}{\Re A_{\lambda}}

\newcommand{\TanBeta}{\tan\beta}
\newcommand{\TanBetas}{\tan^2\beta}
\renewcommand{\vev}{v}

\def\Hu{{H_u}}

\def\Hd{{H_d}}

\newcommand{\mhpm}{m_{H^\pm}}

\newcommand{\mueff}{\mu_{\text{eff}}}   

\newcommand{\QEW}{\ensuremath{Q_{\text{EW}}}}%
\newcommand{\Qmatch}{\ensuremath{Q_{\text{match}}}}%
\newcommand{\Qinp}{\ensuremath{Q_{\text{inp}}}}%
\newcommand{\smMSbar}{{\ensuremath{\SM,\MSbar}}}%
\newcommand{\nmssmDRbar}{{\ensuremath{\NMSSM,\DRbar}}}%
\newcommand{\nmssmMSbar}{{\ensuremath{\NMSSM,\MSbar}}}%
\newcommand{\email}[1]{\href{mailto:#1}{\nolinkurl{#1}}}

\numberwithin{equation}{section}

\begin{document}

\renewcommand*{\thefootnote}{\fnsymbol{footnote}}

\begin{flushright}
    KA-TP-10-2024\\
    DESY-24-093\\
    FR-PHENO-2024-005
\end{flushright}

\begin{center}
	{\LARGE \bfseries Higgs Mass Predictions in the CP-Violating\\ High-Scale NMSSM \par}

	\vspace{.7cm}
	Christoph Borschensky\textsuperscript{a,}\footnote{\email{christoph.borschensky@kit.edu}},
	Thi Nhung Dao\textsuperscript{b,}\footnote{\email{nhung.daothi@phenikaa-uni.edu.vn}},
	Martin Gabelmann\textsuperscript{c,}\footnote{\email{martin.gabelmann@desy.de}},\\
	Margarete M\"uhlleitner\textsuperscript{a,}\footnote{\email{margarete.muehlleitner@kit.edu}},
	Heidi Rzehak\textsuperscript{d,}\footnote{\email{heidi.rzehak@physik.uni-freiburg.de}}

	\vspace{.3cm}
	\textit{
		\textsuperscript{a }Institute for Theoretical Physics, Karlsruhe Institute of Technology, Wolfgang-Gaede-Str.\ 1, 76131 Karlsruhe, Germany\\[.2em]
		\textsuperscript{b }Phenikaa Institute for Advanced Study, PHENIKAA University, Hanoi 12116, Vietnam\\[.2em]
		\textsuperscript{c }Deutsches Elektronen-Synchrotron DESY, Notkestr.~85, 22607 Hamburg, Germany\\[.2em]
		\textsuperscript{d }Albert-Ludwigs-Universit\"at Freiburg, Physikalisches Institut, Hermann-Herder-Str.\ 3, 79104 Freiburg, Germany
	}
\end{center}

\renewcommand*{\thefootnote}{\arabic{footnote}}
\setcounter{footnote}{0}

\vspace*{0.3cm}
\begin{abstract}
In a supersymmetric theory, large mass hierarchies can lead to large
uncertainties in fixed-order calculations of the Standard Model (SM)-like Higgs mass.
A reliable prediction is then obtained by performing the calculation
in an effective field theory (EFT) framework, involving a matching to
the full supersymmetric theory at the high scale to include
contributions from the heavy particles, and a subsequent
renormalisation-group running down to the low scale. We report on the
prediction of the SM-like Higgs mass within the CP-violating
Next-to-Minimal Supersymmetric extension of the SM (NMSSM) in a
scenario where all non-SM particles feature TeV-scale masses. The
matching conditions are calculated at full one-loop order using two
approaches. These are the matching of the quartic Higgs couplings as well as of the SM-like Higgs pole masses of the low- and high-scale theory. A comparison between the two methods allows us to estimate the size of terms suppressed by the heavy mass scale that are neglected in a pure EFT calculation as given by the quartic-coupling matching.
Furthermore, we study the different sources of uncertainty which enter our calculation as well as the effect of CP-violating phases on the Higgs mass prediction. The matching calculation is implemented in a new version of the public program package \NMSSMCALC.
\end{abstract}

\thispagestyle{empty}
\vfill
\pagebreak

{\hypersetup{linkcolor=black}
\tableofcontents
}

\section{Introduction}
Since the discovery of the Standard Model Higgs
 boson with a mass of about 125 GeV by the ATLAS \cite{Aad:2012tfa} and the CMS 
 collaboration
 \cite{Chatrchyan:2012ufa} at the Large Hadron Collider (LHC) at CERN,  
 there has been no clear indication of new degrees of
  freedom in the range of the weak scale (which can be identified with energy scales around the mass of the W boson, $\sim M_W $) to a few TeV scale. 
  Taking into account the constraints from the Higgs data and 
  experimental searches for new degrees of freedom,
 the parameter space of each Standard Model extension (SM) should be reinvestigated in terms of
 whether these models can still satisfy the experimental constraints and give possibly detectable new physics signals. 
 Among them, the Minimal Supersymmetric extension of the SM (MSSM) is one of the most studied ones. 
 By imposing a symmetry between  bosonic and fermionic degrees of freedom, 
 the particle content is more than doubled with respect to the SM.
An interesting feature of this model is related to the Higgs sector. 
It contains two Higgs doublets due to the requirement of the cancellation of gauge 
anomalies on the one hand  as well as for the generation of
non-vanishing masses for all quarks on the other hand.
Furthermore, the quartic couplings in the Higgs sector are completely
determined by the gauge and Yukawa couplings. 
As a consequence, the Higgs boson masses can be predicted and one of the Higgs boson 
masses has an 
upper limit of about 140 GeV including higher-order corrections \cite{Hahn:2013ria}.
This Higgs boson can be identified with the discovered scalar particle.
 Similar features  occur also in the next-to-MSSM (NMSSM),
an extension of the MSSM that includes an extra Higgs singlet
superfield. However, the upper mass bound can be shifted to  higher values due to extra contributions from
 the Higgs singlet-doublet coupling \cite{Ellis:1988er, Drees:1988fc,Ellwanger:2009dp, Maniatis:2009re, Slavich:2020zjv}.
 
 Three approaches  for the computation of the Higgs boson masses including higher-order 
   corrections have been used. These are the fixed-order (FO), the effective field theory
(EFT), and the hybrid technique.
 For the first one, we have to compute Higgs boson self-energies
at fixed loop order and diagonalize  the loop-corrected Higgs mass matrix.
 This calculation involves the full particle spectrum and couplings  in the  broken phase of the electroweak (EW) symmetry of the model.
  The corrections will contain terms which are proportional to 
  {$\ln \ti M_x^2/M_x^2$} \cite{Hahn:2013ria}, where $M_x$ and $\ti M_x$ are a
  masses of an SM particle $x$ and its superpartner, respectively.
  These terms are particularly important for the top/stop sector, since the top Yukawa 
  coupling is the largest Yukawa coupling. Therefore, if there is large hierarchy between the stop and the top mass the FO
  calculation breaks down. In such a case, one needs to resum these large logarithms to obtain reliable results.
  In the EFT calculation, the  couplings of the high-energy extension of the SM
are matched to the corresponding ones in the effective
low-energy field theory such that at the matching
scale the physics described by the two models is the same. If only the
SM-like particles of the SM extension are light and all the Beyond-SM
(BSM) particles are heavy, then the SM is a suitable EFT. {In this case, the loop-corrected SM quartic Higgs coupling at the
matching scale can be identified with a loop-corrected BSM quartic Higgs coupling at the same scale. All SM-type contributions {proportional to powers of $\ln \Qmatch^2 /M_x^2$} are cancelled out, where $\Qmatch$ is the matching scale. The running SM quartic Higgs coupling is now defined from the BSM quartic Higgs coupling which contains logarithmic terms of the form {$\ln \ti M_x^2/\Qmatch^2$.}} Then the logarithmic terms become small
    when $\ti M_x$ is close to 
    $\Qmatch$. The remaining dependence on {$\ln \Qmatch^2/M_x^2$} can
    be resummed with the help of  the SM 
    renormalisation group equations (RGEs) for the  {quartic Higgs coupling}. 
    In the literature, there exist two ways to match the
    loop-corrected BSM quartic Higgs coupling. The first one is called 
    quartic coupling matching which is based on the 
     computation of loop corrections to the four-point vertex using the spectrum and 
    couplings in the limit of the unbroken EW symmetry, $v\to 0$, with $v$ being the 
    vacuum expectation value. The second one is 
    called pole-mass matching which is based on 
    the assumption that also in the BSM theory the
    SM-like quartic Higgs 
    coupling has a relation with the SM-like Higgs 
    pole mass. By computing the BSM contribution to the SM-like Higgs
    mass, we can extract  
    information on the quartic Higgs coupling. The
    computation needs to be done in the broken EW symmetry phase and
    leads to  contributions of ${\cal O}(v^2/\ti M_x^2)$.  The third approach is
    called the hybrid technique which combines fixed-order calculations with the EFT approach where 
    the leading and next-to-leading logarithms are resummed and care
    is taken that no double counting occurs. The gain is
     two-fold. On the one hand, the theory uncertainty at high
     supersymmetric (SUSY) masses is reduced in comparison to a pure fixed-order
   result~\cite{Hahn:2013ria}. On the other hand, 
   {${\cal O}(v^2/\ti M_x^2)$} effects are taken into
   account in comparison to the pure EFT approach, 
   where they are integrated out.   
       
   A lot of effort has been devoted to the precise calculation of the Higgs  
 boson masses in the NMSSM using fixed-order
calculations. Leading one-loop, full one-loop and leading two-loop contributions
 were presented in
\cite{Ellwanger:1993hn,Elliott:1993ex,Elliott:1993uc,Elliott:1993bs,Pandita:1993hx,Pandita:1993tg,King:1995vk,Ellwanger:2005fh, Degrassi:2009yq,Staub:2010ty, Drechsel:2016jdg,
Ham:2001kf,Ham:2001wt,Ham:2003jf,Funakubo:2004ka,Ham:2007mt,Cheung:2010ba,Goodsell:2016udb,Domingo:2017rhb} 
where the $\DRb$ renormalisation scheme was applied, except for
\cite{Drechsel:2016jdg}, which also applied a mixed
$\DRb$-on-shell (OS) renormalisation scheme. At two-loop level, all
contributions have been computed  
in the gaugeless limit and using the zero external momentum approximation.  The QCD 
corrections have been discussed in \cite{Degrassi:2009yq}
 and {$\mathcal{O}((\alpha_\lambda+\alpha_\kappa+\alpha_t)^2)$ corrections} in \cite{Goodsell:2014pla}. Our group has also contributed to 
 the progress of precision predictions for the masses. The full one-loop corrections with momentum
 dependence were presented in \cite{Ender:2011qh,Graf:2012hh} 
 and the two-loop corrections of ${\cal O}{(\al_t\al_s)}$ in
 \cite{Muhlleitner:2014vsa},  of
${\cal O} (\al_t^2)$ in \cite{Dao:2019qaz}, and of ${\cal O} ((\al_t + \al_\lambda +\al_\kappa)^2)$ in \cite{Dao:2021khm} for both the CP-conserving and the CP-violating NMSSM.
We were the first ones to apply a mixed $\DRb$-OS
scheme in the NMSSM with the possibility to choose 
between either $\DRb$ or OS conditions in the renormalisation scheme for the top/stop
sector.  We implemented our FO  
calculations at one-loop and two-loop ${\cal O}(\alpha_t \alpha_s)$,
${\cal O}(\alpha_t^2)$, and ${\cal O}((\al_t +
\al_\lambda +\al_\kappa)^2)$ level in the program package
\NMSSMCALC~\cite{Baglio:2013iia} which also computes the 
Higgs boson decay widths and branching ratios both for the CP-conserving and
CP-violating case. The code furthermore includes the computation of the
  loop-corrected trilinear Higgs self-couplings at one-loop \cite{Nhung:2013lpa}
  and at two-loop ${\cal O}(\alpha_t \alpha_s)$
  \cite{Muhlleitner:2015dua} and ${\cal O}(\alpha_t^2)$
  \cite{Borschensky:2022pfc} as well as the loop 
  corrections to the $\rho$ parameter and the $W$ boson mass
  \cite{Dao:2023kzz}.
There are also other public codes such as  
{\tt NMSSMTools} \cite{Ellwanger:2004xm,Ellwanger:2005dv}, {\tt SARAH/SPheno} \cite{Staub:2009bi,Staub:2010jh,Staub:2012pb,Staub:2013tta,Porod:2003um,Porod:2011nf}, {\tt SOFTSUSY} \cite{Allanach:2001kg, Allanach:2017hcf}, {\tt FlexibleSUSY}
\cite{Athron:2014yba,Athron:2017fvs} which are
dedicated to the computation of the 
NMSSM spectrum, decay widths and other observables.

There have been much less activities regarding the EFT approach in the precision
 Higgs mass calculation
 in the NMSSM. A  discussion of EFT in generic SUSY models including also the NMSSM
  has been presented in
  \cite{Athron:2016fuq}. There the loop-corrected quartic Higgs coupling is obtained from the 
  loop-corrected mass of the lightest Higgs boson after subtracting
  the corresponding part of the SM.
  This matching condition has been implemented in {\tt
    FlexibleEFTHiggs} and later in  {\tt SARAH/SPheno} \cite{Staub:2017jnp}. In
\bib{Bagnaschi:2022zvd}, the authors have used the matching condition where
 the loop-corrected quartic Higgs coupling is obtained from the loop-corrected four-point vertex.
  They have combined a full one-loop
computation with the QCD two-loop contributions for
 the quartic Higgs coupling. This computation has been performed in the limit of 
 the unbroken EW symmetry where the Higgs doublet vacuum expectation values $v_u, v_d \to 0$
 while the singlet vacuum expectation value $v_s$ is kept non-vanishing and large so that the singlet Higgs masses are very heavy and 
can  then be integrated out.   
  
Our purpose in this paper is threefold. First, we implement both
matching conditions discussed in Refs.~\cite{Athron:2016fuq} and 
\cite{Bagnaschi:2022zvd}. For the pole-mass matching condition we make use of 
our FO computations of the full one-loop corrections to the Higgs boson masses, which have been 
implemented in our computer code \NMSSMCALC. For this, we modify the renormalisation scheme from the mixed OS-$\DRb$ to a pure
$\DRb$ scheme for all parameters except the tadpoles for which we still make
use of OS-like conditions as in the previously used mixed OS-$\DRb$ scheme.
Using $\DRb$/$\MSbar$ quantities
    conveniently enables us to make use of higher-order results in the 
renormalisation group evolution from the literature. For the
four-point vertex matching condition, we compute the 
full one-loop corrections in the {limit of the} unbroken phase of the EW symmetry in
the \DRbar scheme and discuss subtleties related to the $v\to0$ limit and
finite tadpole corrections. We then compare the effect of the two 
matching methods on the Higgs mass prediction in a large 
region of the parameter space where the scale of the SUSY particle
masses ranges from TeV to hundred TeV.  We also compare the EFT
approach and the FO calculation in the mixed OS-$\DRb$ scheme being
available  in \NMSSMCALC where the renormalisation scale is chosen to
be the matching scale in the EFT approach. Second, we discuss the
effect of the CP-violating phases in the EFT approach which has not  
been done in the previous publications. Third, we provide an updated
version of \NMSSMCALC that gives a better treatment in the case where
a large mass hierarchy between BSM and SM-like particles occurs.   

The paper is organized as follows. 
In \cref{sec:NMSSMtree-level} we introduce the NMSSM, set up the notation
and derive expressions for the tree-level mass matrices and transformations
into the mass basis in the limit of a vanishing electroweak VEV.
\cref{sec:calculation} discusses the general ingredients for a Higgs mass
calculation using an EFT approach. In the first two subsections {the quartic-coupling and
 pole-mass} matching approaches are explained in detail while the
third subsection describes the estimate of theoretical uncertainties.
\cref{sec:numerical} is dedicated to the numerical analysis which validates
our results numerically with results from the literature and studies the
different EFT approaches and their uncertainties as well as compares to the
FO calculation. We conclude in \cref{sec:conclusions}. The
  appendix contains the derivation of the tadpole-expansion around a small VEV
and details on the implementation in the program \NMSSMCALC.

%%%%%%%%%%%%%%%%%%%%%%
\section{The High-Scale NMSSM at Tree-Level}
\label{sec:NMSSMtree-level}
%%%%%%%%%%%%%%%%%%%%%%
We briefly review the basic ingredients of the complex NMSSM to set up our notation for 
later use.  The model is specified by a scale invariant superpotential $\mathcal{W}_{\text{NMSSM}}$, 
\begin{align}
    \mathcal{W}_{\text{NMSSM}} = 
	&
		\left[y_e \hat{H}_d\ccdot \hat{L} \hat{E}^c 
		+ y_d  \hat{H}_d \ccdot \hat{Q} \hat{D}^c 
		- y_u \hat{H}_u \hat{Q} \hat{U}^c
		\right]  - \lambda \hat{S} \hat{H}_d\ccdot \hat{H}_u 
		+ \frac{1}{3} \kappa \hat{S}^3  \;,
    \label{eq:wnmssm}
\end{align}
where $ \hat{H}_d$ and $ \hat{H}_u$ are the Higgs doublet superfields,
$\hat{S}$ the Higgs singlet superfield and 
  $\hat{L}$, $\hat{Q}$, as well as $\hat{E}$, $\hat{D}$, and $\hat{U}$
  the left-handed lepton and quark doublet superfields as well as the
  right-handed lepton, down-type, and up-type quark singlet
  superfields, respectively. In the following, we will denote the
  scalar part of the Higgs superfields and the fermion part of the
  lepton and quark superfields with the same letter without a hat. The
  lepton, down-type, and up-type quark Yukawa couplings are $y_e$,
  $y_d$, $y_u$ which are $3 \times 3$ matrices that we assume to be
  diagonal. The summation over generation indices is implicit. The
  coupling between the Higgs doublet and Higgs singlet superfields is
  governed by $\lambda$ and the Higgs singlet superfield self-coupling
  is $\kappa$. Both $\lambda,\kappa$ are considered to be complex
  parameters with corresponding phases
  $\varphi_\lambda,\varphi_\kappa$.  All Yukawas couplings are
    taken to be real by rephasing the left- and right-handed
    Weyl-spinor fields accordingly.
The soft-SUSY breaking Lagrangian comprises the soft-SUSY breaking parameters,
\begin{eqnarray}
{\cal L}_{\text{soft},\text{ NMSSM}} &=& -m_{H_d}^2 H_d^\dagger H_d - m_{H_u}^2
H_u^\dagger H_u -
m_{\tilde{Q}}^2 \tilde{Q}^\dagger \tilde{Q} - m_{\tilde{L}}^2 \tilde{L}^\dagger \tilde{L}
- m_{\tilde{u}_R}^2 \tilde{u}_R^* 
\tilde{u}_R - m_{\tilde{d}_R}^2 \tilde{d}_R^* \tilde{d}_R 
\nonumber \\\nonumber
&& - m_{\tilde{e}_R}^2 \tilde{e}_R^* \tilde{e}_R - (\epsilon_{ij} [y_e A_e H_d^i
\tilde{L}^j \tilde{e}_R^* + y_d
A_d H_d^i \tilde{Q}^j \tilde{d}_R^* - y_u A_u H_u^i \tilde{Q}^j
\tilde{u}_R^*] + \mathrm{h.c.}) \\
&& -\frac{1}{2}(M_1 \tilde{B}\tilde{B} + M_2
\tilde{W}_i\tilde{W}_i + M_3 \tilde{G}\tilde{G} + \mathrm{h.c.}) \nonumber
\\ 
\label{eq:lagrangiansoft}
&&- m_S^2 |S|^2 +
(\epsilon_{ij} \lambda 
A_\lambda S H_d^i H_u^j - \frac{1}{3} \kappa
A_\kappa S^3 + \mathrm{h.c.}) \;,
\end{eqnarray}
where $\tilde{B}$, $\tilde{W}_i$, $\tilde{G}$ are the fermionic $U(1)$
bino, $SU(2)$ wino and $SU(3)$ gluino fields and $A_e$, $A_d$, $A_u$
are the soft-SUSY-breaking trilinear couplings, which are $3 \times 3$
matrices and  assumed to be diagonal in this paper; again, the
summation over the generation indices is implicit here.  The
soft-SUSY-breaking mass parameters of the sfermions and Higgs fields, 
 $m_{\tilde{Q}}^2$, $m_{\tilde{L}}^2$, $m_{\tilde{u}_R}^2$, $m_{\tilde{d}_R}^2$, $m_{\tilde{e}_R}^2$,
 $m_{H_d}^2$, $m_{H_u}^2$, and $m_S^2$ are real and positive while
 the gaugino masses $M_1$, $M_2$, $M_3$ and the soft SUSY breaking
 trilinear couplings are complex in general.
The scalar Higgs potential is obtained from the superpotential in \eqref{eq:wnmssm}, the soft-SUSY breaking part of the 
Lagrangian \eqref{eq:lagrangiansoft} and the $D$ terms originating
from the gauge sector of the Lagrangian. Requiring the scalar
potential to be minimal at non-vanishing vacuum expectation values (VEVs) of the two Higgs doublets leads to spontaneous breaking of the EW gauge
symmetry. Allowing also for the possibility of a singlet VEV, the  
three Higgs boson fields can be expanded about their VEVs $v_u$,
$v_d$, and $\vs$, respectively, as 
\begin{equation}
    H_d = \doublet{\frac{v_d + h_d +i a_d}{\sqrt 2}}{h_d^-}, \,\, 
    H_u = e^{i\varphi_u}\doublet{h_u^+}{\frac{v_u + h_u +i a_u}{\sqrt 2}},\,\,
    S= \frac{e^{i\varphi_s}}{\sqrt 2} (\vs + h_s + ia_s)\, ,
   \label{eq:vevs}
\end{equation}
with the CP-violating phases $\varphi_{u,s}$.   
 
For a more comprehensive introduction of the model and its mass spectrum of all sectors
  at tree-level in the broken phase we refer the reader to our paper \cite{Baglio:2019nlc}. We follow the same convention as in  \cite{Baglio:2019nlc}.

Since for the quartic coupling matching we need expressions for masses
and mixings in the unbroken phase of the EW symmetry, in the
following, we present the spectrum of the model in the limit $v_u,
v_d\to 0$ but with a fixed ratio of the two VEVs,   
\be \tan \beta = \fr{v_u}{v_d},\ee
and a non-vanishing singlet VEV $\vs$. We denote the SM-like VEV as
$v$, which is related to the two Higgs doublet VEVs as
\be v^2 = v_u^2 + v_d^2. \ee

\paragraph{Higgs Bosons}
In the limit $\vev\to 0$ the mass matrices of the CP-violating NMSSM
take a particularly simple form which allows for an analytical 
diagonalisation. First, we solve the tadpole equations of $t_{h_u},\,
t_{h_d},\, t_{h_s}$ and $t_{a_d},\, t_{a_s}$ for the soft-SUSY
breaking squared masses $m_{\Hu}^2$, $m_{\Hd}^2$, $m_S^2$ and the
imaginary parts of the parameters $A_\lambda$, $A_\kappa$, see e.g. Ref.~\cite{Dao:2019qaz}. This is
done without taking the limit $\vev\to0$ since the solution to the
tadpoles may contain terms $\order{\vev^{-n}}$ which are multiplied
with terms $\order{\vev^{+m}}$ when inserting them into the mass
matrices.\footnote{{Note that this procedure differs from the one
    discussed in e.g.\ \cite{Bagnaschi:2022zvd,Bagnaschi:2014rsa} which only
relies on the use of tadpole equations for the singlet BSM fields. 
We, however, have found very good agreement with
\cite{Bagnaschi:2022zvd} in the CP-conserving case, \cf Section 4.3 
and also the discussion in \cref{sec:calculation:quartic:oneloop}. This suggests 
that while the intermediate results differ, the final finite result of the matching condition remains unchanged.}}  \\
Using the tree-level tadpole solutions in the 
tree-level mass matrices and then performing the limit $\vev\to0$, we
obtain the following:
The squared mass matrix of the charged Higgs boson has one vanishing
eigenvalue corresponding to the Goldstone boson and one non-zero
eigenvalue corresponding to the physical charged Higgs
boson,\footnote{We use small letters $m$ to denote tree-level
    masses and capital letters $M$ to denote loop-corrected or
    on-shell input masses.} 
\begin{subequations}
    \begin{align}
        m_{G^\pm}^2 & = 0 \\
        {\mhpm^2} & = 
       \frac{\abslam (1 + \TanBetas) \vs}{2\TanBeta \cos(\phiw - \phiy)}\left(
        \sqrt2 \ReAlam + \abskap \vs \cos\phiw \right)  \;,
    \end{align}\label{eq::treemassesC}
\end{subequations}
where
\beq   \phiw&=&  3 \varphi_s + \varphi_\kappa \\
        \phiy&=& 2 \varphi_s+ \varphi_\kappa  - \varphi_\lambda
        -\varphi_u \;.   \eeq
Thus, we can trade $\ReAlam$ for ${\mhpm^2}$.
The squared mass matrix for the neutral Higgs bosons takes a 
block-diagonal form where, after the use of the tadpole solutions,
neither the CP-even and the CP-odd components nor the doublet and
singlet components mix. 
The mixing matrix which diagonalizes the neutral Higgs mass matrix
transforming the basis $(h_d,h_u,h_s,a_d,a_u,a_s)^T$ to the basis $(h, H, S, G^0, A, A_s)$, reads
\begin{equation}
    \mathcal{R}(\beta) = \left(\begin{array}{cc}
            \mathcal{R}^{H}(\beta) & \textbf{0} \\
            \textbf{0}      & \mathcal{R}^{G}(\beta)
    \end{array}\right) \label{eq:neutralmix}
\end{equation}
with
\begin{equation}
    \mathcal{R}^{H}(\beta) = \left(\begin{array}{ccc}
            \cb & \sbb & 0 \\
            -\sbb & \cb & 0 \\
            0 & 0 & 1
        \end{array}\right)\,, \quad
    \mathcal{R}^{G}(\beta) = \left(\begin{array}{ccc}
            -\cb & \sbb & 0 \\
            \sbb & \cb & 0 \\
            0 & 0 & 1
        \end{array}\right)\,,
        \label{eq:rot}
\end{equation}
Diagonalizing the neutral Higgs mass matrix with the help of the
mixing matrix in \eqref{eq:neutralmix} results in a diagonal matrix
with the entries
\begin{subequations}
    \begin{align}
        m_{h}^2(= m_{h_u}^2)    =&  0 \, \\
        m_{H}^2(= m_{h_d}^2)    =& {\mhpm^2}     \\       
        {m_{H_s}^2}(= m_{h_s}^2)    =&  \frac{\abskap \vs (4 \abskap \vs + \sqrt2 \frac{\ReAkap}{\cos\phiw})}{2}\, \\
        m_{G^0}^2 (= m_{a_u}^2) =&  0 \, \\
        m_{A}^2   (= m_{a_d}^2)  =& {\mhpm^2}\, \,  \\
        {m_{A_s}^2}\, (= m_{a_s}^2) =&  -\frac{3\abskap \ReAkap\vs}{\sqrt2\cos\phiw}  \,.
    \end{align}
        \label{eq:treemassesN}
\end{subequations}
The masses in parenthesis denote the dominant
gauge-eigenstates masses. It turns out that for the parameter
  points that are compatible with all applied constraints and that we
  discuss in our numerical analysis, the $h$ ($H$) is mostly $h_u$-like
  ($h_d$-like) and the $A$ ($G^0$) is mostly $a_d$-like ($a_u$-like).
 The two vanishing eigenvalues correspond to the neutral Goldstone boson and 
 the SM-like Higgs boson mass. These are the only two neutral scalar
states that belong to the EFT, and all remaining heavy neutral Higgs bosons are integrated out. The 
third eigenvalue {$m_{H_s}^2$} corresponds to the mass of the
scalar singlet. The second and fifth
eigenvalues, $m_H^2$ and $m_A^2$, are degenerate and coincide
with the mass of the charged Higgs boson, {$m_{H^\pm}^2$}. 
The last eigenvalue, {$m_{A_s}^2$}, corresponds to the pseudoscalar
singlet. 

\paragraph{Fermions} Using the approximation $v\to0$ the top quark as
well as all other SM fermions are massless and do not mix with each other. 
Considering the fermionic supersymmetric partner particles and assuming
 the Weyl basis $\psi^0 =  (\tilde{B},\tilde{W}_3, \tilde{H}^0_d,\tilde{H}^0_u,  \tilde{S})^T$
and $\psi^- =(\tilde{W}^-, \tilde{H}_u^-)^T$ where $\tilde{H}^0_d$,
$\tilde{H}^0_u$, $\tilde{H}_u^-$, $\tilde{S}$ are the neutral and
charged doublet Higgsino and singlino fields,  respectively,  for the
neutral and the charged fields results in the following neutralino and
chargino mass matrices, 
\begin{align}
M_N = \begin{pmatrix} 
M_1               & 0        & 0 &   
                                 0
               & 0\\
0                 & M_2      &    0    & 0 
               & 0\\
0 & 0 & 0                &
                 - \lambda \frac{\vs}{\sqrt{2}} e^{i \phis}   & 
               0
  \\
 0 &  0 & 
                            - \lambda \frac{\vs}{\sqrt{2}}e^{i \phis} & 0 &
                         0 
   \\
0   & 0     & 0 & 
                            0 &
                             \sqrt{2} \kappa \vs e^{i \phis} 
\end{pmatrix}  \,,  \quad   
M_C = \begin{pmatrix} M_2 & 0\\
   0 & \lambda \frac{\vs}{\sqrt{2}} e^{i \phis}
    \end{pmatrix}\,. \label{eq:chaMass}     
\end{align}
The neutralinos and
charginos have a non-zero mass. Their mass-matrices can be
diagonalized analytically. We find the following mass eigenvalues:
\begin{subequations}
    \begin{align}
        m_{\chi^0_1} & =\abs{M_1}\\
        m_{\chi^0_2} & =\abs{M_2}\\
        m_{\chi^0_3} & = m_{\chi^\pm_2}\\
        m_{\chi^0_4} & = m_{\chi^\pm_2}\\
        m_{\chi^0_5} & = {\sqrt{\nicefrac{m_{A_s}^2}{3} + m_{H_s}^2}}\\
        m_{\chi^\pm_1} & =\abs{M_2}\\
        m_{\chi^\pm_2} &= \frac{\abslam}{\abskap}
                         \frac{\sqrt{{\nicefrac{m_{A_s}^2}{3}+m_{H_s}^2}}
                         }{2} \;,
    \end{align}
\end{subequations}
where all complex phases have been absorbed into the rotation matrices
and $\vs$ is replaced by 
\begin{eqnarray}
  \vs =\frac{{\sqrt{m_{A_s}^2/3 + {m_{H_s}^2}}}}{\sqrt{2}|\kappa|} \;.
\end{eqnarray}

\paragraph{Sfermions} In the limit $\vev\to0$ the squared sfermion
mass matrices are only given in terms of the  soft-SUSY
breaking parameters, and the mixing between left- and right-handed scalars vanishes. Thus 
their interaction eigenstates coincide with
the mass eigenstates, 
\begin{equation}
    m^2_{\tilde{X}} = \left(\begin{array}{cc}
            \bm{m}_{\tilde{X}_L}^2 & 0 \\
            0                  & \bm{m}_{\tilde{X}_R}^2   \\
        \end{array}
    \right)\,, \qquad X=u,d,l\, ,
\end{equation}
assuming minimal flavour violation (\ie diagonal $\bm{m}_{\ti X_L}^2$, $\bm{m}_{\ti X_R}^2$).  The 
above diagonal mass matrix has two eigenstates $\tilde{X}_{1,2}$ with masses
$m_{\tilde{X}_{1,2}} = \{ m_{\tilde{X}_{L}}, m_{\tilde{X}_{R}} \}$
for  the superpartners of each SM fermion generation. Note that
only the 3rd generation of quarks and leptons has significant effects
on the Higgs boson masses.  
 
%%%%%%%%%%%%%%%%%%%%%%%%%%%%%%%%%%%%%%%%%%%%%%%%%%%%%%%%%%%%%%%%
\section{The Loop-Corrected Higgs Mass in the EFT Approach}
\label{sec:calculation}
%%%%%%%%%%%%%%%%%%%%%%%%%%%%%%%%%%%%%%%%%%%%%%%%%%%%%%%%%%%%%%
In the scenario that we are considering in this paper, all the soft-SUSY-breaking mass parameters of 
the sfermions and gauginos, $m_{\tilde{Q}}^2$, $m_{\tilde{L}}^2$, $m_{\tilde{u}_R}^2$, $m_{\tilde{d}_R}
^2$, $m_{\tilde{e}_R}^2$, and $M_1$, $M_2$, $M_3$, together with
{$m_{H_s},m_{A_s}, m_{H^\pm}$} are
much larger than the SM EW scale. These masses are similar to
    $\sim \msusy$ where $
\msusy \gg \vev$. Since all SM-like particles have a mass $\msm \propto \vev$, this means $\msusy\gg \msm$.

In such a scenario, a fixed-order approach will lead to large logarithms of 
$\ln(\vev^2/\msusy^2)$ which will destroy the
perturbative expansion and hence a precise prediction. Therefore, we
follow the EFT approach where we match the
NMSSM to the effective field theory (that we identify with the SM in
this paper) at the scale $\msusy$ in such a way that at $\msusy$
both theories lead to the same physical predictions. The contributions
containing large logarithms of $\ln(\vev^2/\msusy^2)$ will be resummed
through the SM renormalisation group equations (RGE). In the NMSSM
calculation, we will denote the SM-like Higgs that is matched to the
Higgs boson of the EFT as the one which is predominantly made up of
the $h_u$ component of \eqref{eq:vevs}. 

\begin{figure}[tp]
\begin{center}
\includegraphics[width=1.0\textwidth]{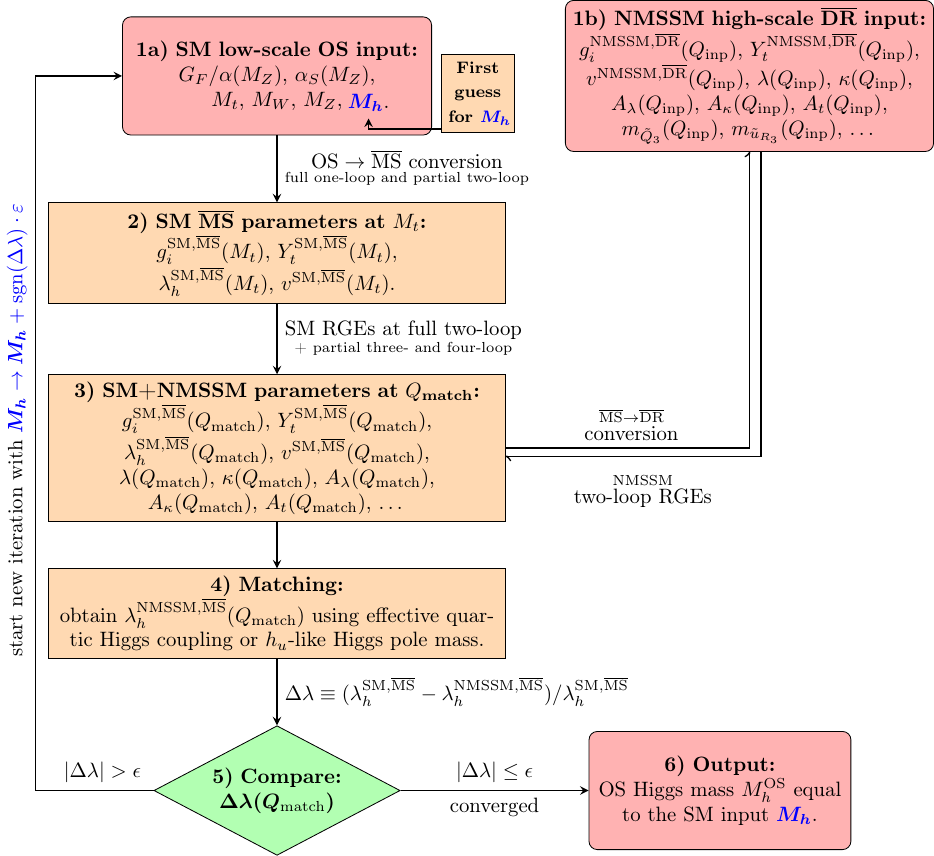}
\end{center}
\caption{Schematic procedure for the computation of the
  loop-corrected Higgs mass in the EFT approach implemented in
\NMSSMCALC.}
\label{fig:HiggsMassEFT}
\end{figure}

For the matching procedure, we are following two matching condition
schemes which allow us to relate the effective quartic Higgs coupling
$\lambda_h^\sm$ in the SM\footnote{Note that we explicitly distinguish
  between $\lambda$, which is the NMSSM superpotential parameter, and
  the quartic Higgs coupling in the SM, $\lambda_h^\sm$.} to the one
in the NMSSM. For the 
    quartic interaction term we use the normalisation $-\lambda_h^\sm
    |H|^4$, with the neutral component of the SM
    Higgs doublet $H$ given by $H^0 = \frac{1}{\sqrt{2}}(v^\sm + h +
    iG^0)$.
\begin{itemize}
	\item In the first scheme, we directly require the four-point
          functions, \ie matrix elements, with four external Higgs
          bosons in the SM and the SM-like Higgs bosons in the NMSSM
          to be the same at the matching scale. We can then
          calculate loop corrections to 
          the quartic Higgs coupling within the NMSSM,
          $\lambda_h^\nmssm$, that we then identify with the
          loop-corrected SM one $\lambda_h^\sm$. In the following, we
          will refer to this scheme as \emph{quartic-coupling
            matching}. 
	\item In the second scheme, we demand that the pole masses of
          the Higgs boson in the SM and the SM-like Higgs boson in the
          NMSSM are equal at the matching scale. The SM $\MSb$ mass at
          the matching scale is then computed from this matching condition.
          Here, it has to be ensured that all large logarithms of
        $\ln(\vev^2/\msusy^2)$ are canceled by applying a proper expansion. 
        The extraction of the effective
      quartic Higgs coupling via \eqref{eq:treesmlameff} is discussed
      in Sec.~\ref{sec:calculation:pole} in more detail. We will 
        refer to this scheme as \emph{pole-mass matching}. 
\end{itemize}
The major difference between the two matching schemes consists in the
diagrams to be evaluated (\cf \cite{Bahl:2020jaq} for a detailed discussion): While the quartic-coupling matching requires
four-point functions to be calculated in the limit of $v \to 0$, the
pole-mass matching requires at most only two-point functions, \ie self
energies, to be evaluated, at the expense of having to carry out the
calculation in the EW-broken phase and then expanding systematically
in $v^2/\Msusy^2$. In both matching schemes, we obtain a value for the
effective quartic Higgs coupling of the SM, capturing the effects of
the heavy particles with masses $\sim \msusy$ and resumming all large
logarithms $\ln(\vev^2/\msusy^2)$ consistently via RGEs. Before presenting
the calculation of the effective quartic Higgs couplings in the two
matching schemes, we show in \figref{fig:HiggsMassEFT} our procedure
for the computation of the loop-corrected Higgs mass in the EFT
approach, implemented in the 
new version of \NMSSMCALC, and describe it in detail in 
the following. 

%%%%%%%%%%%%%

%%%%%%

We start with the six SM input parameters (\textbf{box 1a} of
\figref{fig:HiggsMassEFT}) which can be either 
\begin{equation}\label{eq:gfscheme}
	G_F,~ \alpha_S(M_Z),~ M_t,~ M_W,~ M_Z,~ M_h
\end{equation}
or
\begin{equation}\label{eq:amzscheme}
	\alpha(M_Z),~ \alpha_S(M_Z),~ M_t,~ M_W,~ M_Z,~ M_h,
\end{equation}
where all masses are considered to be the pole masses. 
We call the choice of input parameters of
\eqref{eq:gfscheme} the ``$G_F$ scheme'', while we denote the choice
of \eqref{eq:amzscheme} as the ``$\alpha_{M_Z}$ scheme''. We then have
to convert all OS input parameters to their corresponding $\MSbar$ parameters at the
scale $M_t$. For the $\alpha_{M_Z}$ scheme we use the  conversion formulae 
which are already available in \NMSSMCALC. These conversion formulae have been given in 
the appendix D of \cite{Dao:2019qaz},
but we use them now at the scale $M_t$ instead of $M_Z$. For the $G_F$ scheme, we use the conversion formulae at the scale $M_t$, presented in \cite{Buttazzo:2013uya}. {Although \cite{Buttazzo:2013uya} provides 
the full two-loop corrections to the conversion formulae, we took
into account only one-loop EW corrections for all parameters
and two- and three-loop QCD corrections for the conversion of $M_t$. 
In addition, we converted between $M_h$ and $\lambda_h^{\text{SM},\overline{\text{MS}}}$ at ${\cal O}(\alpha_t(\alpha_t+\alpha_s))$. 
This is the same level of approximation as in the $\alpha_{M_Z}$ scheme. } 

As running $\MSbar$ parameters in the SM, we choose (\textbf{box 2})
\begin{equation}\label{eq:smrunparams}
    g_1^{\text{SM},\,\MSb},~ g_2^{\text{SM},\,\MSb},~
    g_3^{\text{SM},\,\MSb},~ Y_t^{\text{SM},\,\MSb},~ v^{\text{SM},\,\MSb},~
    \lambda_h^{\text{SM},\,\MSb}\,.
\end{equation}
As usual, we denote by $g_1$, $g_2$, $g_3$ the three gauge couplings
of the corresponding three gauge symmetry groups $U(1)_Y$, $SU(2)_L$
and $SU(3)_C$, while $Y_t$ is the top Yukawa coupling. After obtaining
these $\MSbar$ parameters at $M_t$, we apply the SM 
RGEs\footnote{We employ for $g_1$ the GUT normalisation $g_1^{\text{GUT}} =
\sqrt{5/3} g_1$ commonly used in SM RGEs.} including full two-loop and
partial three- and four-loop
contributions
\cite{Staub:2010jh,Schienbein:2018fsw,Chetyrkin:2012rz,Chetyrkin:2016ruf}
to run up to the matching scale which is denoted by $\Qmatch$ with $\Qmatch \gg M_t$.

For the NMSSM calculation at the high-energy scale $\Qinp$
(\textbf{box 1b}), we have the following input parameters:
\begin{equation}\label{eq:susyinput}
	\vs,~\tan\beta,~ m_{\tilde{Q}_3},~ m_{\tilde{u}_{R_3}} ,~
        M_1,~ M_2,~ M_3,~ \lambda,~ \kappa,~ A_{t},~\mbox{Re}
        A_\lambda,~\mbox{Re} A_\kappa,~\varphi_u,~\varphi_s
      \end{equation}
as well as the corresponding parameters of \eqref{eq:smrunparams} in
the $\DRbar$ scheme with the exception of the quartic Higgs coupling
$\lambda_h$, which is not an input parameter in the NMSSM. We remind
that the $M_{1,2,3}$ and $A_t$ are complex, their
phases are included in the running from $\Qinp$ to $\Qmatch$, while
the imaginary parts of $A_\lambda$ and $A_\kappa$ are eliminated through the
tadpole equations at $\Qmatch$. {The phases of the other complex
  parameters, \ie $\lambda$ and $\kappa$, do not run since their UV-counterterms vanish \cite{Dao:2021khm}.}
Note, that for the sfermion contributions we only take into account
corrections from the stops, \ie the top-quark Yukawa coupling is the only
non-zero Yukawa coupling.
{The first
    five parameters in \textbf{box 1b)}, $g^{\nmssmDRbar}_i$, $Y^{\nmssmDRbar}_t$ and
$v^{\nmssmDRbar}$, are not fixed as (user) input parameters but actually depend on
the values of the running SM-parameters in \textbf{box 3)}. To solve this
two-scale problem, the full
set of running SUSY parameters is determined by
an iterative RGE running between $\Qinp$ and $\Qmatch$ with
$X^{\nmssmDRbar}(\Qinp)=X^{\sm,\DRbar}(\Qmatch)$ ($X=g_i,Y_t,v$) as a first guess, which is symbolised by
the double-arrow in \cref{fig:HiggsMassEFT}.}
The $\MSbar \to \DRbar$ conversion formulae for $g_1$, $g_2$ and $g_3$ between
high-scale and low-scale parameters at 1-loop level  are given by
\begin{equation}\label{eq:gaugematching}
	g_i^{\nmssm,\DRbar} = g_i^{\sm,\MSbar} + \delta g_i^{\text{reg}} + \delta g_i^{\text{thr}}\,,
\end{equation}
where the $\delta g_i^{\text{reg}}$ denote $\MSbar$--$\DRbar$ shifts
related to the difference in the regularization schemes \cite{Martin:1993yx},
\begin{equation}\label{eq:gaugemsdr}
	\delta g_1^{\text{reg}} = 0\,,\qquad \delta g_2^{\text{reg}} = \frac{g_2^3}{48\pi^2}\,,\qquad \delta g_3^{\text{reg}} = \frac{g_3^3}{32\pi^2}\,,
\end{equation}
and the $\delta g_i^{\text{thr}}$ are the threshold
corrections\footnote{The threshold corrections for the gauge couplings
  can e.g.\ be obtained from matching the $Z$ and $W^\pm$ boson pole
  masses as well as the running electromagnetic and strong couplings
  \cite{Braathen:2018htl,Athron:2017fvs}.} for $v \to 0$ including the
effects of the heavy particles which are integrated out in the EFT
\cite{Bagnaschi:2014rsa}. They read
\begin{align}
	\begin{split}
		\delta g_1^{\text{thr}} &= -\frac{g_1^3}{512\pi^2}\left[12\ln\frac{{\abs{\mueff^2}}}{Q^2} + 3\ln\frac{{m_{H^\pm}^2}}{Q^2} + \sum_{i=3}\left(3\ln\frac{m_{\tilde L_i}^2}{Q^2} + 6\ln\frac{m_{\tilde e_{R,i}}^2}{Q^2}\right)\right.\\
		&\qquad\qquad\qquad \left. {}+ \sum_{i=1}^3\left(\ln\frac{m_{\tilde Q_i}^2}{Q^2} + 8\ln\frac{m_{\tilde u_{R,i}}^2}{Q^2} + 2\ln\frac{m_{\tilde d_{R, i}}^2}{Q^2}\right)\right],
	\end{split} \label{eq:g1threshold}\\
	\delta g_2^{\text{thr}} &= -\frac{g_2^3}{192\pi^2}\left[8\ln\frac{M_2^2}{Q^2} + 4\ln\frac{{\abs{\mueff^2}}}{Q^2} + \ln\frac{{m_{H^\pm}^2}}{Q^2} + \sum_{i=3}\left(\ln\frac{m_{\tilde L_i}^2}{Q^2} + 3\ln\frac{m_{\tilde Q_i}^2}{Q^2}\right)\right], \label{eq:g2threshold}\\
	\delta g_3^{\text{thr}} &= -\frac{g_3^3}{192\pi^2}\left[12\ln\frac{M_3^2}{Q^2} + \sum_{i=1}^3\left(2\ln\frac{m_{\tilde Q_i}^2}{Q^2} + \ln\frac{m_{\tilde u_{R,i}}^2}{Q^2} + \ln\frac{m_{\tilde d_{R,i}}^2}{Q^2}\right)\right], \label{eq:g3threshold}
\end{align}
where $Q = \Qmatch$, and we introduced the effective $\mu$ parameter,
\begin{equation}
	\mueff = \lambda \frac{\vs}{\sqrt{2}} e^{i\phis},
\end{equation}
and {$m_{H^\pm}^2$} is given in \eqref{eq::treemassesC}. For the top Yukawa
coupling, we only require the matching relation at tree level for a
consistent calculation of the effective quartic Higgs coupling at
1-loop order,
\begin{equation}
	Y_t^{\nmssm,\DRbar} = Y_t^{\sm,\MSbar}/\sin\beta.
\end{equation}
The matching of the VEV will be discussed in
Sec.~\ref{sec:calculation:pole}, as it is not needed for the quartic
coupling matching in the unbroken phase, but for the pole mass matching.
With the NMSSM $\DRbar$
parameters $g_i^{\nmssm,\DRbar}$, $Y_i^{\nmssm,\DRbar}$, expressed
through their low-scale $\MSbar$ counterparts, and the other SUSY
input parameters of \eqref{eq:susyinput} at the scale $\Qmatch
\sim \msusy$ (\textbf{box 3}), we can then compute the loop
corrections in the NMSSM to the quartic Higgs coupling
$\lambda_h^{\nmssmMSbar}(\Qmatch)$ of the $h_u$-like Higgs
boson\footnote{By writing $\lambda_h^{\nmssmMSbar}$, we mean that,
  while the SUSY calculation is performed in the $\DRbar$ scheme, we
  express the Yukawa and gauge parameters via the low-scale $\MSbar$
  quantities.}, or to the pole mass of the $h_u$-like Higgs boson,
subtracting the SM corrections consistently and keeping only the pure
NMSSM contributions (\textbf{box 4}). 

Note that in our implementation, we allow the matching scale $\Qmatch$ to be
different from the input scale $\Qinp$ at which the SUSY parameters
of \eqref{eq:susyinput} are 
given. In the case of $\Qmatch \ne \Qinp$, we use the two-loop NMSSM RGEs as
calculated by
\SARAH~\cite{Staub:2009bi,Staub:2010jh,Staub:2012pb,Staub:2013tta,Machacek:1983tz,Machacek:1983fi,Machacek:1984zw,Sperling:2013eva,Sperling:2013xqa}
to run the SUSY parameters from $\Qinp$ to $\Qmatch$. As the Yukawa and gauge couplings are given at $\Qmatch$ (and not $\Qinp$) via their low-scale inputs, we thus have to implement the running of the SUSY parameters via an iterative procedure until all parameters converge at the matching scale. We note that due to the RGE running, a CP-violating phase of one of the soft-SUSY-breaking parameters typically induces CP-violating phases also for the other SUSY parameters, so that the CP-violating effects cannot be limited to only one sector of the model.

The obtained loop-corrected $\lambda_h^{\nmssmMSbar}(\Qmatch)$ is then
compared to the quartic Higgs coupling of the SM
$\lambda_h^{\smMSbar}(\Qmatch)$ of \eqref{eq:smrunparams} at the scale
$\Qmatch$ (\textbf{box 5}). If the absolute value of the relative
difference between the two 
quartic couplings, $\Delta \lambda \equiv (\lambda^{\sm,\MSbar} -
\lambda^{\nmssm,\MSbar})/\lambda^{\sm,\MSbar} $, is larger than
$\epsilon=10^{-5}$, we change the SM input $M_h$ and, starting again
from the top of \figref{fig:HiggsMassEFT}, iterate the procedure until
the precision goal $|\Delta\lambda|<\epsilon$ is reached.\footnote{Note that our iterative procedure is
slightly different from the one used in \cite{Athron:2016fuq,Staub:2017jnp}
where in the fifth step, the authors have set $\lambda_h^{\smMSbar}(\Qmatch)
= \lambda_h^{\nmssmMSbar}(\Qmatch)$ and then use the SM RGEs to run
$\lambda_h^{\smMSbar}$ down to the EW scale, where they compare to their
input value for the quartic Higgs coupling. Our procedure is, however, quite
similar to the one used in \cite{Bagnaschi:2022zvd}.} In order to efficiently
scan over different values of $M_h$, we use the bisection method for which
the procedure converges in logarithmic time. The found value of $M_h$ for
which the SM and NMSSM quartic Higgs couplings have the same value at the
matching scale $\Qmatch$ (within the precision goal) and at the considered loop order is then identified
with our predicted loop-corrected SM-like on-shell Higgs mass in the EFT
approach (\textbf{box 6}). 

To compare the two approaches for the matching conditions, in
the numerical discussion in Sec.~\ref{sec:numerical}, we will denote the
obtained values for the SM-like on-shell Higgs mass by $\MHqcm$ for
the quartic-coupling matching and by $\MHpmm$ for the pole-mass
matching, \ie the Roman superscript specifies which scalar $n$-point
function was used in the matching. 

%%%%%%%%%%%%%%%%%%%%%%%%%%%%%%%%%%%%%%%%%%%%%%%%%%%%%%%%%%%%
\subsection{Quartic-Coupling Matching Conditions}
\label{sec:calculation:quartic}
%%%%%%%%%%%%%%%%%%%%%%%%%%%%%%%%%%%%%%%%%%%%%%%%%%%%%%%%%%%%%
We present here our computation of the effective quartic Higgs
coupling at the tree and one-loop level after subtracting the SM
contributions, \ie the contributions from all particles which appear
in the SM EFT Lagrangian. To improve our prediction, we have included
two-loop QCD and mixed QCD-EW corrections {as well as the ${\cal
    O}(\alpha_t^2)$ corrections} in the limit of the
CP-conserving MSSM which are available in {\tt SUSYHD} \cite{PardoVega:2015eno} and Ref.~\cite{Bagnaschi:2019esc}, respectively\footnote{Note that the MSSM results in
  \cite{PardoVega:2015eno} assume a normalisation of the quartic
  interaction term in the SM Lagrangian of
  $-\frac{\lambda_h^\sm}{2}|H|^4$, so that we have to multiply all
  MSSM terms by a factor of $\frac12$ for our choice of
  normalisation.}. Our $\lambda_h^{\nmssmMSbar}(\Qmatch)$ can then be
written as the sum of the tree-level, one-loop, and MSSM two-loop parts,
\be\label{eq:qcmfull}
	\lambda_h^{\nmssmMSbar}(\Qmatch) = \lambda_h^{\nmssm,\text{tree}} + \Delta\lambda_h^{\NMSSM,\text{1l}} + \Delta\lambda_h^{\text{MSSM},\text{2l}}\,.
\ee
Note, that $\Delta\lambda_h^{\text{MSSM},\text{2l}}$ is not sensitive to the
CP-violating phases entering $\lambda_h^{\nmssm,\text{tree}}$ and
$\Delta\lambda_h^{\NMSSM,\text{1l}}$.
\subsubsection{Tree Level}

\begin{figure}[tp]
\centering
\begin{tabular}{ccc}
	\includegraphics[height=2.2cm]{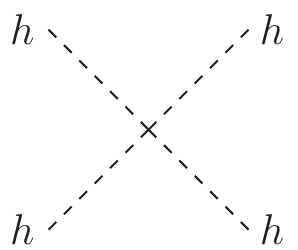} & \includegraphics[height=2.2cm]{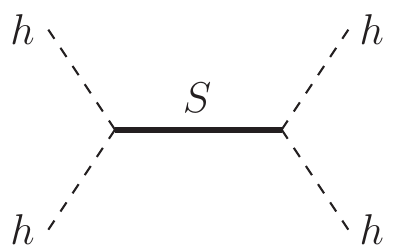} & \includegraphics[height=2.2cm]{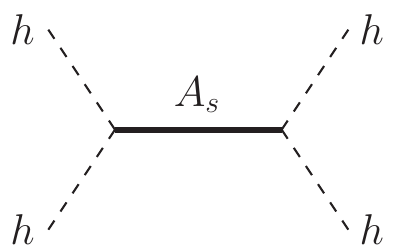}
\end{tabular}
\caption{Exemplary tree-level diagrams which contribute to the
  tree-level quartic-coupling matching. Thick (thin) lines denote
  heavy (light) particles which are (not) integrated out in the
  EFT. The dashed external lines correspond to the SM-like
  Higgs eigenstate {$h$}. The second and third diagrams
  exhibiting a heavy CP-even ($S$) or CP-odd ($A_s$) singlet in the
  propagator have to be taken into account also for $t$- and
  $u$-channel propagators. Note that the third diagram with the $A_s$ exchange is absent in the CP-conserving case.}
\label{fig:feynmandiagramstree}
\end{figure}

At the tree-level the {four-$h$ vertex receives contributions} from the
Feynman diagrams sketched in \figref{fig:feynmandiagramstree}. Taking
into account all tree-level contributions to the effective quartic
Higgs coupling, we get the following expression, 
\begin{subequations}
\begin{align}
	\lambda_h^{\nmssm,\text{tree}} &= \underbrace{\frac{1}{8}(g_1^2+g_2^2)\cos^22\beta}_{\text{MSSM $D$-terms}} {}~\,+{} \underbrace{\frac{1}{4}\abslam^2\sin^2 2\beta}_{\text{NMSSM $F$-terms}} \label{eq:treelevel_EFT1} \\
	\begin{split}
		&\quad {}- \frac{1}{48 |\kappa|^{2} {m_{H_s}^{2}} (3 {m_{H_s}^2  + m_{A_s}^2})} \Bigg(3|\kappa|^{2} {m_{H^\pm}^2} \left(1 - \cos4\beta\right) \\
		&\quad \underbrace{\qquad\qquad\qquad\qquad\quad {}+ (3 {m_{H_s}^2  +m_{A_s}^2})\left(|\kappa| |\lambda| \cos\varphi_y  \sin2 \beta - 2|\lambda|^{2}\right)\Bigg)^{2}}_{\text{$s$/$t$/$u$-channel $S$}}
	\end{split} \label{eq:treelevel_EFT2} \\
	&\quad {} \underbrace{-\frac{3}{16 {m_{A_s}^{2}}} |\lambda|^{2} (3 {m_{H_s}^2  + m_{A_s}^2})\sin^{2}2 \beta \sin^{2}\varphi_{y}}_{\text{$s$/$t$/$u$-channel $A_s$}}\,, \label{eq:treelevel_EFTCL}
\end{align}
\end{subequations}
where the origin of each term is explained by the corresponding text underneath. While the two contributions of \eqref{eq:treelevel_EFT1} directly originate from the scalar potential in the NMSSM, the terms of Eqs.~(\ref{eq:treelevel_EFT2}) and (\ref{eq:treelevel_EFTCL}) arise when the heavy CP-even and CP-odd singlets, appearing as intermediate states in the $s$-, $t$-, and $u$-channels, are integrated out. {For simplicity of the expressions in Eq.~(\ref{eq:treelevel_EFT2}) and Eq.~(\ref{eq:treelevel_EFTCL}) we have used the masses
    $m_{H_s}^2, m_{A_s}^2, m_{H^\pm}^2$ instead of $v_s, \text{Re}A_\kappa$ and  $\text{Re}A_\lambda$.} Compared to the CP-conserving case presented in \cite{Bagnaschi:2022zvd}, there are additional contributions from the CP-odd singlet field $A_s$, corresponding to the term in \eqref{eq:treelevel_EFTCL}. This term will vanish if $\sin\varphi_{y} = 0$, \ie if there is no CP-violation at tree-level in the Higgs sector.

\subsubsection{One Loop}
\label{sec:calculation:quartic:oneloop}
\begin{figure}[tp]
\centering
\begin{tabular}{cccc}
	\includegraphics[height=2cm]{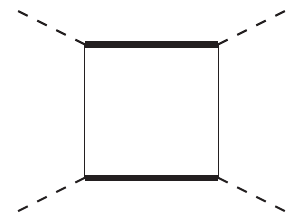} &
    \includegraphics[height=2cm]{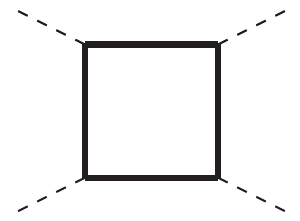} &
    \includegraphics[height=2cm]{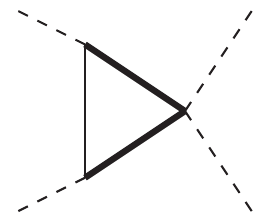} &
    \includegraphics[height=2cm]{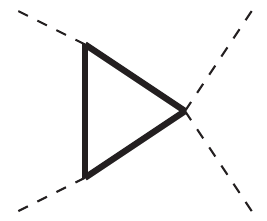} \\
    a) & b) & c) & d) \\
    \includegraphics[height=2cm]{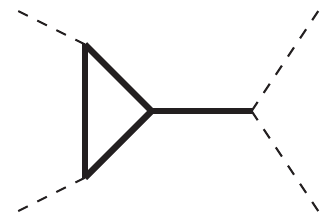} &
    \includegraphics[height=2cm]{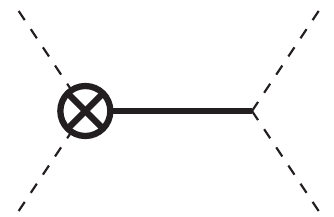} &
    \includegraphics[height=2cm]{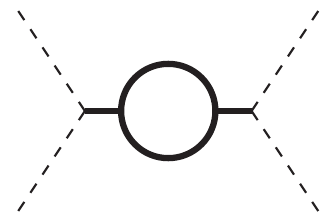} &
    \includegraphics[height=2cm]{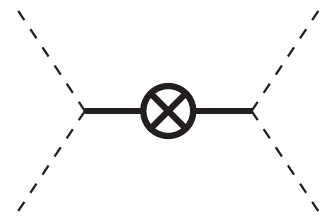}\\
    e) & f) & g) & h)
\end{tabular}
\caption{Selection of one-loop diagrams contributing to the one-loop
  quartic-coupling matching. Thick (thin) lines denote heavy (light)
  particles which are (not) integrated out in the EFT. The diagrams
  a)-e) and g) show exemplary box, vertex correction and self-energy
  diagrams, while the diagrams f) and h) represent the vertex and mass
  counterterm diagrams, which implicitly contain contributions from tadpoles.}
\label{fig:feynmandiagramsoneloop}
\end{figure}
At the one-loop level, the matching condition of the
quartic Higgs coupling receives corrections from
diagrams involving at least one heavy SUSY particle.
In \cref{fig:feynmandiagramsoneloop} we show example diagrams, where thick
lines correspond to heavy SUSY particles and thin lines to light SM fields.
We divide the one-loop corrections
into the following six pieces,
\be\label{eq:lamthrescorrection}
	\Delta\lambda_h^{\NMSSM,\text{1l}} = \Delta\lambda_{\Box} + \Delta\lambda_{\triangle}+ \Delta\lambda_{\rm SE}+ \Delta\lambda_{\rm CT} + \Delta\lambda_{\text{reg}} + \Delta\lambda_{\text{gauge-thr}}\,.
\ee
The first four terms correspond to the box, vertex correction, self-energy and
counterterm contributions, respectively. The last two terms correspond to the
shift induced by the different regularization schemes used in the
NMSSM and the SM
calculations and the contributions from the matching of the gauge couplings.
In all above contributions, diagrams with only SM particles (light states) in
the internal lines are discarded, since they belong to the SM contributions
and would cancel in the matching condition.
All diagrams which contain at least one SUSY particles (heavy state) in the
internal lines are kept. 
Since the momenta of the external Higgs bosons are set to be zero, all four-,
three- and two-point one-loop integrals can be reduced to vacuum
integrals.
For the calculation of the diagrams we make use of the mass- and mixing-matrices 
of the NMSSM in the unbroken phase of the EW symmetry as specified in \cref{sec:NMSSMtree-level}.

Box diagrams such as shown in
  \cref{fig:feynmandiagramsoneloop} a) to d) are
of similar structure as those which are encountered in the MSSM with the
difference that the additional NMSSM degrees-of-freedom are present in the
loop. They constitute a separately UV-finite subset. 
An entirely new type of correction arises in the (complex) NMSSM due to
the presence of the non-local contributions at tree-level, \cf
\cref{fig:feynmandiagramstree} second and third diagrams. At the one-loop level, these
diagrams receive vertex corrections $\Delta
 \lambda_{\triangle}$, propagator corrections $\Delta
\lambda_{\text{SE}}$, and corresponding counterterm corrections $\Delta
\lambda_{\text{CT}}$ shown exemplary in
\cref{fig:feynmandiagramsoneloop} e) to h). 
The vertex corrections originate from
the exchanges of a CP-even or a CP-odd singlet which can be written
as 
\be \Delta\lambda_{\triangle} = -\fr{g_{hhS}}{ {m_{H_s}^2}} \Delta g_{hhS} -\fr{g_{hhA_s}}{ {m_{A_s}^2}} \Delta g_{hhA_s}   \ee   
where $\Delta g_{hhS}, \Delta g_{hhA_s}$ are genuine one-loop
contributions to the triple Higgs vertices $h-h-S$ and $h-h-A_s$, 
respectively, and the trilinear couplings $g_{hhS}, g_{hhA_s}$ are
given in \eqref{eq:trilinearCLsa} and 
\eqref{eq:trilinearCLsb}, respectively.
The propagator corrections
come from the one-loop self-energy diagrams of the heavy CP-even and CP-odd singlet states. They can be expressed as
\be
\Delta\lambda_{\rm SE} = -\frac12 \braket{\fr{g_{hhS}}{ {m_{H_s}^2}}}^2 \Sigma_{SS}(0) -
\frac12 \braket{\fr{g_{hhA_s}}{ {m_{A_s}^2}}}^2 \Sigma_{A_sA_s}(0) -
\fr{g_{hhA_s}g_{hhS}}{ {m_{A_s}^2 m_{H_s}^2}} \Sigma_{SA_s}(0)
\ee
where $\Sigma_{xy}(0) $ ($x,y=S,A_s$) are the self-energies of the transitions $x\to y$ 
evaluated at zero external momentum in the limit $v\to0$.
{For the counterterm contributions, we note that all
    parameters} {appearing explicitly in the tree-level expression in
    \eqref{eq:treelevel_EFT1} to \eqref{eq:treelevel_EFTCL} are
 renormalised in the $\DRb$ scheme. For the tadpoles, we instead use OS-like
 renormalisation conditions, such that the minimum of the scalar tree-level
 potential is identical to the loop-corrected one. At the same
 time, this choice removes all heavy-light mixing contributions
 on the external legs of \cref{fig:feynmandiagramstree}. In the case of the
 $h$-$H$ mixing, this is equivalent to working directly in the unbroken phase
 and using an OS-like counterterm for $\tan\beta$ for the removal of the
heavy-light mixing (\cf \cite{Bagnaschi:2014rsa}). In the CP-violating case,
one also has to deal with $h$-$A$ mixing which is removed via an OS-like
condition for the $t_{a_d}$ tadpole.
}
 As a consequence, the counterterm of
 $\lambda_h^{\NMSSM}$ gets UV-finite contributions
 only from the {diagonal} wave-function renormalisation constant of the external
 Higgs fields, the singlet tadpoles and the tadpole of the field
 $a_d$. The latter enters via
 the counterterm of $\text{Im}(A_{\lambda})$. We can express
 $\Delta\lambda_{\rm CT}$ as
\begin{align}\
    \Delta\lambda_{\rm CT} &= 2\lambda_h^{\NMSSM,\text{tree}} \delta^{(1)}Z_{h} 
  - \fr{\delta^{(1)} t_{h_s}}{2v_s} \braket{\fr{g_{hhS}^2}{ {m_{H_s}^4}} + \fr{g_{hhA_s}^2}{ {m_{A_s}^4}}  }\nonumber\\
                         & + \fr{\delta^{(1)} t_{a_s}}{2v_s} \braket{3\fr{g_{hhA_s}^2}{ {m_{A_s}^4}} \tan\varphi_w 
                         - {4}\fr{g_{hhA_s}g_{hhS} }{{m_{A_s}^2 m_{H_s}^2}}  - \fr{g_{hhS}^2 }{{m_{H_s}^4}}\tan\varphi_w } \nonumber\\
                         & \left.+\delta^{(1)}\text{Im}(A_\lambda)\left( \frac{g_{hh{A_s}}}{{m_{A_s}^2}}\frac{\partial g_{hh{A_s}}}{\partial \text{Im}(A_\lambda)} +  \frac{g_{hhS}}{{m_{H_s}^2}}\frac{\partial g_{hhS}}{\partial \text{Im}(A_\lambda)}\right)\right|_{\text{min}}
   \label{eq:CTcontribution}
\end{align}
where the tadpole counterterms $\delta^{(1)}t_{h_s,a_s}$ originate from
mass-counterterm inserted diagrams, \cref{fig:feynmandiagramsoneloop} h), and $\delta^{(1)}\text{Im}A_\lambda$ from
the vertex-counterterm inserted diagrams in
\cref{fig:feynmandiagramsoneloop} f). The subscript 'min' indicates
  that the expression is evaluated at the minimum of the potential
  (where Im$(A_\lambda)$ is no input anymore), \ie
  using the solutions for the tadpole equations.
The Higgs field wave-function renormalisation constant at one-loop order is given by
\be \delta^{(1)}Z _{h} = - \fr{d\Sigma_{hh}}{dp^2}\bigg|_{p^2=0} \ee
and the trilinear Higgs couplings related to the singlet states as
well as their partial derivatives w.r.t.~to $\text{Im}(A_\lambda)$ are
\begin{subequations}
    \begin{align}
        g_{hhS} & = \frac{1}{2v_s}\left(
        |\kappa||\lambda|v_s^2\sin(2\beta)\cos\varphi_y -2|\lambda|^2v_s^2 +
    {m_{H^\pm}^2}\sin^2(2\beta)\right)\,, \label{eq:trilinearCLsa} \\
            g_{hhA_S} & = -\frac{3}{2} |\kappa||\lambda| v_s
                        \sin(2\beta)\sin\varphi_y\,, \label{eq:trilinearCLsb}
      \\
        \frac{\partial g_{hhS}}{\partial \text{Im}(A_\lambda)} & =
        -\frac{|\lambda|}{{\sqrt{2}}}\sin(2\beta)\sin(\varphi_w-\varphi_y)\,,\\
        \frac{\partial g_{hhA_S}}{\partial \text{Im}(A_\lambda)} & =
        -\frac{|\lambda|}{{\sqrt{2}}}\sin(2\beta)\cos(\varphi_w-\varphi_y) \,.
    \label{eq:trilinearCLs}
    \end{align}
\end{subequations}
All other counterterm diagrams are of $\mathcal{O}(v^2/\msusy^2)$ or higher and neglected in
the quartic coupling matching. 
Note, that the couplings in
\cref{eq:trilinearCLsa}-\cref{eq:trilinearCLs} are given at the
minimum of the tree-level potential while the derivatives of the couplings have to be
evaluated before using the tadpole solutions. Finally, the counterterm
of $\text{Im}(A_\lambda)$ is related to the counterterm of the
$a_d$-tadpole via the tree-level tadpole solution as
\begin{equation}
    \delta^{(1)}\text{Im}(A_\lambda) = \frac{\sqrt{2}}{|\lambda|v_s
    \cos(\varphi_w-\varphi_y)\sin\beta}\frac{\delta^{(1)}t_{a_d}}{v}\,.
        \label{eq:dImAlam}
\end{equation}
The above contributions have been obtained by two independent calculations.
One calculation relies on \SARAH \cite{Gabelmann:2018axh} to compute the
expression for the effective quartic Higgs self-coupling and the other one uses {\tt FeynArts-3.11} \cite{Kublbeck:1990xc,Hahn:2000kx} and {\tt FeynCalc-9.3} 
\cite{FeynCalc,Shtabovenko:2016sxi,Shtabovenko:2020gxv}. Note that in
\cite{Bagnaschi:2022zvd}, the authors found that in the old version of
\SARAH, a term related to the singlet tadpole was missing. After implementing
generic tadpoles into a private version of \SARAH and computing the singlet
tadpole contributions, the results from the two calculations were found to
agree. {In addition, we performed a non-trivial check of the scale dependence of 
$\lambda_h^{\nmssm,\text{tree}} + \delta\lambda_h^{\NMSSM,\text{1l}}
$. We tested that the slope of its scale dependence
is the same as the slope of the scale dependence of the SM quartic coupling at one-loop level.}

At the  beginning of this section, we discussed that the quartic-coupling
    matching is performed in the limit of the unbroken phase, $v\to0$. This
    is also the general strategy employed in \cite{Braathen:2018htl,Gabelmann:2018axh}.
    However, from \cref{eq:dImAlam} we can see, that the actual limit $v\to0$
    has to be taken with care in the CP-violating case and requires the expansion 
    of tadpoles up to $\mathcal{O}(v)$. The explicit expansion of
    $\delta^{(1)}t_{a_d}/v$ up to $\mathcal{O}((v^2/\msusy^2)^0)$ is derived in
    \cref{app:tadexpand}.
    It should be stressed, that this situation was not
    encountered before in e.g. calculations within the CP-violating MSSM: In
    the MSSM all diagrams that contain $\delta^{(1)}t_{i}/v$-terms are
    suppressed by additional powers of $v$.
    {Therefore, the crucial difference to the pure
        CP-conserving case is that,
    in the CP-violating case, the effects of the $h$-$A$ mixing cannot be
    entirely removed from the matching condition
    via the introduction of a finite counterterm. In the language of
Ref.~\cite{Bagnaschi:2014rsa} this means, while contributions from the
counterterm of $\tan\beta$ exactly cancel the $h$-$H$ mixing in all
diagrams (no $\delta\tan\beta$ is left over), the counterterm
contributions from the mixing angle that describes the $h$-$A$ mixing have a
left-over finite effect on the matching condition via vertex counterterms as in \cref{fig:feynmandiagramsoneloop} f).}

Finally, we have to take into account the shift due to the NMSSM calculation
being done in the $\DRbar$ scheme and the SM contributions being
calculated in the $\MSbar$ scheme, 
\begin{equation}\label{eq:regshift}
	\Delta\lambda_{\text{reg}} = \frac{1}{64\pi^2}\left[\frac{g_2^4}{3}\cos^2 2\beta - \frac12\left(g_1^4 + 2g_1^2 g_2^2 + 3g_2^4\right)\right].
\end{equation}
There are two contributions to this shift: 
The first term in \eqref{eq:regshift} accounts for the $\DRbar$--$\MSbar$
conversion of the gauge couplings given in \eqref{eq:gaugemsdr},
since we express all gauge and Yukawa
parameters in the threshold corrections to the quartic coupling in
\eqref{eq:lamthrescorrection} by their $\MSbar$ values of the low-energy EFT.
An additional contribution arises due to diagrams involving quartic couplings between two
Higgs and two gauge bosons \cite{Gabelmann:2018axh}. As explained above, we discard all
diagrams containing only SM-fields \ie we implicitly subtract these pieces in the
$\DRbar$ scheme. However, the subtraction-term strictly would need to be computed in the
\MSbar scheme using dimensional regularization rather than dimensional
reduction. The second
term of \eqref{eq:regshift} remedies this mismatch between the two schemes in
the subtracted SM contributions.

In addition to the regularisation-scheme shift to the gauge couplings, we also take into
account the one-loop gauge thresholds from
the matching of the gauge couplings between the NMSSM and the SM, 
\begin{align}
    \Delta\lambda_{\text{gauge-thr}} & = \lambda^{\nmssm,\text{tree}}_h(g_i\to
                                       g_i + \delta g_i^{\text{thr}})
- \lambda^{\nmssm,\text{tree}}_h(g_i)
                                       \nonumber \\ 
                                       &= \frac14\left(g_1\delta
                                       g_1^{\text{thr}} + g_2\delta
                                   g_2^{\text{thr}}\right)\cos^2 2\beta\, +
                                   \order{(\delta g_i)^2},
    \label{eq:deltalamgaugethresh}
\end{align}
that arise when integrating out all heavy degrees of freedom of the
NMSSM. The $\delta g_i^{\text{thr}}$ are defined in
Eqs.~(\ref{eq:g1threshold}) and (\ref{eq:g2threshold}). As the singlet states in the NMSSM do not influence the gauge couplings, the shifts of Eqs.~(\ref{eq:regshift}) and (\ref{eq:deltalamgaugethresh}) are identical to the ones of the MSSM \cite{Bagnaschi:2014rsa}.

%%%%%%%%%%%%%%%%%%%%%%%%%%%%%%%%%%%%%%%%%%%%%%%%%%%%%%%%%%%%%%%%%%%%%%%%%%%
\subsection{Pole-Mass Matching Conditions}
\label{sec:calculation:pole}
%%%%%%%%%%%%%%%%%%%%%%%%%%%%%%%%%%%%%%%%%%%%%%%%%%%%%%%%%%%%%%%%%%%%%%%%%%
The pole-mass matching scheme is defined by the condition that the pole-mass
of the SM-like Higgs mass eigenstate in the NMSSM\footnote{We remind that we
consider the Higgs state to be SM-like if it is predominantly made up of the
$h_u$ component.} is equal to the SM one,
\begin{equation}\label{eq:polemasscondition}
	(M_h^\sm)^2 \stackrel{!}{=} (M_h^\nmssm)^2\,.
\end{equation}
The defining equation for the pole-mass in the SM reads
\begin{equation}\label{eq:polemasssm}
    0=p^2-(m_h^\sm)^2 + \mbox{Re}\hat\Sigma_h^\sm(p^2=(M_h^\sm)^2)\,.
\end{equation}
Here, $m_h^\sm$ denotes the running $\MSbar$ mass of the SM Higgs boson, \ie
the tree-level mass expressed through $\MSbar$ parameters, and
$\hat\Sigma_h^\sm$ is the renormalised one-loop self-energy
calculated at a fixed order in the $\MSbar$ renormalisation
scheme. The solution  $p^2 = (M_h^\sm)^2$, which fulfills
\cref{eq:polemasssm} in general has to be found iteratively.
The calculation of the pole mass of the SM-like Higgs
boson in the NMSSM on the right-hand side of \eqref{eq:polemasscondition} is
more complicated due to the appearance of multiple Higgs states and their
mixing. In general, the pole masses of the Higgs bosons in the NMSSM are the
eigenvalues of the loop-corrected Higgs mass matrix
$\mathcal{M}_H$,\footnote{Here and above for the SM, we use the same sign
    convention for the self-energy corrections as in
\cite{Ender:2011qh,Graf:2012hh,Muhlleitner:2014vsa,Dao:2019qaz,Dao:2021khm}.}
\begin{equation}\label{eq:polemassnmssm}
	\big(\mathcal{M}_H\big)_{ij} =  (m_{h_i}^\nmssm)^2 \delta_{ij}- \mbox{Re}\hat\Sigma_{ij}^\nmssm(p^2)\qquad\text{for } i,j = 1, \ldots, 5\,,
\end{equation}
where $m_{h_i}^\nmssm$ is the tree-level mass  of $h_i$ (expressed
through the running $\DRbar$ parameters). In \NMSSMCALC, the squared
tree-level masses are obtained after
  factorising the Goldstone boson and then diagonalizing the
  tree-level mass matrix. The eigenvalues
    are the squared masses that are ordered by ascending mass values. The
$\hat\Sigma_{ij}^\nmssm(p^2)$ in \eqref{eq:polemassnmssm} denote the
$\DRbar$-renormalised self-energies of the transitions $h_i\to h_j$ at the momentum squared $p^2$. Similarly to the SM, we take only the real part of the renormalised self-energy for our following discussions. The $i$th loop-corrected pole mass, $(M_{h_i}^\nmssm)^2$,
 is then obtained
by iteratively diagonalizing the mass matrix $\mathcal{M}_H$
until $p^2$ approaches $(M_{h_i}^\nmssm)^2$.
However, both the diagonalisation of the loop-corrected mass matrix
and the iterative procedure mix different orders of perturbation theory.
This mixing can spoil the cancellation of large logarithms by inducing higher powers of
$\ln(v^2/\msusy^2)$-terms. Thus, an iterative procedure may induce a large theory
uncertainty.

In order to obtain a consistent one-loop expansion which is free of
any powers of $\ln(v^2/\msusy^2)$, we approximate the loop-corrected
SM-like Higgs pole mass: We work in the tree-level mass
basis as in \eqref{eq:polemassnmssm}. In the following, we assume that
the SM-like Higgs state always corresponds to $h_1$ with the
tree-level mass $m_h^\nmssm \equiv m_{h_1}^\nmssm$,
\begin{equation}\label{eq:polemassnmssm1l}
    (M_h^\nmssm)^2 \equiv \big(\mathcal{M}_H\big)_{11} = (m_{h}^\nmssm)^2 -
    \mbox{Re}\hat\Sigma_{11}^\nmssm\big(p^2\big)\,,
\end{equation}
\ie we consider only the diagonal element corresponding to the
lightest state.\footnote{For the pole-mass matching implemented in
  {\tt NMSSMCALC} the SM-like Higgs is not required to be the lightest
Higgs state ($h_1$), but could also be a heavier state. However, in
such scenarios our EFT approach may not be valid any more and the
result has to be taken with care.}
At the one-loop level, it is consistent to ignore all mixing self-energy
contributions since the diagonalisation of the loop-corrected mass matrix 
only involves terms proportional to the product of two or more one-loop self-energies.\footnote{It can
    be seen that, when diagonalizing \eqref{eq:polemassnmssm} and then
    expanding in the self-energies, the off-diagonal self-energy corrections
with $i \ne j$ only contribute at two-loop order or higher.}
In order to avoid further mixing of orders in the iteration, we take only the
first iteration of the pole-mass equation where the momentum squared is set to be equal to 
the tree-level mass squared,
$(M_h^X)^2=(m_h^{X})^2-\mbox{Re}\hat{\Sigma}_h^X(p^2=(m_h^{X})^2)$ for
$X=\{\sm,\nmssm\}$. 
The matching of the pole-masses in the two theories,
\cref{eq:polemasssm,eq:polemassnmssm1l},
is then performed successively: We first evaluate the matching condition at the tree-level
which yields $(m_h^\sm)^2 = (m_h^\nmssm)^2$.
Using this in the one-loop matching condition, \cref{eq:polemasscondition},
we find
\begin{equation}\label{eq:polemasscondition1l}
	(m_h^\sm)^2 - \mbox{Re}\hat\Sigma_h^\sm\big((m_h^\nmssm)^2\big) \stackrel{!}{=} (m_h^\nmssm)^2 - \mbox{Re}\hat\Sigma_h^\nmssm\big((m_h^\nmssm)^2\big)\,,
\end{equation}
where we write $\hat\Sigma_h^\nmssm \equiv \hat\Sigma_{11}^\nmssm$ to
simplify the notation. For a consistent expansion of the
matching condition in $v^2/\Msusy^2$, {the real part of the self-energies can 
furthermore be expanded} around small arguments, 
\begin{equation}\label{eq:sigmavexpand}
	\mbox{Re}\hat{\Sigma}_h^X\big((m_h^\nmssm)^2\big) = \hat{\Sigma}_h^X(0)
        + (m_h^\nmssm)^2\, \hat{\Sigma}^{X\prime}_h(0) +
        \mathcal{O}\big((m_h^\nmssm)^4\big)\,.
      \end{equation}
Using the expansion of Eq.~(\ref{eq:sigmavexpand}) in
\eqref{eq:polemasscondition1l}, the large  
logarithms $\ln(\Msusy/v)$ at the matching scale $\Qmatch \sim \Msusy$ are
the same on the left- and right-hand sides, and the matching condition as a
whole is thus free of these logarithms.      
We finally note that, contrary to e.g.\ \cite{Athron:2016fuq}, no explicit
tadpole contributions are appearing in \eqref{eq:polemasscondition1l}, as we
define the minimum of our scalar potential to correspond to the tree-level
one at all orders (``on-shell tadpole scheme''). Thus, the tadpole
contributions are taken into account implicitly via the mass-matrix
counterterm included in the renormalised self-energies (see \eg Appendix~G of
\cite{Dao:2019qaz}, with all counterterms, other than the ones for the
tadpoles, set to zero due to the $\MSbar/\DRbar$ scheme used in our
calculation).

The tree-level relation between the $\MSbar$ mass $m_h^\sm$ and the quartic
coupling parameter of the SM Lagrangian in the $\MSbar$ scheme reads
\begin{equation}\label{eq:treesmlameff}
	(m_h^\sm)^2 = 2(v^\sm)^2\,\lambda_h^{\sm,\pmii}\,,
\end{equation}
where $v^\sm$ is the VEV of the SM in the $\MSbar$
scheme.\footnote{{The $v^\sm$ is gauge dependent,
    which, however, we have not investigated in this study.
    We work in the 't Hooft-Feynman gauge throughout all parts of the
    calculation. Some {discussions and investigations of the
    gauge dependence can be found in} \cite{Dao:2019nxi,Domingo:2020wiy}.}} Using this
relation, \eqref{eq:polemasscondition1l} implicitly defines a matching
condition for the quartic coupling and therefore allows to extract a
prediction for the effective quartic coupling of the SM-like Higgs in the
NMSSM at the matching scale, which we will denote for consistency with the
above notation as $\lambda_h^{\nmssmMSbar,\pmii}$. Contrary to the
quartic-coupling matching, the appearance of the VEV in
\eqref{eq:treesmlameff} prevents us from setting $v \to 0$ right from the
beginning. The pole-mass matching thus requires a double expansion in the
loop order as well as in $v^2/\Msusy^2$. Solving \eqref{eq:polemasscondition1l} for
the quartic coupling appearing in \eqref{eq:treesmlameff}, the pole-mass
matching condition can then be cast in a similar form as the quartic-coupling
matching:
\begin{equation} \label{eq:lambdapolemass}
	\lambda_h^{\nmssmMSbar,\pmii} = \lambda_h^{\nmssm,\text{tree},\pmii} + \Delta \lambda_h^{\nmssm,\text{1l},\pmii} + \Delta \lambda_h^{\text{MSSM},\text{2l}}\,,
\end{equation}
where we again improve our result by adding the two-loop MSSM
corrections of \cite{PardoVega:2015eno,Bagnaschi:2019esc} as in the case of the
quartic-coupling matching in \eqref{eq:qcmfull}. We introduce the
additional superscript $\pmii$ in order to distinguish the effective
quartic coupling obtained via the pole-mass matching approach from the
corresponding one of the quartic-coupling matching approach in
\eqref{eq:qcmfull}, since the former includes also partial $v^2/\Msusy^2$
terms. 

The pole mass obtained in the NMSSM depends on the VEV as defined in the high-energy theory, $v^\nmssm$. As we want to express the matching condition only in terms of either $v^\nmssm$ or $v^\sm$, it is thus also required to match the VEV and take into account the shift between the two,
\begin{equation}\label{eq:vevshift}
	(v^\sm)^2 = (v^\nmssm)^2 + \delta v^2 = (v^\nmssm)^2\left(1 + \frac{\delta v^2}{v^2}\right)\,.
\end{equation}
In the last term of \eqref{eq:vevshift}, we do not distinguish between $v^\sm$ and $v^\nmssm$, as the difference is of higher order. The threshold correction $\delta v^2$ can be obtained from matching \eg the $Z$-boson pole mass at one loop in the SM and the NMSSM, which can  futhermore be related through Ward identities to the wave-function renormalisation of the Higgs boson \cite{Braathen:2018htl},
\begin{equation}\label{eq:vevshiftward}
	\frac{\delta v^2}{v^2} = \left[\hat{\Sigma}^{\nmssm\prime}_h(0) - \hat{\Sigma}^{\sm\prime}_h(0)\right] + \mathcal{O}(v^2/\Msusy^2)\,,
\end{equation}
where $\hat{\Sigma}^{X\prime}_h$ denotes the first derivative of the
self energy with respect to the four-momentum squared.

Analogously to the quartic-coupling matching, we express the gauge and Yukawa
couplings entering the NMSSM self-energies in terms of the
$\MSbar$ quantities of the low-energy effective theory. Thus,
we use a tree-level matching of the Yukawa couplings due
to their appearance only starting from one loop, and the one-loop matching
for the gauge couplings of \eqref{eq:gaugematching}. {If we were to simply
plug \eqref{eq:gaugematching} into the tree-level mass term}
$(m_h^\nmssm)^2$ of \eqref{eq:polemasscondition1l}, we would induce partial
two-loop contributions and higher (possibly spoiling the cancellation of
large logarithms). In order to include the gauge shifts
consistently at the one-loop order, we expand the tree-level mass in
$\delta g_i = \delta g_i^{\text{reg}}+\delta g_i^{\text{thr}}$ ($i=1,2$)
to first order
\begin{align}
    \left(m_h^{\nmssm}\right)^2 &\equiv \left(m_h^\nmssm(g_i^{\nmssmDRbar} \to g_i^{\smMSbar} + \delta g_i)\right)^2 \nonumber \\
                                & = \left(m_h^{\nmssm,\text{tree}}(g_i^{\smMSbar})\right)^2 + \delta^{\text{gauge}} m_h^2 + \order{(\delta g_i)^2} 
\end{align}
The pole-mass matching involves a rotation into the mass basis,
\begin{align}
    \delta^{\text{gauge}} m_h^2   &= \left(\mathcal{R}^{H}(v)\delta^{\text{gauge}}\bm{M}_H\mathcal{R}^{H^T}(v)\right)_{11}\,,
\label{eq:massgaugeshifts1l}
\end{align}
where $\mathcal{R}^{H}(v)$ are the rotation matrices that diagonalise the
squared neutral Higgs mass matrix,
$\mathcal{R}^{H}(v)\bm{M}_H\mathcal{R}^{H^T}(v)=(m_{h_i}^{\nmssm})^2\delta_{ij}$, in the broken phase (\ie not
as in \cref{eq:rot} but
for the case of non-zero $v$) and 
\begin{equation}
    \delta^{\text{gauge}}\bm{M}_H = \sum_{i=1,2}\left. (\delta g_i^{\text{thr}} + \delta
        g_i^{\text{reg}})\frac{\partial }{\partial g_i^{\nmssmDRbar}} \bm{M}_H
    \right|_{g_i^{\nmssmDRbar}\to g_i^{\smMSbar}} \,.
\end{equation}
With this treatment we guarantee that all logarithms of the form $\ln
v/\Qmatch$ appearing in the electroweak corrections can
cancel in the pole-mass matching while we still correctly take into account
the leading corrections from the $\DRbar-\MSbar$ conversion. However, in the numerical analysis we found
that these effect are numerically small compared to \eg the stop
contributions.

\subsubsection{Tree Level}
Keeping only the lowest-order terms of \eqref{eq:polemasscondition1l} and setting the self-energy corrections to zero, we obtain together with \eqref{eq:treesmlameff} the condition
\begin{equation}\label{eq:treematch}
    (m_h^\sm)^2 = 2(v^\sm)^2\, \lambda_h^{\sm,\pmii} \stackrel{!}{=} (m_h^\nmssm)^2\,.
\end{equation}
At lowest order, we do not have to take into account the threshold
corrections to the VEV, so we can set $v^\sm = v^\nmssm$ with $\delta
v^2 = 0$. Furthermore, the right-hand side of \eqref{eq:treematch} is
expressed via the SUSY and the low-energy $\MSbar$ gauge parameters
only, so that we also set $\delta m^2_{h,\text{gauge}}$ of
\eqref{eq:massgaugeshifts1l} to zero. Equation (\ref{eq:treematch}) thus becomes the tree-level matching relation for the effective quartic coupling:
\begin{equation}\label{eq:treepolequartic}
    \lambda_h^{\sm,\pmii} \stackrel{!}{=} \frac{(m_h^\nmssm)^2}{2(v^\nmssm)^2} \equiv \lambda_h^{\nmssm,\text{tree},\pmii}.
\end{equation}
We have checked that, by analytically diagonalizing the tree-level
Higgs mass matrix in the NMSSM with full VEV dependence to obtain
$m_h^\nmssm$ and then expanding \eqref{eq:treepolequartic} in
$v^2/\Msusy^2$, the same expression as in
Eqs.~(\ref{eq:treelevel_EFT1})--(\ref{eq:treelevel_EFTCL}) is obtained
at the lowest order $\mathcal{O}((v^2/\Msusy^2)^0)$.

\subsubsection{One Loop}
At one-loop order, we take into account the one-loop self energies in \eqref{eq:polemasscondition1l}, and obtain after plugging in \eqref{eq:treesmlameff}:
\begin{equation}
	2(v^\sm)^2\, \lambda_h^{\sm,\pmii} - \hat\Sigma_h^\sm\big((m_h^\nmssm)^2\big) \stackrel{!}{=} (m_h^\nmssm)^2 - \hat\Sigma_h^\nmssm\big((m_h^\nmssm)^2\big)\,,
\end{equation}
which results in the expression for the effective quartic coupling:
\begin{equation}\label{eq:onelooppolequartic}
	\begin{split}
		\lambda_h^{\sm,\pmii} &\stackrel{!}{=} \frac{1}{2(v^\sm)^2}\left[(m_h^\nmssm)^2 - \hat\Sigma_h^\nmssm\big((m_h^\nmssm)^2\big) + \hat\Sigma_h^\sm\big((m_h^\nmssm)^2\big)\right]\\
		&\equiv \lambda_h^{\nmssm,\text{tree},\pmii} + \Delta\lambda_h^{\nmssm,\text{1l},\pmii} \,.
	\end{split}
\end{equation}
To extract the leading terms in the expansion of $v^2/\Msusy^2$, we replace $v^\sm$ by $v^\nmssm$ according to Eqs.~(\ref{eq:vevshift}) and (\ref{eq:vevshiftward}), and we expand the self-energies according to \eqref{eq:sigmavexpand}, so that eventually, we obtain for the one-loop contribution to the matching condition:
\begin{equation}\label{eq:polemh1}
	\Delta\lambda_h^{\nmssm,\text{1l},\pmii} = -\frac{1}{2(v^\nmssm)^2}\left[\Delta\hat\Sigma_h + 2(m_h^\nmssm)^2 \Delta\hat\Sigma^{\prime}_h\right],
\end{equation}
where we have introduced the abbreviation $\Delta\hat\Sigma^{(\prime)}_h \equiv \hat\Sigma^{\nmssm(\prime)}_h(0) - \hat\Sigma^{\sm(\prime)}_h(0)$. The last term of \eqref{eq:polemh1} can for $v \to 0$ immediately be identified with the first term of \eqref{eq:CTcontribution}, corresponding to the wave-function-renormalisation contribution. In the tree-level piece of \eqref{eq:onelooppolequartic}, given via \eqref{eq:treepolequartic}, we apply the replacement of \eqref{eq:massgaugeshifts1l} in order to take into account the $\DRbar$-$\MSbar$ and gauge threshold shifts consistently at the one-loop order. Then, \eqref{eq:onelooppolequartic} is again expressed through the SUSY input parameters as well as the $\MSbar$ parameters of the low-energy theory only.

The self-energies and wave-function renormalisation contributions required for the calculation of the pole-mass matching at one loop are identical to the ones used for fixed-order calculations of the pole masses, and we can therefore reuse the available expressions in the \NMSSMCALC code after modifying the counterterms such that the self-energies, which are given in the code in a mixed on-shell--$\DRbar$ scheme, are renormalised purely in the $\DRbar$ scheme.\footnote{We note that this procedure thus requires the use of $\operatorname{Re}A_\lambda$ as a $\DRbar$ input parameter instead of the on-shell input for the charged Higgs mass {$m_{H^\pm}$}.}

As the pole-mass matching procedure depends non-trivially on the value of the
VEV due to the tree-level mass diagonalisation, and relies on numerical
cancellations between different terms, the VEV cannot be set to zero exactly,
and the suppressed $v^2/\Msusy^2$ terms are thus always included.\footnote{We
want to note that we include the dominant $v^2/\Msusy^2$ terms, neglect, however,
some numerically small $v^2/\Msusy^2$ contributions arising from \eg the matching
of the gauge couplings, which we do for an exactly vanishing VEV $v \to 0$.}
As a cross check of the consistent one-loop implementation of the pole-mass
matching procedure, we have numerically evaluated the matching procedure for
an artificially small value of $v \sim 1$ GeV to decrease the size of the
$v^2/\Msusy^2$ terms, and found in general very good agreement with the
quartic-coupling matching approach of Sec.~\ref{sec:calculation:quartic}, see
the discussion in Sec.~\ref{sec:CPV:treelevel}.

Finally, we also include two-loop MSSM corrections \cite{PardoVega:2015eno,Bagnaschi:2019esc} to the matching condition, see \eqref{eq:lambdapolemass}.
However, unlike the one-loop results obtained with the pole-mass
matching, the two-loop results are not sensitive to $v^2/\msusy^2$ terms
or CP-violating phases.

%%%%%%%%%%%%%%%%%%%%%%%%%%%%%%%%%%%%%%%%%%%%
\subsection{Uncertainty Estimate}
\label{sec:calculation:uncertainty}
%%%%%%%%%%%%%%%%%%%%%%%%%%%%%%%%%%%%%%%%%%%%
In this section we describe the method used to estimate different theoretical
uncertainties entering the Higgs mass prediction.
For a review of commonly considered uncertainties see \eg
Ref.~\cite{Slavich:2020zjv}.
It is useful to distinguish between two sources of uncertainty
which originate in relations used at the low-energy electroweak scale (SM uncertainty)
and at the high-energy matching scale (SUSY uncertainty).

\paragraph{\underline{SM uncertainties}:}
 {The uncertainty
  originating from relations at the low-energy scale} contains different components:
\begin{itemize}
    \item Missing electroweak corrections in the extraction of SM \MSbar
        parameters are estimated by choosing either the Fermi constant
        $G_F$ or the fine structure constant $\alpha_{M_Z}$ as an input and
        adapting the renormalisation of the electroweak sector accordingly
        using either the $G_F$-scheme \cite{Dao:2023kzz} or the $\alpha_{M_Z}$-scheme
        \cite{Dao:2021khm}. Note that we treat the level a
        We denote the difference in the Higgs mass prediction between the two renormalisation schemes by
        \be \Delta^{\SM}_{G_F/\alpha_{M_Z}} = \abs{M_h^{G_F} - M_h^{\alpha_{M_Z}}} .\ee
    \item To estimate missing higher-order corrections in the relation
        between $\lambda^{\SM,\, \MSbar}$ and the Higgs pole-mass beyond the gauge-less
        limit, we take the \MSbar parameters (obtained in step 2 of
        \cref{fig:HiggsMassEFT}), run them to $\QEW=M_t/2$ and $2M_t$, respectively, using
        SM RGEs and compute the Higgs 
        pole-mass at the two-loop order in the \MSbar scheme by solving
        \begin{equation}
            0 = p^2 - 2\lambda_h^{\smMSbar}(\QEW) v^2(\QEW) +
            \mathrm{Re}\Sigma_{h}(p^2,\QEW)|_{\mathrm{UV-fin}} 
            \label{eq:SMpole}
        \end{equation}
        iteratively for $p^2=M_h^{\MSbar \mathrm{, pole}}(\QEW)$. In
        \cref{eq:SMpole} we evaluate the UV-finite part of the Higgs
        self-energy in the \MSbar scheme at the full one-loop level
        and take into account the leading two-loop $\order{\alpha_t(\alpha_t+\alpha_s)}$ corrections, obtained with \texttt{FeynCalc} and \texttt{TARCER}. 
        As a reference point we use the OS Higgs
        pole-mass from step 1a and estimate the uncertainty as
\be \Delta^{\SM}_{Q_{\rm EW}} = \max\{
    \abs{M_h^{\mathrm{OS}} - M_h^{\overline{\mathrm{MS}} \mathrm{, pole}}(2M_t)}
    ,\abs{M_h^{\mathrm{OS}} - M_h^{\overline{\mathrm{MS}} \mathrm{,pole}}(M_t/2)}
    \}.
    \label{eq:smscaleunc}
    \ee
Since these shifts are not symmetric around $M_t$, we take the maximum of the two 
differences. The estimate is performed for a fixed electroweak scheme
which can be chosen 
in the \texttt{SLHA} input file ($\alpha_{M_Z}$ or $G_F$).
    \item The third component computes the Higgs boson mass while
        adding/removing three-loop (and higher-order) corrections to the
        $\MSbar$ top quark Yukawa coupling:
        \be \Delta^{\SM}_{Y_t} = M_h({Y_t^{\order{\alpha_s^2}}}) -  M_h({Y_t^{\order{\alpha_s^3}}}).\ee
        This shift has been 
        obtained numerically using the code \texttt{mr} for $m_h=\unit[125.1]{GeV}$ and 
        $M_t=\unit[172.76]{GeV}$. The three-loop shift is negative and
        typically causes a decrease of
        the effective SM  Higgs mass of about $\sim$\unit[800]{MeV}. 
\end{itemize}
It should be noted that the three types of uncertainties are not completely independent
from each other with the exception of
$\Delta^{\SM}_{G_F/\alpha_{M_Z}} $ and $\Delta^{\SM}_{Y_t}$, which can be considered to be
independent.

\paragraph{\underline{SUSY uncertainties}:}
For the estimate of the high-scale uncertainty, we generated the two-loop
RGEs for the CP-violating NMSSM using \SARAH and implemented them in \NMSSMCALC.
As stated before, the matching scale and the SUSY
scale (defining the scale of the SUSY \DRbar input parameters) do not need to be the same.
We change the matching scale in the range
of $[\msusy/2, 2\msusy]$, and then compare to the result obtained
with $\Qmatch= \msusy$.
It should be noted that these shifts are not symmetric around $Q_{\rm match}=
\msusy$
and therefore we take the maximum of the two 
differences as 
\be \Delta^{\rm SUSY}_{Q_{\rm match}} = \max\{ \abs{M_h^{\msusy/2} - 
M_h^{\msusy}}, \abs{M_h^{2\msusy} - 
M_h^{\msusy} }\}.\ee
To estimate the uncertainty of missing higher-order corrections to the
matching condition which are not scale-dependent, one typically changes the
definition of the top-quark Yukawa coupling entering the matching condition.
The structure of the NMSSM-specific component of this type of uncertainty was
already discussed in Ref.~\cite{Bagnaschi:2022zvd}. Since we plan to include
exactly this type of missing higher-order corrections via a pole-mass
matching using the results of Ref.~\cite{Dao:2021khm} in the near future, we
also leave the corresponding uncertainty estimate for future work.

\paragraph{\underline{Combined uncertainty}:} The total uncertainty is
computed by assuming independent individual uncertainties,
\begin{equation}
    \Delta \MHpmm = \left[
          \left(\Delta^{\rm SM}_{G_F/\alpha_{M_Z}}\right)^2
        + \left(\Delta^{\rm SM}_{Q_{\rm EW}}\right)^2 
        + \left(\Delta^{\rm SM}_{Y_t}\right)^2
        + \left(\Delta^{\rm SUSY}_{Q_{\rm match}}\right)^2
        \right]^{\frac{1}{2}}
        \,.
        \label{eq:uncert}
\end{equation}
As stated above, not all uncertainties are independent of each other.
Therefore, the total uncertainty computed by \NMSSMCALC is a rather
conservative estimate. Equation (\ref{eq:uncert}) is used
to estimate the uncertainty of the Higgs mass prediction if the pole-mass
matching was chosen in the \texttt{SLHA} input. If the quartic coupling
matching was chosen, the uncertainty is given by
\begin{equation}
    \Delta \MHqcm = \Big[\left(\Delta
        \MHpmm\right)^2
    +(\underbrace{\MHpmm-\MHqcm}_{
         \equiv \Delta^{\rm SUSY}_{v^2/\msusy^2}
    })^2
        \Big]^{\frac{1}{2}}
        \,,
        \label{eq:uncertqm}
\end{equation}
which takes the missing $\nicefrac{v^2}{\msusy^2}$-terms into account and
is labeled as a third uncertainty\footnote{{Note that,
    $\MHpmm$, $\MHqcm$ are both computed at the same order of
    approximation. In an abuse of language 
we call the difference between them ``uncertainty''
  to address the missing ${v^2}/{\msusy^2}$ effect of the quartic coupling
matching result.}}, the \textit{EFT uncertainty} $\Delta^{\rm
SUSY}_{v^2/\msusy^2}$, in
Ref.~\cite{Slavich:2020zjv}.
%%%%%%%%%%%%%%%%%%%%%%%%%%%%%%%%%%%%%%%%%%%%
\section{Numerical Results}
\label{sec:numerical}
%%%%%%%%%%%%%%%%%%%%%%%%%%%%%%%%%%%%%%%%%%%%
In this section we investigate the results for the SM-like Higgs 
boson mass prediction numerically using the implementation in \NMSSMCALC.
We first perform a numerical cross-check of our result by comparing with the
findings of Ref.~\cite{Bagnaschi:2022zvd}.
Furthermore, we investigate the size of the $\nicefrac{v^2}{\msusy^2}$ corrections in different corners of the parameter space by comparing
results obtained with either the pole-mass or quartic-coupling matching and
investigate the size of the individual uncertainty components.
Finally we also discuss the effects of CP-violating phases on the Higgs mass
prediction.
\subsection{Setup and Applied Constraints}
The physical SM input parameters used in step 1a of \cref{fig:HiggsMassEFT}
are
\begin{eqnarray}
  \begin{array}{lcllcl}
G_F &=&1.1663788\times 10^{-5} \, \gev^{-2}, &\quad  
  \alpha(M_Z)& =& 1/127.955\,,\\
\alpha_s(M_Z) & =& 0.1181\,, &\quad M_t&=& 172.69\,\gev\,, \\
m_b^{\MSb} & =&  4.18 \, \gev\,, &\quad M_\tau &=& 1.77682\, \gev\,, \\
M_W &=& 80.377 \, \gev \,, &\quad    M_Z & =& 91.1876 \, \gev\,,
   \end{array} 
\end{eqnarray}
where either $G_F$ or $\alpha_{M_Z}$ is used as an input depending on the
renormalisation scheme choice, \cf
\cref{sec:calculation}.\footnote{The bottom and $\tau$ masses are
  needed in the fixed-order calculations.}

In order to investigate the difference between the two matching 
procedures and the FO calculation (in the \DRbar scheme) and to assess 
the reliability of each calculation in different corners of the NMSSM parameter space,
we have performed a parameter scan varying the NMSSM input parameters
uniformly in the following ranges,
\begin{align}
    \unit[100]{GeV} \le M_1,M_2 \le \unit[1.5]{TeV},  &\qquad
    \unit[100]{GeV} \le\mueff\le\unit[1.5]{TeV},  \non\\
    \unit[1]{TeV}\le m_{\tilde{Q}_{L_3}},m_{\tilde{t}_{R_3}} \le  \unit[2.5]{TeV},  &  \qquad
    \msusy=\sqrt{m_{\tilde{Q}_{L_3}}m_{\tilde{t}_{R_3}}}, \non\\
    M_3 = \mathrm{max}\{ \msusy, \unit[2.3]{TeV}\},  & \qquad 
    \unit[-2.5]{TeV} \le A_\kappa \le \unit[100]{GeV},  \label{eq:scanranges}\\
    \unit[-2.5]{TeV} \le A_\lambda \le \unit[2.5]{TeV},  &\qquad 
    -\sqrt6 \le \hat{X}_t\le \sqrt6,  \non\\
    1 \le \tan\beta \le 20, &\qquad 
    0.05 \le \lambda,\kappa \le 1.0\, .\non
\end{align}
All soft-SUSY-breaking trilinear couplings are set equal to $A_i=\hat{X}_t
\msusy + \nicefrac{\mueff}{\tan\beta}$, whereas all left- and
right-handed soft-SUSY-breaking sfermion masses are set equal to
$m_{\tilde{Q}_{L_3}}$ and $m_{\tilde{t}_{R_3}}$, respectively.
The input scale is set to $Q_{\rm inp} = \msusy$. In order to simplify
the scan we set all CP-violating phases to zero and instead study the
influence of CP-violation for individual parameter points in \cref{sec:CPV}.
Note that within this scan, we do not restrict the masses of the SUSY particles
to be very heavy such that parameter regions suitable for a fixed-order as
well as for the EFT calculation (and intermediate regions) are contained in
the sample. However, we put a lower bound on the masses of SUSY particles
according to the null search results at LEP and LHC 
\cite{ParticleDataGroup:2022pth} as follows:
\begin{equation}
  M_{\chi^0,\chi^\pm }  > \unit[200]{GeV},\quad  
    M_{\ti t_1} > \unit[1310]{GeV}, \quad M_{\ti g} > \unit[2300]{GeV}, \quad
    M_{H^+}>\unit[500]{GeV}.
\end{equation}
The constraints on all other sfermion masses are automatically fulfilled
since the sfermions are approximately mass-degenerate in the chosen parametrisation.
We demand that the lightest neutral CP-even Higgs boson is the
SM-like Higgs boson (by requiring an $h_u$ component of at least 50\%).
Its mass is required to lie in the range 
\beq
\unit[122]{GeV} \le \MHpmm \le \unit[128]{GeV} \;.
\eeq
It should be noted that scenarios where the SM-like Higgs boson is not the
lightest scalar state are not excluded by current measurements. However, in
these scenarios the SM is not the right EFT (rather a singlet-extended SM
needs to be considered) and therefore we exclude them from the scan.
We use {\tt HiggsTools} \cite{Bahl:2022igd}, which contains {\tt HiggsBounds}
\cite{Bechtle:2020pkv}, to check if the parameter points pass all the exclusion
limits from the searches at LEP, Tevatron and the LHC, and {\tt HiggsSignals}
\cite{Bechtle:2020uwn} to check if the points are consistent with the LHC data
for a \unit[125]{GeV} Higgs boson within 2$\sigma$. We do so by requiring 
$\Delta \chi^2=\left| \chi^2_{\msusy\to\infty} - \chi^2_{i}\right|<6.18$, where
$\chi^2_i$ is the $\chi^2$-value computed by {\tt HiggsSignals} (assuming a
fix mass-uncertainty of $\pm\unit[3]{GeV}$ for all Higgs boson masses) for the
specific parameter point and $\chi^2_{\msusy\to\infty}$ is a
SM-reference point obtained in the decoupling limit.\footnote{With the current
{\tt HiggsSignals} dataset we find
$\chi^2_{\msusy\to\infty}\approx 152.1$, which is reasonably close to
$\chi^2\approx152.5$ found with the built-in reference model
\texttt{SMHiggsEW} of {\tt HiggsTools}.} We furthermore require that
$\lambda^2+\kappa^2\le1$, which slightly relaxes the requirement of
perturbative unitarity below the GUT-scale \cite{Masip:1998jc}.

Concerning the concrete setup in \NMSSMCALC we chose to apply the constraints
on the spectrum computed with the pole-mass matching since this method
promises to give precise results for both low and high SUSY masses. In addition,
\NMSSMCALC computes and provides individual results for $M_h$ using the quartic-coupling
matching and the old fixed-order calculation, \cf \cref{app:implementation}.

In the following we also discuss three individual parameter points \BP{\{1,2,3\}}.
We list their input parameters in \cref{tab:inputs} and a subset of the
resulting mass spectra in \cref{tab:higgsmasses,tab:masses}.
The benchmark points \BP1 and \BP2 are taken from Refs.~\cite{Bagnaschi:2022zvd}
and~\cite{Slavich:2020zjv} and have a BSM mass spectrum which is at or
above \unit[2.5]{TeV}.
The parameter point \BP3 is part of the scan sample
described above and features a rather light singlet-light state,
$M_{h_2}\approx\unit[300]{GeV}$,  which mixes
to approximately 4\% with the SM-like Higgs boson.
\BP1 and \BP3 feature relatively large $\lambda$ and
$\kappa$ while \BP2 is given in the MSSM-limit.
This choice of parameters 
enables us to compare with the literature as well as to study NMSSM-specific
scenarios in the EFT-context not considered before.

\setlength{\tabcolsep}{0.15em}
\begin{table}[tp]
    \centering
    \begin{tabular}{|c|c|c|c|c|c|c|c|c|c|c|c|c|c|}
    \hline
         & $\tan\beta$ & $\lambda$ & $\kappa$ & $M_1$ & $M_2$ & $M_3$ & $A_t$  & $A_\lambda$ & $A_\kappa$ & $\mu_{\text{eff}}$ & $m_{\tilde{Q}_{L_3}}$ & $m_{\tilde{t}_{R_3}}$ & Ref.\\ \hline
    \BP1 & 3.0         &   0.6     &  0.6     &  1.0  & 2.0   &  2.5  & 12.75  & 0.3         & -2.0       &   1.5        &   5.0                 &  5.0                  & \cite{Bagnaschi:2022zvd}   \\ \hline
    \BP2 & 20.0        &   0.05    &  0.05    &  {4.5}  &
    {4.5}   &  {4.5}  & {-10.79}
                           & {-4.28}       &
    {-1.5}       &   {4.5}        &   {4.5}                 &  {4.5}                  & \cite{Slavich:2020zjv} \\ \hline
    \BP3 & 1.27        &   0.73    &  0.62    &  0.24 & 1.18  &  2.3  & -0.39  & 0.06        & -1.44      &   0.49       &  1.79                 &  1.51                 & this work \\ \hline
    \end{tabular}
    \caption{The most relevant input parameters for the Higgs boson mass
    prediction chosen for three benchmark points considered in this work
    (rounded to two digits). All parameters with mass dimension one are given
in units of TeV. \BP3 is part of the scan sample obtained in this work, \BP1
was taken from Ref.~\cite{Bagnaschi:2022zvd}, Fig.2
and \BP2 from Ref.~\cite{Slavich:2020zjv}, Fig.5. The complete set
of input parameters can be found on the webpage of the program.}
\label{tab:inputs}
\end{table}

\newcommand{\ignore}[1]{}

\begin{table}[tp]
    \centering
\begin{tabular}{|c|c|c|c|c|c|c|c|}
    \hline
                 & \MHpmm         & \MHqcm         & $M_{h_2}$      & $M_{h_3}$       & $M_{A_1}$      & $M_{A_2}$      & $M_{H^+}$ \\ \hline
    \texttt{BP1} & 124.29 $(h_u)$ & 124.31 $(h_u)$ & 2407.6 ($h_s$) &  2971.8 ($h_d$) & 2905.7 ($a$)   & 3000.2 ($a_s$) & 2967.1     \\ \hline
    \texttt{BP2} & 125.{15} $(h_u)$ & 125.{15} $(h_u)$ &
    {4486.6} ($h_d$) &  {8616.7} ($h_s$) &
    {4510.9} ($a_s$) & {4984.0} ($a$)   &
    {4995.0}     \\ \hline
    \texttt{BP3} & 127.1{6} $(h_u)$ & 129.4{6} $(h_u)$ & 305.5 ($h_s$)  &  659.5 ($h_d$)  & 663.8 ($a$)    & 1308.7 ($a_s$) & 658.4      \\ \hline
\end{tabular}

\caption{Neutral and charged Higgs boson masses derived from the input parameters in
\cref{tab:inputs} using \NMSSMCALC. All values are given in units
of GeV. For neutral CP-even/odd Higgs bosons we indicate the
dominant gauge-eigenstate admixture in brackets. The lightest neutral
Higgs boson was computed using the pole-mass (\MHpmm) and
quartic-coupling (\MHqcm) matching approaches. All other BSM Higgs boson masses are computed in the
\DRbar scheme with the highest-available but fixed order in \NMSSMCALC.}
\label{tab:higgsmasses}
\end{table}
\begin{table}[tp]
    \centering
\begin{tabular}{|c|c|c|c|c|c|c|c|c|c|}
    \hline
                 &  $m_{\tilde{t}_1}$ & $m_{\tilde{t}_2}$  & $m_{\chi^0_1}$ & $m_{\chi^0_2}$  & $m_{\chi^0_3}$  & $m_{\chi^0_4}$  & $m_{\chi^0_5}$ & $m_{\chi^+_1}$ & $m_{\chi^+_2}$ \\ \hline
    \texttt{BP1} &      4829.6        & 5168.2             & 997.2          & 1491.5          & 1502.4          & 2010.5          & 3003.3         &  1490.2        &  2010.5        \\ \hline
    \texttt{BP2} &      {4335.2}        &
    {4662.4}             & {4432.6}         &
    {4500.0}          & {4500.0}          &
    {4567.8}          & {9000.0}         &
    {4407.2}        &  {4559.9}        \\ \hline
    \texttt{BP3} &      1514.2        & 1799.1             & 232.8          & 484.1           & 498.2           & 835.4           & 1192.7         &  477.3         &  1192.6        \\ \hline
\end{tabular}
\caption{Excerpt of the SUSY mass spectrum derived from the input parameters in \cref{tab:inputs} using \NMSSMCALC. All values are given in units
of GeV.}
\label{tab:masses}
\end{table}

\subsection{Uncertainties}
\label{sec:numerical:uncertainty}
In \cref{tab:uncertainties} we list the individual uncertainties contributing
to the total uncertainty as defined in \cref{sec:calculation:uncertainty} for
the benchmark points \BP{\{1,2,3\}}. The two dominant sources are the
SUSY scale-uncertainty and missing higher-orders in the extraction of the SM
top-quark coupling followed by the SM scale-uncertainty which are all between
about \unit[200-800]{MeV} (in absolute values). The SUSY scale-uncertainty is particularly large for the
point \BP3 which is due to its BSM mass spectrum being spread across
both 
the electroweak and the TeV-scale. The uncertainty due to the scheme choice
between $\alpha_{M_Z}$ and $G_F$, $\Delta^{\rm SM}_{G_F/\alpha_{M_Z}}$, is
always smaller than \unit[100]{MeV} indicating that the missing two-loop electroweak
corrections in the SM-part of \NMSSMCALC are rather small.

If the quartic-coupling matching is considered, the missing
$\nicefrac{v^2}{\msusy^2}$ corrections, $\Delta^{\rm SUSY}_{v^2/\msusy^2}$, also
contribute to the total uncertainty $\Delta\MHqcm$. These corrections are particularly important
for the parameter point \BP3 as it features a rather light singlet. We find
that the total uncertainty of \BP3 is shifted from
$\order{\unit[900]{MeV}}$
to about $\order{\unit[2]{GeV}}$ when using the quartic-coupling matching
instead of the pole-mass matching. The effect of the light singlet and the
interplay with the $\nicefrac{v^2}{\msusy^2}$ corrections is studied in
\cref{sec:numerical:voMSUSY} in more detail.

\begin{table}[tp]
    \centering
\begin{tabular}{|c|c|c|c|c|c||c|c|}
    \hline
       & $\Delta^{\rm SM}_{Y_t}$     & $\Delta^{\rm SM}_{Q_{\rm EW}}$     & $\Delta^{\rm SM}_{G_F/\alpha_{M_Z}}$       & $\Delta^{\rm SUSY}_{Q_{\rm match}}$      & $\Delta^{\rm SUSY}_{v^2/\msusy^2}$  & $\Delta \MHpmm$ & $\Delta \MHqcm$ \\ \hline
  \BP1 & -738 & 208 & -19  & 376  & -21   &  854   & 836     \\ \hline
  \BP2 & -6{85} & {208} & -69  &
  {189}  & {-5}   &
  {743}   & {743}       \\ \hline
  \BP3 & -401 & 19{8} & 2{0}   & {694}  & -2294 &  {826}   & 24{15}    \\ \hline
\end{tabular}
\caption{Individual (first five columns) and total (last two columns)
    uncertainty estimate of the SM-like Higgs boson mass prediction
    for the three parameter points defined in \cref{tab:inputs}. All values are given
    in units of MeV. The uncertainty
$\Delta^{\rm SUSY}_{v^2/\msusy^2}$ only contributes to the total
uncertainty of the quartic coupling matching, \ie to $\Delta \MHqcm$.}
\label{tab:uncertainties}
\end{table}

\subsection{Numerical Validation and Comparison with Previous Works}
\label{sec:numerical:validation}
In this section we numerically validate the calculation and implementation 
of the two matching procedures in \NMSSMCALC.
The one-loop matching condition for the quartic coupling has previously been
computed in \eg Ref.~\cite{Bagnaschi:2022zvd}
and combined with the tool \mr \cite{Kniehl:2016enc} which performs an OS-\MSbar conversion and RGE
running of all SM parameters incorporating all state-of-the-art higher-order
corrections
\cite{Bednyakov:2012rb,Bednyakov:2012en,Bednyakov:2013eba,Kniehl:2015nwa,vanRitbergen:1997va,Vermaseren:1997fq}
(see also \cite{Bezrukov:2012sa,Jegerlehner:2001fb,Jegerlehner:2002em,Jegerlehner:2003py}, where the OS-\MSbar conversion formulae for the SM Higgs, top Yukawa, and gauge couplings have been computed at $\order{\alpha \alpha_s}$).
In contrast, \NMSSMCALC implements only the full one-loop and leading
two-loop corrections in the extraction of the SM \MSbar parameters.
Therefore, we implemented an optional link of \NMSSMCALC to the program \mr
which replaces the in-house calculations performed in steps 1a) to 3)
in \cref{fig:HiggsMassEFT} with the predictions of \mr. This ensures that we
use the very same running SM \MSbar parameters as Ref.~\cite{Bagnaschi:2022zvd} 
at a given scale $\Qmatch$ for a given set of SM input parameters.
Alternatively, we provide a similar link to the tool \SMDR
\cite{Martin:2019lqd} which uses a different treatment of the Higgs tadpole
and works in the \MSbar scheme but goes similarly beyond the corrections
computed by \NMSSMCALC \cite{Martin:2005qm,Martin:2014cxa,Martin:2015lxa,Martin:2015rea,Martin:2016xsp,Martin:2016bgz,Martin:2017lqn,Martin:2022qiv}.
It should be noted that both, \mr and \SMDR, increase the runtime of
\NMSSMCALC significantly such that their use within a parameter scan
effectively becomes unviable.
However, in \cite{Alam:2022cdv}, the authors provide interpolation formulae based on \SMDR covering a 5$\sigma$ range of the SM input parameters. We implemented these interpolation formulae in \NMSSMCALC as an alternative approach to extract the SM \MSbar parameters, see \cref{app:implementation}.

\begin{figure}[tp]
    \centering
    \includegraphics[width=1.0\textwidth]{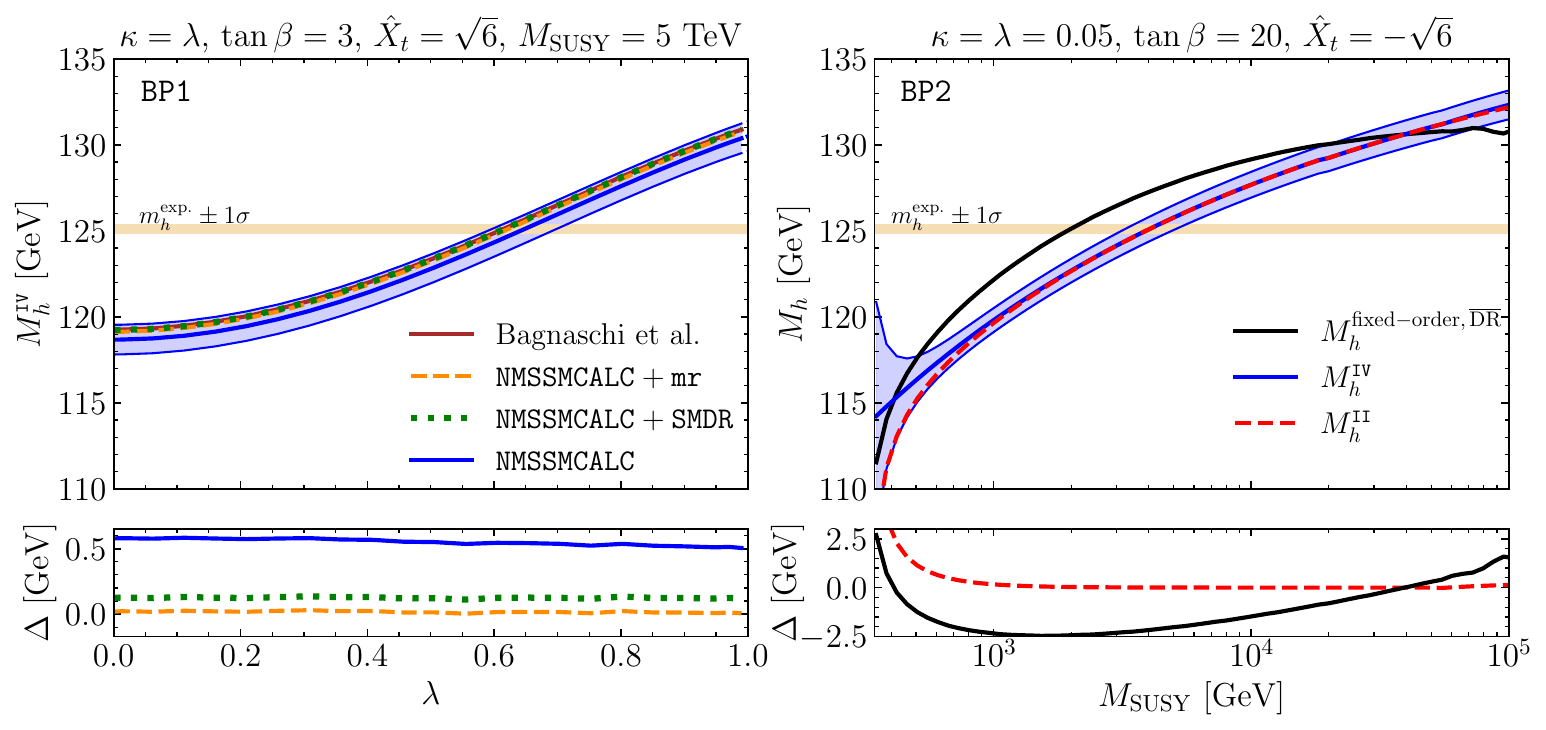}
    \caption{\textbf{Left:} Higgs mass prediction for the parameter point
        \BP1 as a function of $\lambda$ using the quartic coupling matching.
        Shown are the results for $\MHqcm$ and its uncertainty
        estimate defined in \cref{eq:uncertqm}, 
        obtained with \NMSSMCALC (blue solid) or with a modified \NMSSMCALC
        which uses \texttt{mr} (orange dashed) or \texttt{SMDR} (green dotted)
        for the SM \OS-\MSbar conversion and SM RGE-running. In brown we
        re-print the result taken from Fig.1 of
        Ref.~\cite{Bagnaschi:2022zvd}. The lower panel shows the differences
        of the result found in Ref.~\cite{Bagnaschi:2022zvd} and the
        results obtained with \NMSSMCALC. 
        \textbf{Right:} Higgs mass prediction for \BP2 using the fixed-order
        (black), quartic coupling matching (blue), pole-mass matching (red
        dashed) calculation of \NMSSMCALC. The blue band is the uncertainty of the
        quartic-coupling matching defined in \cref{eq:uncertqm}. The lower
        panel shows the differences with the result obtained in the
        quartic-coupling matching.}
    \label{fig:comparison1}
\end{figure}

In \cref{fig:comparison1}~(left) we show the Higgs mass prediction
applying the quartic coupling matching for $\MHqcm$ for the
parameter point \BP1 as a function of $\lambda$. The parameter
$\kappa=\lambda$ is varied simultaneously. The brown-solid line is a 
reprint the one-loop curve found in
Ref.~\cite{Bagnaschi:2022zvd}~(Fig.1) 
while the orange-dashed, green-dotted and blue-solid lines are obtained with
\NMSSMCALC when using \mr, \SMDR or the in-built SM calculation,
respectively. The blue band shows the uncertainty estimate (\cf
\cref{sec:calculation:uncertainty}) of the pure \NMSSMCALC result as
defined in \cref{eq:uncertqm}. 
In the lower panel we plot the difference $\Delta =
\left.\MHqcm\right|_{\texttt{\cite{Bagnaschi:2022zvd}}} - \left.\MHqcm\right|_{\NMSSMCALC}$ for each
individual \NMSSMCALC result.
We find very good agreement between \NMSSMCALC and
Ref.~\cite{Bagnaschi:2022zvd} within the numerical accuracy if \mr is used
for the SM calculation (orange-dashed) which is a strong numerical cross-check of our quartic coupling matching.
If we use \SMDR instead of \mr, the Higgs mass prediction is moved downwards
by $\sim\unit[100]{MeV}$. The \NMSSMCALC result differs by $\sim\unit[500-600]{MeV}$
throughout the shown range of $\lambda$ but is in agreement with the
other three results within the estimated uncertainty. Since the SM RGEs in \NMSSMCALC are of the same order as in \mr (full two-loop and leading three- and four-loop), the difference between the \NMSSMCALC and \mr result is mainly caused by missing higher-order corrections in the $\OS-\MSbar$ conversion performed by \NMSSMCALC. 

Furthermore, the availability of the pole-mass matching also enables us to
perform another cross-check. The pole-mass and quartic-coupling matching only
differ by terms of $\order{\nicefrac{v^2}{\msusy^2}}$ and consequently should converge to
each other for large \msusy if all large logs appearing in the pole-mass
calculation are cancelled properly. The two parameter points \BP1 and \BP2
are suitable for such a comparison as the BSM particle spectrum is
of the order of the TeV-scale. In particular the stop masses, which control the 
numerically largest loop corrections, are above \unit[2.5]{TeV} and hence the related
  uncertainties $\Delta^{\rm 
SUSY}_{v^2/\msusy^2}\sim\order{\unit[10]{MeV}}$ (\cf \cref{tab:uncertainties}). This behaviour is demonstrated
in \cref{fig:comparison1}~(right) for the parameter point \BP2 where
all SUSY particle masses are varied simultaneously with \msusy. The
blue-solid line shows the Higgs mass prediction obtained using the
quartic-coupling matching, \MHqcm, while the red-dashed lines shows the
result when using the pole-mass matching, \MHpmm. In addition, we show the
fixed-order result (black-solid) obtained in the \DRbar scheme at
$\order{\alpha_t(\alpha_t+\alpha_s)}$.\footnote{Since {\tt BP2} is in
  the MSSM limit, this is the most precise 2-loop order available in {\tt NMSSMCALC}.} The lower panel
in~\cref{fig:comparison1}~(right) shows the difference $\Delta=
\MHqcm-M_h^i$ between the quartic coupling matching and the other two
results.
For large $\msusy$, starting from $\msusy>
\unit[2]{TeV}$, we find perfect agreement between the 
pole-mass and quartic-coupling matching while for low \msusy they can differ
by several GeV. 
  The blue uncertainty-band for \MHqcm also includes the
differences to the pole-mass matching result, thereby demonstrating the
importance of the $\nicefrac{v^2}{\msusy^2}$ corrections in this regime.
On the other hand, the fixed-order result and the pole-mass matching show
{better} agreement for {small \msusy} while for larger \msusy
the fixed-order line features a different shape.
{One might naively expect better agreement between the
    result obtained in the pole-mass matching and the fixed-order calculation
    if $\msusy<\unit[1]{TeV}$ (see e.g. \cite{Slavich:2020zjv}).
    However, the two-loop corrections to the matching condition taken from the literature 
    do not contain any $\nicefrac{v^2}{\msusy^2}$ terms while the two-loop
    fixed-order calculation has the full dependence on the electroweak VEV.
    Thus, the discrepancy between the two approaches for very small $\msusy$
    can be explained by the missing $\nicefrac{v^2}{\msusy^2}$-terms in the
    two-loop part of the EFT calculation. Indeed we verified numerically,
    that the fixed-order and EFT-pole-mass predictions agree better at low
    scales if only one-loop corrections are taken into account.}
    Therefore, the pole-mass
matching procedure implemented in \NMSSMCALC possesses features of a hybrid approach taking into account resummed logs as well as pieces of $\nicefrac{v^2}{\msusy^2}$ as the 
\textit{hybrid} approaches in \texttt{FlexibleEFTHiggs} \cite{Athron:2014yba,Athron:2016fuq,Athron:2017fvs,Kwasnitza:2020wli} and
\texttt{FeynHiggs} \cite{Hahn:2013ria,Bahl:2016brp,Bahl:2018ykj,Bahl:2019hmm,Bahl:2017aev}.  {This method} gives precise predictions for $M_h$
across a large range of \msusy (see \cite{Slavich:2020zjv} for a
complete list of references). 
However, parameter points like \BP2 which are
rather MSSM-like, often can only pass experimental constraints from stop
searches, as well as the theoretical constraint of predicting
$M_h\approx\unit[125]{GeV}$,  by having \msusy larger than a few TeV and are
therefore often saturated in the energy regime where a quartic coupling matching is
sufficient. In the next section we show that this is not the case
for the NMSSM, as it can predict a light singlet, which can greatly benefit
from a pole-mass matching.

\subsection{The Case of a Light Singlet}
\label{sec:numerical:voMSUSY}
\begin{figure}[tp]
    \centering
    {\hspace{-2mm}%
    \includegraphics[width=0.52\textwidth]{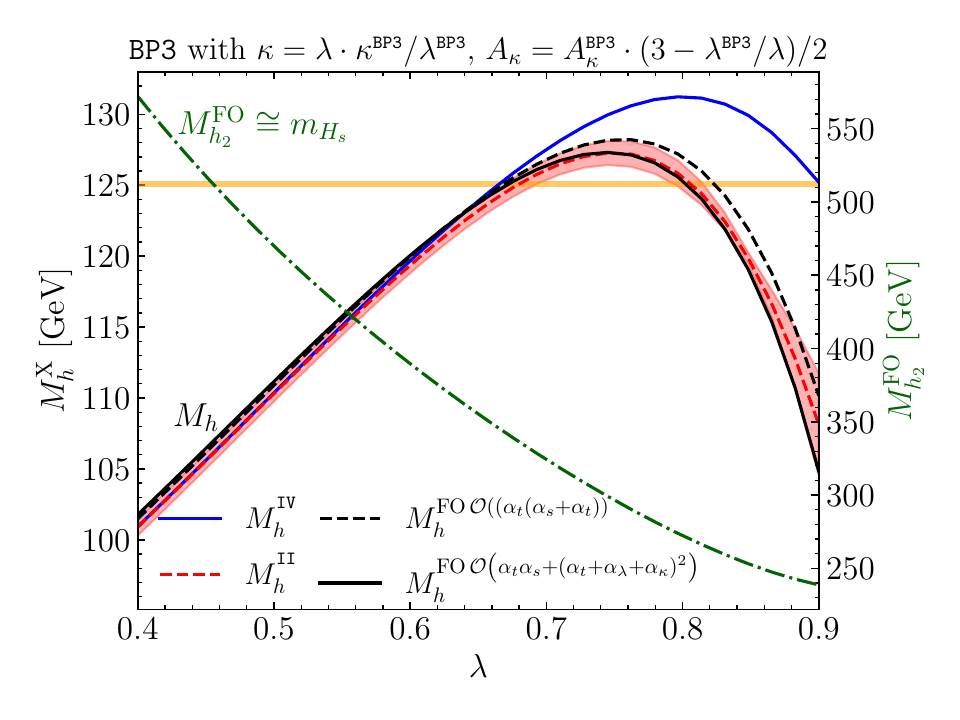}%
    \includegraphics[width=0.47\textwidth]{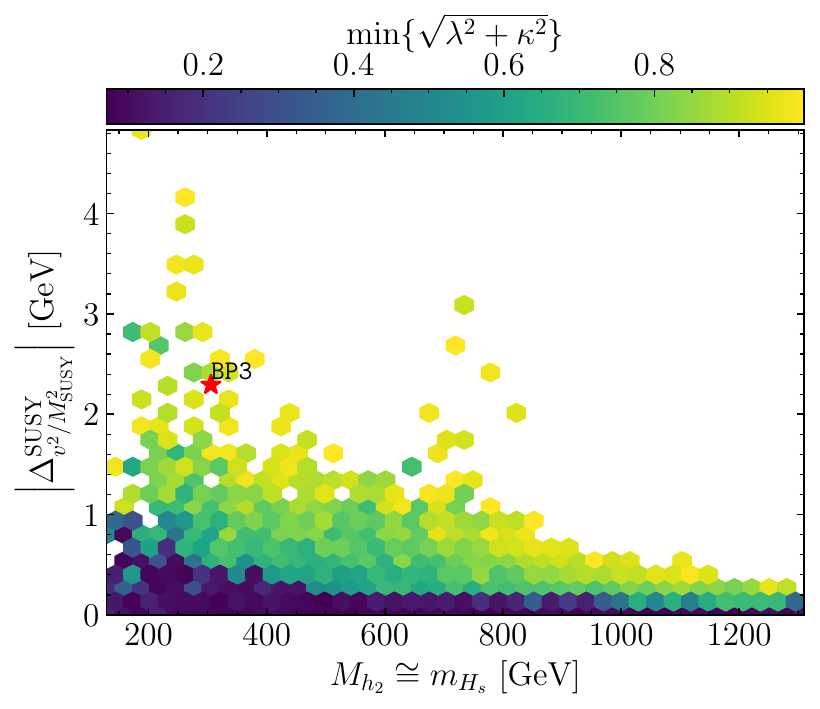}}
    \caption{
        \textbf{Left:} The SM-like Higgs boson mass computed using the
        quartic-coupling matching (blue), pole-mass matching (red dashed),
        fixed-order $\order{\alpha_t(\alpha_t+\alpha_s)}$ (black dashed) and the fixed-order
        $\order{\alpha_t\alpha_s+(\alpha_t+\alpha_\lambda+\alpha_\kappa)^2}$
        (black solid)
        calculation as a function of $\lambda$ for the parameter point \BP3.
        The green line (to be read on the right axis), shows the
        mass-prediction of the singlet-like state obtained at fixed
        order $\order{\alpha_t\alpha_s+(\alpha_t+\alpha_\lambda+\alpha_\kappa)^2}$.
        The red band is the uncertainty obtained for the pole-mass matching
    result, as defined in Eq.~(\ref{eq:uncert}). 
        \textbf{Right:} Size of the $v^2/\msusy^2$-type corrections as a function of the
        { second-lightest Higgs mass} for all parameter points { with a $h_2$ featuring at least 50\% singlet admixture} obtained in the random scan.
    Each hexagonal bin shows the minimum value of
    $\sqrt{\lambda^2+\kappa^2}$ found in that bin.}
    \label{fig:BP3lambda}
\end{figure}

{We now consider the scenario of a rather light singlet which is realised by
the parameter point \BP3. Note that due to the appearance of a light state, 
we do not expect the SM to be the correct EFT parametrisation for this parameter point 
(and instead, the SM extended by a singlet scalar should be used), but we nonetheless want to study its features and compare how well the two matching approaches perform.} In~\cref{fig:BP3lambda}~(left) we show the
Higgs boson mass prediction using the quartic-coupling matching
(blue-solid), the pole-mass matching 
approach (red-dashed, including the red uncertainty band, applying Eq.~(\ref{eq:uncert})) and the fixed-order calculation at
$\order{\alpha_t(\alpha_t+\alpha_s)}$ (black dashed) and
$\order{\alpha_t\alpha_s+(\alpha_t+\alpha_\lambda+\alpha_\kappa)^2}$
(black solid) as a function of $\lambda$. The green line (to be read on the
right axis) shows the fixed-order prediction for the mass $m_{H_s}$ of the singlet-like
state. We scale the other
NMSSM-parameters according to $\kappa=\lambda\cdot \kappa^{\texttt{\small
BP3}}/\lambda^{\texttt{\small BP3}},\, A_\kappa=A_\kappa^{\texttt{\small
BP3}}\cdot (3-\lambda^{\texttt{\small BP3}}/\lambda)/2$.
This parametrisation allows us to vary $\lambda$ throughout a large range
while maintaining a decreasing singlet mass with increasing $\lambda$
and it 
furthermore avoids the presence of tachyonic tree-level masses. For
$\lambda=\lambda^{\texttt{\small BP3}}$ we recover the original parameter
point.
For small $\lambda\approx 0.4$ the singlet-like mass is about \unit[500]{GeV} 
while for large $\lambda\approx0.9$ it can be as light as
\unit[250]{GeV}.
{The SM-Higgs boson mass varies between 100 and 131 GeV within the
  considered $\lambda$ range.} 
  In the small-$\lambda$ (and large $m_{H_s}$) region we observe good agreement
  between all four methods. However, as $\lambda$ increases ($m_{H_s}$ decreases)
the quartic-coupling matching result starts to deviate, reaching a
  difference w.r.t.~the other results of up to $\sim\unit[15]{GeV}$.
It should be stressed that the
stop masses are always above \unit[1.5]{TeV} for this parameter point.
Therefore, all $\nicefrac{v^2}{\msusy^2}$ contributions from the stop sector
are negligible compared to those originating in the singlet sector.

It remains the question whether missing higher-order
corrections to the matching condition proportional to $\lambda$ and
$\kappa$ can become significant for large $\lambda$.
 {While the two-loop corrections of this type
to the matching condition are not available in \NMSSMCALC, they, however,
have been included in the \NMSSMCALC fixed-order prediction \cite{Dao:2021khm}.
This allows us to estimate the importance of NMSSM-specific corrections by
comparing the three results.} In \cref{fig:BP3lambda}~(left) we
observe, that the relative
size of the NMSSM-specific higher-order corrections in the fixed-order
calculation is always much smaller than the relative size of the missing
$\nicefrac{v^2}{\msusy^2}$ contributions in the EFT-approach. Therefore, the
$\nicefrac{v^2}{\msusy^2}$ corrections in the one-loop matching condition
of the quartic-coupling approach are numerically more significant
than the missing NMSSM-specific higher-order corrections to the matching
condition. Regarding the pole-mass matching the evaluation of the importance of the missing
two-loop corrections is left for future work.

The region in \cref{fig:BP3lambda}~(left) that features very large
$\Delta^{\rm SUSY}_{v^2/\msusy^2}$ is clearly not in agreement with the Higgs
boson mass measurement. {It is interesting to ask 
how large $\Delta^{\rm SUSY}_{v^2/\msusy^2}$ can be for parameter points satisfying the
contraint.
} In \cref{fig:BP3lambda}~(right) we {therefore} study the size of the
$\nicefrac{v^2}{\msusy^2}$ terms by plotting the absolute value of $\Delta^{\rm
  SUSY}_{v^2/\msusy^2}$ 
for all parameter points found in the random scan, which fulfill all
applied constraints and feature a second-lightest singlet-like scalar,
as a function of the singlet mass $m_{H_s}$. The parameter points are
grouped in hexagonal bins whereas the color of each bin indicates the minimum
value of $\sqrt{\lambda^2+\kappa^2}$ found in that bin. We observe that
$\Delta^{\rm SUSY}_{v^2/\msusy^2}$ tends to decrease for increasing $m_{H_s}$
and therefore shows a similar behaviour as the stop sector (note
that \eg the neutralinos could still be lighter than the singlet and also
cause $\nicefrac{v^2}{\msusy^2}$ contributions). However, the size of $\Delta^{\rm
SUSY}_{v^2/\msusy^2}$ is also strongly influenced by the size of $\lambda$ and
$\kappa$. For $\sqrt{\lambda^2+\kappa^2}\lesssim 0.2$ we find $\Delta^{\rm
SUSY}_{v^2/\msusy^2}\lesssim \unit[1]{GeV}$ which is similar to what
we obtain in the MSSM with the present exclusion limits on the
stops. It is remarkable that the two
matching approaches agree with each other reasonably well for small
$\lambda$ and $\kappa$ 
even if the singlet is as light as \unit[200]{GeV} (\cf dark blue points in
the plot). However, for
$\sqrt{\lambda^2+\kappa^2}\gtrsim 0.3$ the quartic-coupling matching would
suffer from large missing $\nicefrac{v^2}{\msusy^2}$ corrections,
which can reach up to $\order{\unit[5]{GeV}}$ for
$\sqrt{\lambda^2+\kappa^2}\gtrsim 0.9$. 

%%%%%%%%%%%%%%%%%%%%%%%%%%%%%%%%%%%%%%%%%%%%
\subsection{CP-Violating Effects in the EFT Calculations}
\label{sec:CPV}
In the following we study the effect of non-vanishing CP-violating phases
onto the Higgs mass prediction at the example of the parameter points \BP1
and \BP3. We distinguish two scenarios: (i) CP in the Higgs sector is
conserved at the tree-level but broken by loop effects from the SUSY fermions
and scalars and (ii) CP is already broken at the tree-level.
%%%%%%%%%%%%%%%%%%%%%%%%%%%%%%%%%%%%%%%%%%%%

\subsubsection{Loop-induced CP-Violation}
\begin{figure}[tp]
    \centering
    \includegraphics[width=0.49\textwidth]{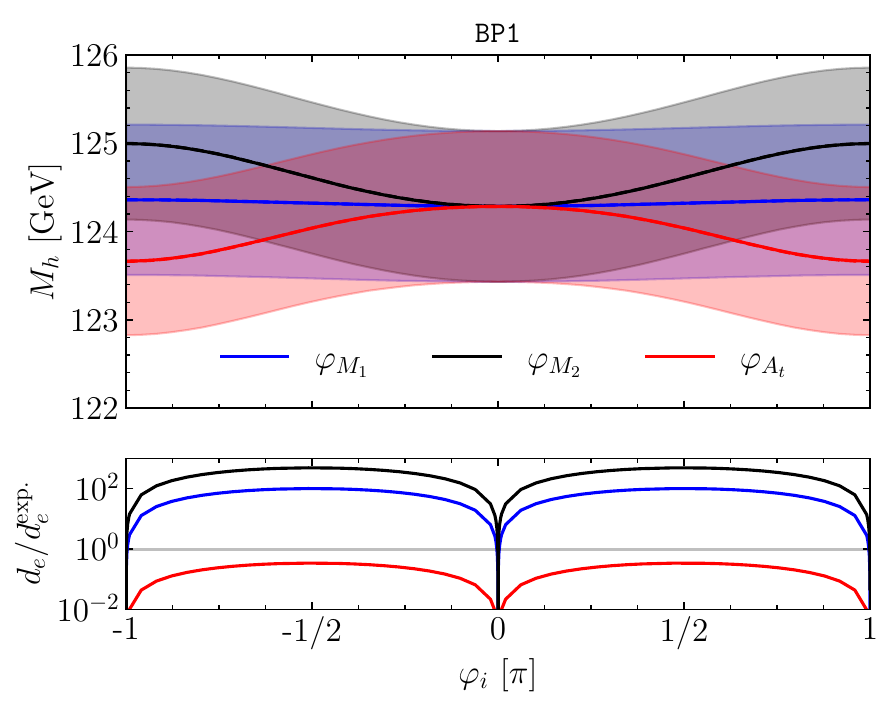}
    \includegraphics[width=0.49\textwidth]{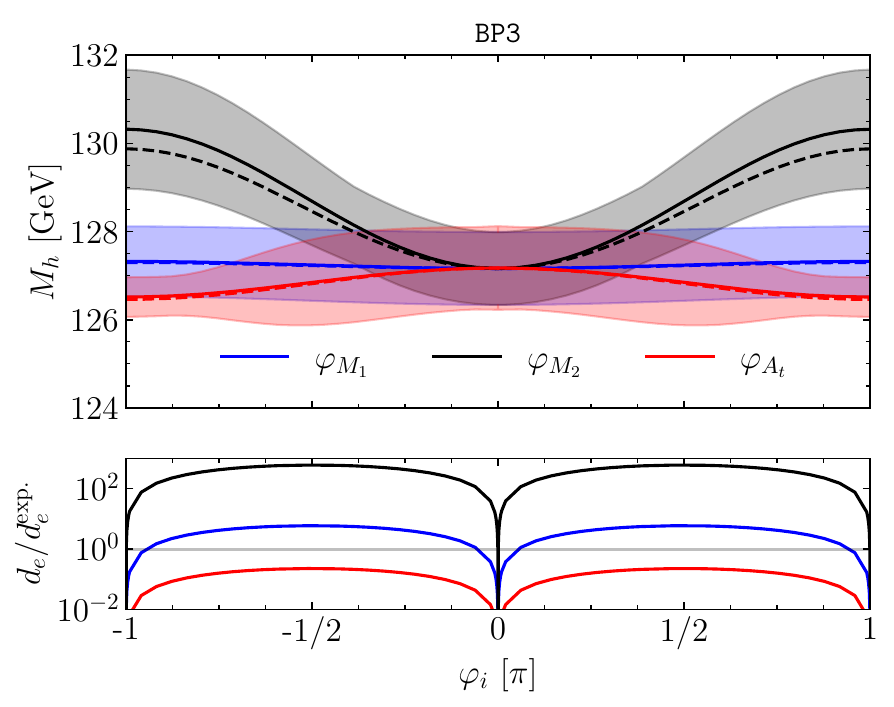}
    \caption{The dependence of the Higgs mass prediction using the pole-mass
    matching (solid) and quartic-coupling matching (dashed) on the
    CP-violating phases of $M_1$ (blue), $M_2$ (black) and $A_t$ (red).
    The quartic-coupling matching results have been shifted by the constant values
    $\left.\Delta^{\rm SUSY}_{v^2/\msusy^2}\right|_{\varphi_i=0}$ from
        \cref{tab:uncertainties}. {Note that the phases in the
            plots are defined relative to the phases, \ie signs, of the
            \mbox{CP-conserving}
            reference points, such that the benchmark scenarios in
            \cref{tab:inputs} are recovered for $\varphi_i=0$.}
    }
    \label{fig:CPV}
\end{figure}

In \cref{fig:CPV} we show the Higgs mass prediction for \BP1 (left) and \BP3
(right) for individually varied phases of $M_1$ (blue), $M_2$ (black) and
$A_t$ (red). The lower panels show the prediction for the electric dipole
moment of the electron (eEDM) obtained with \NMSSMCALC normalised to the current
experimental upper bound
\cite{ACME:2018yjb}. The solid lines show the Higgs mass prediction obtained with the
pole-mass matching while the dashed lines are obtained with the
quartic-coupling matching. In addition, the results of the quartic-coupling
matching have been shifted by the constant difference $\left.\Delta^{\rm
SUSY}_{v^2/\msusy^2}\right|_{\varphi_i=0}$ from \cref{tab:uncertainties} such
that the dashed and solid lines overlap for $\varphi=0$. Therefore, one can
directly read-off additional $\nicefrac{v^2}{\msusy^2}$-effects caused by the
CP-violating phases. For both parameter points we find that $\varphi_{A_t}$
is not constrained by the eEDM while still having an effect on $M_h$ of
$\order{\unit[1]{GeV}}$. The phases $\varphi_{M_{1,2}}$ can have a similar
effect on $M_h$ but are strongly constrained when varied individually. It
should be noted, however, that for some choices of
$\varphi_{M_1}\neq\varphi_{M_2}\neq 0$ the EDM constraints can be avoided
while still achieving non-negligible effects on $M_h$. 

{The $v^2/\msusy^2$ effects caused by the CP-violating phases are
negligible for the parameter point \BP1 since all SUSY particles are heavy in
this scenario. This is not the case for \BP3, which has a rather light Higgsino and a large mixing between wino
 and Higgsino, and therefore a large $\left.\Delta^{\rm
SUSY}_{v^2/\msusy^2}\right|_{\varphi_{M_2}\neq 0}$ contribution.}

\subsubsection{Tree-level CP-Violation}\label{sec:CPV:treelevel}
Considering \cref{eq:treelevel_EFTCL} any phase-combination
$\varphi_y=2\varphi_s+\varphi_\kappa-\varphi_\lambda-\varphi_u\neq 0$ will
immediately introduce CP-violating effects in the Higgs sector thereby having
a strong impact on the EDM prediction. In the following we focus on
$\varphi_{\lambda}$ and $\varphi_{\kappa}$ which were found to have the
smallest impact on the eEDM for the considered parameter points.

\begin{figure}[tp]
    \centering
    \includegraphics[width=0.49\textwidth]{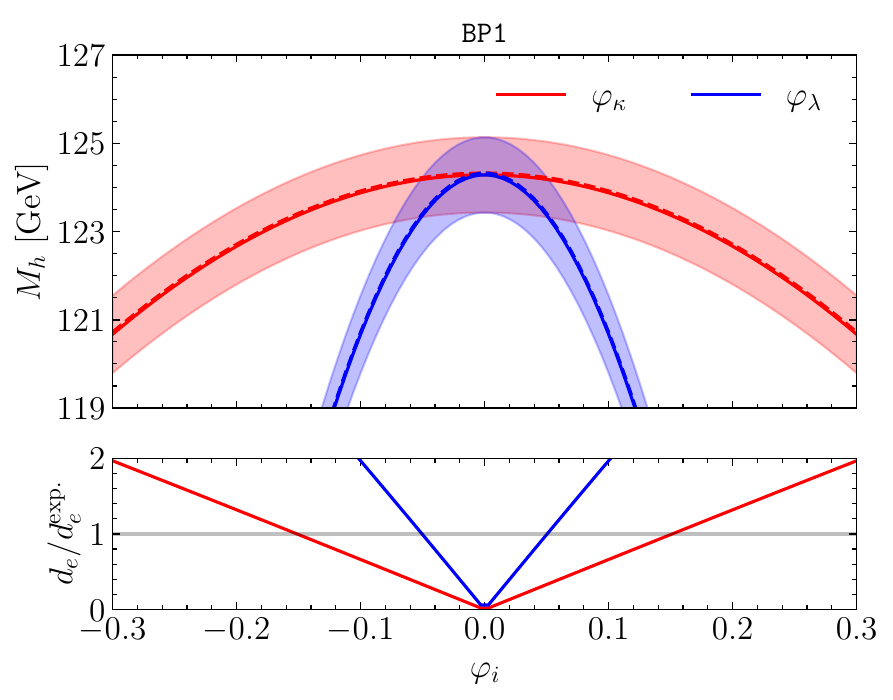}
    \includegraphics[width=0.49\textwidth]{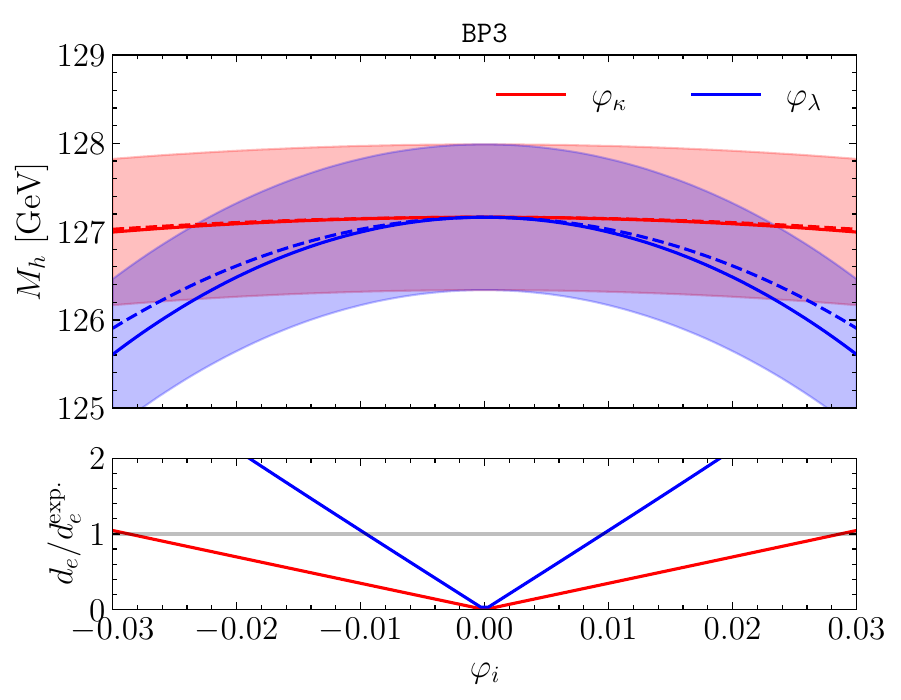}
    \caption{The dependence of the Higgs mass prediction using the pole-mass
    matching (solid) and quartic-coupling matching (dashed) on the
    CP-violating phases of $\lambda$ (blue) and $\kappa$ (red).
    The quartic-coupling matching results have been shifted by the constant values
    $\left.\Delta^{\rm SUSY}_{v^2/\msusy^2}\right|_{\varphi_i=0}$ from \cref{tab:uncertainties}.
    }
    \label{fig:CPVtree}
\end{figure}

In \cref{fig:CPVtree} we show the same quantities as in \cref{fig:CPV} but as
a function of $\varphi_{\kappa}$ (red) and $\varphi_{\lambda}$ (blue). We
find that for \BP1 $|\varphi_{\lambda}|\gtrsim 0.05$
($|\varphi_{\kappa}|\gtrsim 0.15$) and for \BP3
$|\varphi_{\lambda}|\gtrsim 0.01$ ($|\varphi_{\kappa}|\gtrsim 0.03$)
is excluded by the eEDM. However, even in these small ranges the Higgs mass
prediction depends strongly on the CP-violating phases. Concerning the size
of the $\nicefrac{v^2}{\msusy^2}$ contributions, the picture is similar to
the loop-induced CP-case \ie only \BP3 shows a significant difference between
pole-mass and quartic-coupling matching. 

As an additional cross-check we verified that pole-mass and
quartic-coupling matching ({\ie} solid and dashed lines of the same color)
are in agreement for all values considered in \cref{fig:CPV,fig:CPVtree}
once we set the running VEV $v(\Qmatch)$ at the
matching scale to a numerically small value in the pole-mass matching
which effectively turns-off all
$\nicefrac{v^2}{\msusy^2}$ corrections. {However, for very large values of
$|\lambda|,\,|\kappa|$ and
$\varphi_{y}\sim\order{1}$ one may
loop-induce sizeable mixing between CP-even and CP-odd Higgs fields
(such that the SM is no longer the right EFT) even for $v\to 0$.}
Thus, CP-violating cases with very large
CP-even/odd mixing have to be considered with caution. The
  matching to appropriate EFTs that are not the SM but include more
  light degrees of freedom is left for future work.

\section{Conclusions}
\label{sec:conclusions}
In this paper, we presented our new computation of the
higher-order corrections to the SM-like Higgs boson of the
CP-violating NSSM for large mass hierarchies. In this case,
fixed-order computations become unreliable due to the involved
logarithms of large mass hierarchies, requiring the application of an
EFT framework. We chose a scenario where all non-SM particles are very
heavy, so that the low-energy EFT is given by the SM. The matching of
the full NMSSM to the SM at the high-scale is performed at full
one-loop order in the NMSSM, including two-loop corrections in the
MSSM limit. We applied two matching approaches given by the quartic
coupling matching in the unbroken theory and the pole mass matching
after EWSB. The latter includes terms of the order
$v^2/\msusy^2$, so that the comparison between the two methods
allowed us to estimate the importance of these terms that are
neglected in the former approach. We additionally provided an estimate of the
different sources of uncertainty.
Our new computation has been implemented in the public program package  
{\tt NMSSMCALC} and can be downloaded from the url:
\begin{center}
  \url{https://www.itp.kit.edu/~maggie/NMSSMCALC/}
\end{center}
For our numerical analysis, we performed a scan in the NMSSM parameter
space and only kept points that are in agreement with the Higgs signal
data and exclusion constraints on additional Higgs bosons and
supersymmetric particles. We validated our calculation and
implementation of the two matching procedures in {\tt NMSSMCALC}
against existing results in the literature and found good agreement for
the tested parameter point. We subsequently investigated the case of a light
singlet-like Higgs boson in the NMSSM spectrum. As expected, the EFT
approach applying the pole-mass matching shows good agreement with the
fixed-order result  (including partly resummation
in the top/stop sector by applying \DRbar renormalisation \cite{Dao:2019qaz})
within the theoretical uncertainty, while the
quartic coupling matching starts more and more deviating with
decreasing singlet mass due to the missing ${\cal
  O}(v^2/\msusy^2)$ terms. The behavior is confirmed by the
analysis of our entire found parameter sample. With increasing values
of the specific NMSSM coupling parameters $\lambda$ and $\kappa$, the
singlet-like Higgs boson mass decreases, and the 
$v^2/\msusy^2$ effects beome increasingly important,
deteriorating the description by the quartic coupling matching. Within
the uncertainty band of the pole-mass matching, the two matching
approached nevertheless can agree with each other even if the singlet mass
is as light as 200~GeV provided that $\sqrt{\lambda^2+\kappa^2}\lesssim 0.2$.
{We studied for the first time the effects of
CP violation in the EFT approach in the NMSSM. From a conceptual point of view,
we found that care has to be taken in the derivation of the
quartic-coupling matching in order not to miss finite contributions that do not appear in the
CP-conserving case nor the CP-violating MSSM.} This requires the
expansion of the tadpoles up to ${\cal O}(v\times \msusy^2)$. Both for loop-induced
and tree-level CP violation, we find the $v^2/\msusy^2$
contributions to the matching condition to become important for our benchmark point {\tt BP3}
which features a light singlet Higgs boson. 

{In summary, we find good agreement between the pole-mass and the quartic-coupling matching approaches within theoretical uncertainties and our EFT implementation reliably describes} NMSSM scenarios with a heavy
non-SM mass spectrum. Scenarios with light singlet-like states (\ie lighter than \unit[125]{GeV})
require the extension of the approach beyond the SM as effective
low-energy description. This is left for future work. 

\section*{Acknowledgements}
%%%%%%%%%%%%%%%%%%%%%%%%%%%%%%%%%%%%%%%%%%%%%%%%%%%%%%%%%%%%%%
{We thank Dominik St{\"o}ckinger and Johannes W{\"u}nsche for
    discovering a bug in the first release of the code that affected the implementation of the two-loop
$\alpha_t(\alpha_s+\alpha_t)$ results for the matching condition from the literature.}
The research of C.B.\ and M.M.\ was supported by the Deutsche
Forschungsgemeinschaft (DFG, German Research Foundation) under grant
396021762 - TRR 257. T.N.D thanks Phenikaa University for its financial support of this work. 
M.G.\ acknowledges support by the Deutsche Forschungsgemeinschaft (DFG, German Research Foundation) under Germany’s
Excellence Strategy – EXC 2121 “Quantum Universe” – 390833306 and partially by 491245950.
H.R.’s research is funded by the
Deutsche Forschungsgemeinschaft (DFG, German Research Foundation) — project
no. 442089526.

\begin{appendix}
%-------------------------------------------------
\section{Expansion of the One-Loop Tadpoles to $\mathcal{O}(v)$}
\label{app:tadexpand}
In this appendix we derive the leading terms of $\delta^{(1)}t_{a_d}/v$ which
are not suppressed by inverse powers of $\msusy$.
The tadpole counterterm $\delta^{(1)}t_{a_d}$ can be expanded to
$\order{v\times\msusy^2}$ in two equivalent ways.
The first method involves considerations about the dependence of couplings,
mixing matrices and mass eigenvalues on the SM VEV. The second method
relies on a systematic expansion in $v$.

\paragraph{\underline{Method (i):}} The couplings entering the Feynman
diagrams of $\delta^{(1)}t_{a_d}$ in general consist of a linear combination
of $v$-independent factors, which correspond to couplings defined in the
gauge-basis (or unbroken EW phase) and of $v$-dependent
mixing-matrices. The lowest dependence of the mass eigenvalues on $v$
is of ${\cal O}(v^2)$ in the tadpole diagrams. The only linear
dependence on $v$ arises in the mixing matrices that hence need to be expanded to $\order{v}$. The expansion of the mixing matrices can be
performed by using the solutions of the rotation matrices that have been
obtained analytically for the case $v=0$ as an ansatz and
introducing $\order{v}$-terms on the off-diagonal elements. These elements
can be determined by requiring unitarity of the
mixing matrix up to $\order{v^2/\msusy^2}$ and by requiring that the mass matrix
with the full $v$-dependence is diagonalised up to $\order{v^2/\msusy^2}$.
For the example of the stop-mixing matrix $\tilde{U}$ we find at $\order{v/\msusy}$:
\begin{align*}
    \tilde{U}_{ii} &= 1,\,\\
    \tilde{U}_{12}&=-\left(\tilde{U}_{21}\right)^* = \frac{v}{2}
  \frac{\sqrt2 e^{i (\varphi_w - \varphi_y + \varphi_{A_t})}|A_t|
    -|\lambda|v_s\cot\beta
    }{m_{\tilde Q_3}^2-m_{\tilde U_{R,3}}^2}
     e^{\varphi_\lambda+\varphi_s}\,,
\end{align*}
whereas the squared mass-eigenvalues {do not receive additional corrections at $\order{v/\msusy}$ but only at $\order{v^2/\msusy^2}$.}
The resulting stop-contribution to the counterterm reads
\begin{align}
    \delta^{(1)} t_{a_d} = \vev \frac{3 |\lambda| |A_t| \vs
  Y_t^2}{16\pi^2 \sqrt{2}\sin\beta} B(m_{\tilde Q_3},m_{\tilde
  U_{R,3}})\sin(\varphi_w-\varphi_y+\varphi_{A_t}) \,,
  \label{eq:dtadstop}
\end{align}
with
\begin{align*}
B(x,y)&=\frac{A(x) -A(y)}{x^2-y^2} \,,\\
B(x,x)&= \ln\frac{\mu^2}{x^2}\,,\\
A(x) &= x^2\left(1+\ln\frac{\mu^2}{x^2}\right)\,,\\
A(0) &= 0 \,.
\end{align*}
\paragraph{\underline{Method (ii):}} A systematic way of computing the
tadpole without performing an explicit expansion of mixing matrix
elements can be performed by considering the Taylor expansion of the
tadpole counter-term around $v=0$:
\begin{equation}
    \delta^{(1)} t_{a_d} = v \cdot \left. \frac{\partial t_{a_d}}{\partial
        v}\right|_{v=0} + \order{v^2}\,.
\end{equation}
The tadpole itself can be written as the derivative of the one-loop 
effective potential $V^{(1)}$ w.r.t.\ the field $a_d$, \ie $\delta^{(1)} t_{a_d}=
\partial V^{(1)}/\partial a_d$. The derivative of the potential w.r.t.\ the SM-like Higgs VEV
$v$ can be replaced by the derivative w.r.t.\ the Higgs field itself. Thus, we
find
\begin{subequations}
\begin{align}
    \delta^{(1)} t_{a_d} &=v \cdot\left. \frac{\partial^2 V(h,a_d,\dots)}{\partial a_d \partial h}\right|_{h=a_d=0;v=0}+\order{v^2}\\
    &= v\cdot \left.\Sigma_{h\to a_d}(p^2=0)\right|_{v=0}+\order{v^2}\\
    &=v\cdot \left. \left({-\cos\beta}\Sigma_{h\to G}(p^2=0) {+\sin\beta}
      \Sigma_{h\to A}(p^2=0)\right)\right|_{v=0} + \order{v^2} \,,
\end{align}
\end{subequations}
where in the second line we used that the second derivative of the effective
potential is equivalent to the self-energy evaluated at vanishing external
momentum. In the last line we rotated the second index of the $i\to j$
self-energy from the gauge-basis into the mass-basis using \cref{eq:rot}.
Note that the self-energy is again computed in the unbroken phase
and can be conveniently generated with e.g. {\tt SARAH}. Using this result we
find full agreement with the stop contribution derived above. {Furthermore, we
also computed the full result including contributions from
electroweakinos as well as the 
neutral and charged Higgses and found }full agreement (within the numerical
accuracy) with the result obtained when evaluating $\delta^{(1)}t_{a_d}/v$
numerically for $v=\unit[1]{GeV}$.

\section{New Implementation in \NMSSMCALC}
\label{app:implementation}
In this appendix we describe how the new and old Higgs mass prediction
in \NMSSMCALC is controlled using the \texttt{SLHA} interface, and
discuss examples of in- and output files.

The \texttt{SLHA} NMSSM input parameters are interpreted as \DRbar parameters
at the scale $Q_{\rm inp.}$ (cf.~\cref{fig:HiggsMassEFT} box 1b)).
The fixed-order calculation implemented in \NMSSMCALC allows to renormalise
the charged Higgs boson mass {$m_{H^\pm}^{\rm OS}$} on-shell or to alternatively
compute the charged Higgs boson mass at two-loop order and use $\Re
A_\lambda^{\DRbar}$ as input. Furthermore, \NMSSMCALC has two options
for the renormalisation scheme for the top/stop sector: on-shell and
$\DRb$. In contrast, the newly implemented EFT calculation currently
only takes SUSY contributions in the \DRbar scheme into account. Thus, we
recommend to use the \DRbar scheme in combination with the EFT
calculation. However, it is in principle still possible to choose a different
(the OS) scheme in the precision fixed-order
prediction of the BSM masses by using the flags 7 and 8 in the \texttt{MODSEL} block, \cf
  \cref{fig:modsel}. With flag 6, the user chooses the loop-order of
the fixed-order calculation. 
These flags only concern the fixed-order calculations which are
written out into the \texttt{MASS}
block. In addition, there are three new flags, 15, 16 and 17, to control the
calculation of the SM-like Higgs boson mass. If the EFT calculation
was chosen by setting the flag 15 to a value larger than 10, the entry of the SM-like
Higgs boson mass is overwritten with the EFT result while all
other \texttt{MASS} entries are those obtained with the fixed-order
calculation (flags 6,7,8). For scenarios, where the SM-like Higgs is the lightest of the
neutral Higgs bosons, this would be the entry of ``\texttt{MASS
  25}''. If, however, the dominantly $h_u$-like Higgs boson is not the lightest
neutral Higgs boson, the entry of the heavier neutral Higgs state
corresponding to the SM-like Higgs is overwritten with the EFT result. In
this case, however, the EFT approach is not valid any more as it
assumes all other Higgs bosons to be heavy. The
\texttt{NMHMIXC} block, containing the numerical
values of the neutral Higgs mixing matrix, corresponds to the matrix that
diagonalises the loop-corrected mass matrix if the fixed-order calculation
was chosen by the user. If the EFT calculation is chosen,
\texttt{NMHMIXC} corresponds to the tree-level mixing
matrix. Setting flag 15 to -2, the program decides 
whether the fixed-order or the EFT calculation is performed, for scenarios
where the size of the SUSY input scale (given in ``\texttt{EXTPAR 0}'') is larger than \unit[1.5]{TeV}.\footnote{Note that the applied criterion is only a
  rough estimator for the validity of the EFT/fixed-order calculation.}
  The
input scale, if not specified by the user, by default is set to
$\Qinp=\sqrt{m_{\tilde{Q}_3}m_{\tilde{u}_{R_3}}}$, which indicates the size of
the stop masses. Scenarios with \eg a light singlet, where the EFT approach
might be questionable, are therefore not recognised by this estimate. 
If flag 15 is set to -1, the program always
uses the fixed-order rather than the EFT calculation. For flag 15 being
between 11 and 24, the first digit (1 or 2) determines the EFT approach
(pole-mass or quartic-coupling matching) and the second digit turns on
various higher-order corrections in the EFT calculation, \cf
\cref{fig:modsel}. These are the full NMSSM one-loop corrections
including CP-violating effects computed in this work
and the 2-loop corrections in the MSSM limit at ${\cal O}(\alpha_s
\alpha_t)$  {with full mass dependence} \cite{PardoVega:2015eno}, ${\cal O} (\alpha_t^2)$
{in the case of degenerate masses} \cite{PardoVega:2015eno}, and ${\cal O}(\alpha_s
\alpha_{\text{EW}})$ \cite{Bagnaschi:2019esc}. 
Setting flag 16 to 12 or 22 chooses the $\alpha_{M_Z}$ or the
$G_F$-scheme in the extraction of the SM \MSbar parameters.
Alternatively, we implemented the \SMDR interpolation formulae \cite{Alam:2022cdv} for the extraction of the SM \MSbar parameters as discussed in Sec.~\ref{sec:numerical:validation}, which can be enabled by setting flag 16 to 49.
Flag 17 controls the uncertainty estimate of the EFT calculation.
  In \cref{fig:dmass} we show the new \texttt{DMASS} block, which
    contains the theoretical uncertainty of the EFT result, obtained for the
  parameter point \BP3 with the settings given in \cref{fig:modsel}.
  In addition to the total uncertainty (``\texttt{DMASS 25}'', corresponding to
  \cref{eq:uncert}), it also lists the individual uncertainties as well as the
  difference to the fixed-order result obtained with the \texttt{MODSEL}
  flags 6, 7 and 8. The mass uncertainty of the BSM Higgs masses is set to
  the commonly chosen fixed value of
  \unit[3]{GeV}. This is often larger than the overall size of the
  loop-corrections to the scalar BSM masses and therefore a rather conservative
  estimate.
\newcommand{\ind}{\phantom{\_}}
\newcommand{\indd}{\phantom{\_\_}}
\newcommand{\inddd}{\phantom{\_\_\_\_\_\_\_}\#\phantom{\_\_}}
\begin{figure}[tp]
    \centering
\fbox{\parbox{0.8\textwidth}{%
\texttt{Block MODSEL\\
\ind  3\indd1\indd   \# NMSSM\\
\ind  5\indd2\indd   \# 0:CP-conserving; 2:general CP-violation\\
\ind  6\indd3\indd   \# loop level of fixed-order calculations: \\ \inddd1:one-loop; 2:$\order{\alpha_t \alpha_s}$; 3:$\order{\alpha_t \alpha_s + \alpha_t^2}$;\\ \inddd 4:$\order{\alpha_t \alpha_s + (\alpha_t+\alpha_\lambda+\alpha_\kappa)^2}$\\
\ind  7\indd2\indd   \# top/stop sector (in fixed-order calculation):\\ \inddd 2:\DRbar; 3:OS scheme\\
\ind  8\indd0\indd   \# 0:MHpm as OS-input; 1:ReAlambda as \DRbar-input\\
\ind  10 0\indd   \# EDM calculation:0/1/2 (off/on/detailed output)\\
\ind  11 0\indd   \# Atextcolor{red}M calculation:0/1/2 (off/on/detailed output)\\
\ind  12 0\indd   \# compute effective HHH couplings:0/1 (off/on) \\
\ind  13 0\indd   \# loop-corrected W-mass:0/1 (off/on)\\
\ind  15 14  \# fixed-order or EFT.\\ \inddd -2:automatic; -1:always
fixed-order;\\ \inddd 1n/2n:EFT with pole-mass/quartic matching\\
\inddd \phantom{\_\_\_\_\_\_}at n-order with n=1:one-loop (this work); \\ \inddd  \phantom{\_\_\_\_\_}
n=2:$\order{\alpha_s\alpha_t}$ \cite{PardoVega:2015eno};
n=3:$\order{\alpha_t^2}$ \cite{PardoVega:2015eno}; \\ \inddd  \phantom{\_\_\_\_\_} n=4:$\order{\alpha_s\alpha_{\text{EW}}}$~\cite{Bagnaschi:2019esc}\\
\ind  16 22  \# extraction of \MSbar SM parameters using\\ \inddd$\alpha_{M_Z}$-scheme (12) or $G_F$-scheme (22),\\ \inddd\SMDR interpolation formulae \cite{Alam:2022cdv} (49) \\
\ind  17 1\indd   \# uncertainty estimate:0/1 (off/on)}%
}}
\caption{Example of all new and old flags available in the \texttt{MODSEL}-block of the \texttt{SLHA} input-file.}
\label{fig:modsel}
 \end{figure}
 \begin{figure}[tp]
    \centering
\fbox{\parbox{0.96\textwidth}{%
\texttt{
BLOCK DMASS  \# Theoreticalerror\\
\indd       25\indd    9.52016291E-01   \# H1\\
\indd       35\indd    3.00000000E+00   \# H2\\
\indd       36\indd    3.00000000E+00   \# H3\\
\indd       45\indd    3.00000000E+00   \# H4\\
\indd       46\indd    3.00000000E+00   \# H5\\
\indd       37\indd    3.00000000E+00   \# Hc\\
\ind       251\ind    -8.17298889E-01   \# $M_h({\Qmatch=\Qinp})-M_h({\Qmatch = \Qinp/2})$\\
\ind       252\indd    8.38518143E-01   \# $M_h({\Qmatch=\Qinp})-M_h({\Qmatch = 2\Qinp})$\\
\ind       253\ind    -4.01258469E-01   \# $\Delta^{\SM}_{Y_t} = M_h(Y_t^{{\SM},\order{\alpha_s^3}})-M_h(Y_t^{{\SM},\order{\alpha_s^2}})$\\
\ind       254\indd    2.09808350E-02   \# $\Delta^{\SM}_{G_F/\alpha_{M_Z}}= M_h^{G_F}-M_h^{\alpha_{M_Z}}$\\
\ind       255\ind    -2.29334831E+00   \# estimated size of $\nicefrac{v^2}{\msusy^2}$ terms ($\Delta_{v^2/\msusy^2}^{\mathrm{SUSY}}$)\\
\ind       256\ind   -7.32089989E-02   \# ${M_h^{\texttt{EFT}}-M_h^{\texttt{fixed-order}}}$ {\small (not included in total uncertainty)}\\
\ind       257\indd    2.04386369E-01   \# SM scale uncertainty ($\Delta^{\SM}_{\QEW}$) %
}}}
\caption{Example of the \texttt{DMASS} block in the \texttt{SLHA}
output-file. The numerical values correspond to the output obtained for the
benchmark point \BP3 using the pole-mass matching (\texttt{MODSEL 15=14}) and
the $G_F$-scheme (\texttt{MODSEL 16=22}).}
\label{fig:dmass}
 \end{figure}

\clearpage
\end{appendix}

\bibliographystyle{JHEP}
\bibliography{paper}

\providecommand{\href}[2]{#2}\begingroup\raggedright\begin{thebibliography}{100}

\bibitem{Aad:2012tfa}
{\scshape ATLAS Collaboration} collaboration, \emph{{Observation of a new
  particle in the search for the Standard Model Higgs boson with the ATLAS
  detector at the LHC}},
  \href{https://doi.org/10.1016/j.physletb.2012.08.020}{\emph{Phys.Lett.}
  {\bfseries B716} (2012) 1} [\href{https://arxiv.org/abs/1207.7214}{{\ttfamily
  1207.7214}}].

\bibitem{Chatrchyan:2012ufa}
{\scshape CMS Collaboration} collaboration, \emph{{Observation of a new boson
  at a mass of 125 GeV with the CMS experiment at the LHC}},
  \href{https://doi.org/10.1016/j.physletb.2012.08.021}{\emph{Phys.Lett.}
  {\bfseries B716} (2012) 30}
  [\href{https://arxiv.org/abs/1207.7235}{{\ttfamily 1207.7235}}].

\bibitem{Hahn:2013ria}
T.~Hahn, S.~Heinemeyer, W.~Hollik, H.~Rzehak and G.~Weiglein,
  \emph{{High-Precision Predictions for the Light CP -Even Higgs Boson Mass of
  the Minimal Supersymmetric Standard Model}},
  \href{https://doi.org/10.1103/PhysRevLett.112.141801}{\emph{Phys. Rev. Lett.}
  {\bfseries 112} (2014) 141801}
  [\href{https://arxiv.org/abs/1312.4937}{{\ttfamily 1312.4937}}].

\bibitem{Ellis:1988er}
J.R.~Ellis, J.~Gunion, H.E.~Haber, L.~Roszkowski and F.~Zwirner, \emph{{Higgs
  Bosons in a Nonminimal Supersymmetric Model}},
  \href{https://doi.org/10.1103/PhysRevD.39.844}{\emph{Phys.Rev.} {\bfseries
  D39} (1989) 844}.

\bibitem{Drees:1988fc}
M.~Drees, \emph{{Supersymmetric Models with Extended Higgs Sector}},
  \href{https://doi.org/10.1142/S0217751X89001448}{\emph{Int.J.Mod.Phys.}
  {\bfseries A4} (1989) 3635}.

\bibitem{Ellwanger:2009dp}
U.~Ellwanger, C.~Hugonie and A.M.~Teixeira, \emph{{The Next-to-Minimal
  Supersymmetric Standard Model}},
  \href{https://doi.org/10.1016/j.physrep.2010.07.001}{\emph{Phys.Rept.}
  {\bfseries 496} (2010) 1} [\href{https://arxiv.org/abs/0910.1785}{{\ttfamily
  0910.1785}}].

\bibitem{Maniatis:2009re}
M.~Maniatis, \emph{{The Next-to-Minimal Supersymmetric extension of the
  Standard Model reviewed}},
  \href{https://doi.org/10.1142/S0217751X10049827}{\emph{Int.J.Mod.Phys.}
  {\bfseries A25} (2010) 3505}
  [\href{https://arxiv.org/abs/0906.0777}{{\ttfamily 0906.0777}}].

\bibitem{Slavich:2020zjv}
P.~Slavich, S.~Heinemeyer et~al., \emph{{Higgs-mass predictions in the MSSM and
  beyond}}, \href{https://doi.org/10.1140/epjc/s10052-021-09198-2}{\emph{Eur.
  Phys. J. C} {\bfseries 81} (2021) 450}
  [\href{https://arxiv.org/abs/2012.15629}{{\ttfamily 2012.15629}}].

\bibitem{Ellwanger:1993hn}
U.~Ellwanger, \emph{{Radiative corrections to the neutral Higgs spectrum in
  supersymmetry with a gauge singlet}},
  \href{https://doi.org/10.1016/0370-2693(93)91431-L}{\emph{Phys. Lett. B}
  {\bfseries 303} (1993) 271}
  [\href{https://arxiv.org/abs/hep-ph/9302224}{{\ttfamily hep-ph/9302224}}].

\bibitem{Elliott:1993ex}
T.~Elliott, S.~King and P.~White, \emph{{Supersymmetric Higgs bosons at the
  limit}},
  \href{https://doi.org/10.1016/0370-2693(93)91107-X}{\emph{Phys.Lett.}
  {\bfseries B305} (1993) 71}
  [\href{https://arxiv.org/abs/hep-ph/9302202}{{\ttfamily hep-ph/9302202}}].

\bibitem{Elliott:1993uc}
T.~Elliott, S.~King and P.~White, \emph{{Squark contributions to Higgs boson
  masses in the next-to-minimal supersymmetric standard model}},
  \href{https://doi.org/10.1016/0370-2693(93)91321-D}{\emph{Phys.Lett.}
  {\bfseries B314} (1993) 56}
  [\href{https://arxiv.org/abs/hep-ph/9305282}{{\ttfamily hep-ph/9305282}}].

\bibitem{Elliott:1993bs}
T.~Elliott, S.~King and P.~White, \emph{{Radiative corrections to Higgs boson
  masses in the next-to-minimal supersymmetric Standard Model}},
  \href{https://doi.org/10.1103/PhysRevD.49.2435}{\emph{Phys.Rev.} {\bfseries
  D49} (1994) 2435} [\href{https://arxiv.org/abs/hep-ph/9308309}{{\ttfamily
  hep-ph/9308309}}].

\bibitem{Pandita:1993hx}
P.~Pandita, \emph{{One loop radiative corrections to the lightest Higgs scalar
  mass in nonminimal supersymmetric Standard Model}},
  \href{https://doi.org/10.1016/0370-2693(93)90137-7}{\emph{Phys. Lett. B}
  {\bfseries 318} (1993) 338}.

\bibitem{Pandita:1993tg}
P.~Pandita, \emph{{Radiative corrections to the scalar Higgs masses in a
  nonminimal supersymmetric Standard Model}},
  \href{https://doi.org/10.1007/BF01562550}{\emph{Z. Phys. C} {\bfseries 59}
  (1993) 575}.

\bibitem{King:1995vk}
S.~King and P.~White, \emph{{Resolving the constrained minimal and
  next-to-minimal supersymmetric standard models}},
  \href{https://doi.org/10.1103/PhysRevD.52.4183}{\emph{Phys. Rev. D}
  {\bfseries 52} (1995) 4183}
  [\href{https://arxiv.org/abs/hep-ph/9505326}{{\ttfamily hep-ph/9505326}}].

\bibitem{Ellwanger:2005fh}
U.~Ellwanger and C.~Hugonie, \emph{{Yukawa induced radiative corrections to the
  lightest Higgs boson mass in the NMSSM}},
  \href{https://doi.org/10.1016/j.physletb.2005.07.039}{\emph{Phys. Lett. B}
  {\bfseries 623} (2005) 93}
  [\href{https://arxiv.org/abs/hep-ph/0504269}{{\ttfamily hep-ph/0504269}}].

\bibitem{Degrassi:2009yq}
G.~Degrassi and P.~Slavich, \emph{{On the radiative corrections to the neutral
  Higgs boson masses in the NMSSM}},
  \href{https://doi.org/10.1016/j.nuclphysb.2009.09.018}{\emph{Nucl. Phys. B}
  {\bfseries 825} (2010) 119}
  [\href{https://arxiv.org/abs/0907.4682}{{\ttfamily 0907.4682}}].

\bibitem{Staub:2010ty}
F.~Staub, W.~Porod and B.~Herrmann, \emph{{The Electroweak sector of the NMSSM
  at the one-loop level}},
  \href{https://doi.org/10.1007/JHEP10(2010)040}{\emph{JHEP} {\bfseries 1010}
  (2010) 040}.

\bibitem{Drechsel:2016jdg}
P.~Drechsel, L.~Galeta, S.~Heinemeyer and G.~Weiglein, \emph{{Precise
  Predictions for the Higgs-Boson Masses in the NMSSM}},
  \href{https://doi.org/10.1140/epjc/s10052-017-4595-1}{\emph{Eur. Phys. J. C}
  {\bfseries 77} (2017) 42}.

\bibitem{Ham:2001kf}
S.~Ham, J.~Kim, S.~Oh and D.~Son, \emph{{The Charged Higgs boson in the
  next-to-minimal supersymmetric standard model with explicit CP violation}},
  \href{https://doi.org/10.1103/PhysRevD.64.035007}{\emph{Phys.Rev.} {\bfseries
  D64} (2001) 035007} [\href{https://arxiv.org/abs/hep-ph/0104144}{{\ttfamily
  hep-ph/0104144}}].

\bibitem{Ham:2001wt}
S.~Ham, S.~Oh and D.~Son, \emph{{Neutral Higgs sector of the next-to-minimal
  supersymmetric standard model with explicit CP violation}},
  \href{https://doi.org/10.1103/PhysRevD.65.075004}{\emph{Phys. Rev. D}
  {\bfseries 65} (2002) 075004}
  [\href{https://arxiv.org/abs/hep-ph/0110052}{{\ttfamily hep-ph/0110052}}].

\bibitem{Ham:2003jf}
S.~Ham, Y.~Jeong and S.~Oh, \emph{{Radiative CP violation in the Higgs sector
  of the next-to-minimal supersymmetric model}},
  \href{https://arxiv.org/abs/hep-ph/0308264}{{\ttfamily hep-ph/0308264}}.

\bibitem{Funakubo:2004ka}
K.~Funakubo and S.~Tao, \emph{{The Higgs sector in the next-to-MSSM}},
  \href{https://doi.org/10.1143/PTP.113.821}{\emph{Prog.Theor.Phys.} {\bfseries
  113} (2005) 821} [\href{https://arxiv.org/abs/hep-ph/0409294}{{\ttfamily
  hep-ph/0409294}}].

\bibitem{Ham:2007mt}
S.~Ham, S.~Kim, S.~Oh and D.~Son, \emph{{Higgs bosons of the NMSSM with
  explicit CP violation at the ILC}},
  \href{https://doi.org/10.1103/PhysRevD.76.115013}{\emph{Phys.Rev.} {\bfseries
  D76} (2007) 115013} [\href{https://arxiv.org/abs/0708.2755}{{\ttfamily
  0708.2755}}].

\bibitem{Cheung:2010ba}
K.~Cheung, T.-J.~Hou, J.S.~Lee and E.~Senaha, \emph{{The Higgs Boson Sector of
  the Next-to-MSSM with CP Violation}},
  \href{https://doi.org/10.1103/PhysRevD.82.075007}{\emph{Phys.Rev.} {\bfseries
  D82} (2010) 075007} [\href{https://arxiv.org/abs/1006.1458}{{\ttfamily
  1006.1458}}].

\bibitem{Goodsell:2016udb}
M.D.~Goodsell and F.~Staub, \emph{{The Higgs mass in the CP violating MSSM,
  NMSSM, and beyond}},
  \href{https://doi.org/10.1140/epjc/s10052-016-4495-9}{\emph{Eur. Phys. J. C}
  {\bfseries 77} (2017) 46}.

\bibitem{Domingo:2017rhb}
F.~Domingo, P.~Drechsel and S.~Pa{\ss}ehr, \emph{{On-Shell neutral Higgs bosons
  in the NMSSM with complex parameters}},
  \href{https://doi.org/10.1140/epjc/s10052-017-5104-2}{\emph{Eur. Phys. J.}
  {\bfseries C77} (2017) 562}
  [\href{https://arxiv.org/abs/1706.00437}{{\ttfamily 1706.00437}}].

\bibitem{Goodsell:2014pla}
M.D.~Goodsell, K.~Nickel and F.~Staub, \emph{{Two-loop corrections to the Higgs
  masses in the NMSSM}},
  \href{https://doi.org/10.1103/PhysRevD.91.035021}{\emph{Phys. Rev.}
  {\bfseries D91} (2015) 035021}.

\bibitem{Ender:2011qh}
K.~Ender, T.~Graf, M.~Muhlleitner and H.~Rzehak, \emph{{Analysis of the NMSSM
  Higgs Boson Masses at One-Loop Level}},
  \href{https://doi.org/10.1103/PhysRevD.85.075024}{\emph{Phys. Rev. D}
  {\bfseries 85} (2012) 075024}.

\bibitem{Graf:2012hh}
T.~Graf, R.~Grober, M.~Muhlleitner, H.~Rzehak and K.~Walz, \emph{{Higgs Boson
  Masses in the Complex NMSSM at One-Loop Level}},
  \href{https://doi.org/10.1007/JHEP10(2012)122}{\emph{JHEP} {\bfseries 10}
  (2012) 122}.

\bibitem{Muhlleitner:2014vsa}
M.~Mühlleitner, D.T.~Nhung, H.~Rzehak and K.~Walz, \emph{{Two-loop
  contributions of the order $ \mathcal{O}\left({\alpha}_t{\alpha}_s\right) $
  to the masses of the Higgs bosons in the CP-violating NMSSM}},
  \href{https://doi.org/10.1007/JHEP05(2015)128}{\emph{JHEP} {\bfseries 05}
  (2015) 128} [\href{https://arxiv.org/abs/1412.0918}{{\ttfamily 1412.0918}}].

\bibitem{Dao:2019qaz}
T.N.~Dao, R.~Gr\"ober, M.~Krause, M.~M\"uhlleitner and H.~Rzehak,
  \emph{{Two-loop $ \mathcal{O} $ ( $ {\alpha}_t^2 $ ) corrections to the
  neutral Higgs boson masses in the CP-violating NMSSM}},
  \href{https://doi.org/10.1007/JHEP08(2019)114}{\emph{JHEP} {\bfseries 08}
  (2019) 114} [\href{https://arxiv.org/abs/1903.11358}{{\ttfamily
  1903.11358}}].

\bibitem{Dao:2021khm}
T.N.~Dao, M.~Gabelmann, M.~M\"uhlleitner and H.~Rzehak, \emph{{Two-loop $
  \mathcal{O} $((\ensuremath{\alpha}$_{t}$ + \ensuremath{\alpha}$_{\lambda}$ +
  \ensuremath{\alpha}$_{\kappa}$)$^{2}$) corrections to the Higgs boson masses
  in the CP-violating NMSSM}},
  \href{https://doi.org/10.1007/JHEP09(2021)193}{\emph{JHEP} {\bfseries 09}
  (2021) 193} [\href{https://arxiv.org/abs/2106.06990}{{\ttfamily
  2106.06990}}].

\bibitem{Baglio:2013iia}
J.~Baglio, R.~Grober, M.~Muhlleitner, D.~Nhung, H.~Rzehak et~al.,
  \emph{{NMSSMCALC: A Program Package for the Calculation of Loop-Corrected
  Higgs Boson Masses and Decay Widths in the (Complex) NMSSM}},
  \href{https://doi.org/10.1016/j.cpc.2014.08.005}{\emph{Comput.Phys.Commun.}
  {\bfseries 185} (2014) 3372}
  [\href{https://arxiv.org/abs/1312.4788}{{\ttfamily 1312.4788}}].

\bibitem{Nhung:2013lpa}
D.T.~Nhung, M.~Muhlleitner, J.~Streicher and K.~Walz, \emph{{Higher Order
  Corrections to the Trilinear Higgs Self-Couplings in the Real NMSSM}},
  \href{https://doi.org/10.1007/JHEP11(2013)181}{\emph{JHEP} {\bfseries 1311}
  (2013) 181} [\href{https://arxiv.org/abs/1306.3926}{{\ttfamily 1306.3926}}].

\bibitem{Muhlleitner:2015dua}
M.~Mühlleitner, D.T.~Nhung and H.~Ziesche, \emph{{The order $
  \mathcal{O}\left({\alpha}_t{\alpha}_s\right) $ corrections to the trilinear
  Higgs self-couplings in the complex NMSSM}},
  \href{https://doi.org/10.1007/JHEP12(2015)034}{\emph{JHEP} {\bfseries 12}
  (2015) 034}.

\bibitem{Borschensky:2022pfc}
C.~Borschensky, T.N.~Dao, M.~Gabelmann, M.~M\"uhlleitner and H.~Rzehak,
  \emph{{The trilinear Higgs self-couplings at $\mathcal {O}(\alpha _t^2)$ in
  the CP-violating NMSSM}},
  \href{https://doi.org/10.1140/epjc/s10052-023-11215-5}{\emph{Eur. Phys. J. C}
  {\bfseries 83} (2023) 118}
  [\href{https://arxiv.org/abs/2210.02104}{{\ttfamily 2210.02104}}].

\bibitem{Dao:2023kzz}
T.N.~Dao, M.~Gabelmann and M.~M\"uhlleitner, \emph{{The $\mathcal{O}(\alpha
  _t+\alpha _\lambda +\alpha _\kappa )^2$ correction to the $\rho $ parameter
  and its effect on the W boson mass calculation in the complex NMSSM}},
  \href{https://doi.org/10.1140/epjc/s10052-023-12236-w}{\emph{Eur. Phys. J. C}
  {\bfseries 83} (2023) 1079}
  [\href{https://arxiv.org/abs/2308.04059}{{\ttfamily 2308.04059}}].

\bibitem{Ellwanger:2004xm}
U.~Ellwanger, J.F.~Gunion and C.~Hugonie, \emph{{NMHDECAY: A Fortran code for
  the Higgs masses, couplings and decay widths in the NMSSM}},
  \href{https://doi.org/10.1088/1126-6708/2005/02/066}{\emph{JHEP} {\bfseries
  02} (2005) 066} [\href{https://arxiv.org/abs/hep-ph/0406215}{{\ttfamily
  hep-ph/0406215}}].

\bibitem{Ellwanger:2005dv}
U.~Ellwanger and C.~Hugonie, \emph{{NMHDECAY 2.0: An Updated program for
  sparticle masses, Higgs masses, couplings and decay widths in the NMSSM}},
  \href{https://doi.org/10.1016/j.cpc.2006.04.004}{\emph{Comput. Phys. Commun.}
  {\bfseries 175} (2006) 290}
  [\href{https://arxiv.org/abs/hep-ph/0508022}{{\ttfamily hep-ph/0508022}}].

\bibitem{Staub:2009bi}
F.~Staub, \emph{{From Superpotential to Model Files for FeynArts and
  CalcHep/CompHep}},
  \href{https://doi.org/10.1016/j.cpc.2010.01.011}{\emph{Comput.Phys.Commun.}
  {\bfseries 181} (2010) 1077}
  [\href{https://arxiv.org/abs/0909.2863}{{\ttfamily 0909.2863}}].

\bibitem{Staub:2010jh}
F.~Staub, \emph{{Automatic Calculation of supersymmetric Renormalization Group
  Equations and Self Energies}},
  \href{https://doi.org/10.1016/j.cpc.2010.11.030}{\emph{Comput.Phys.Commun.}
  {\bfseries 182} (2011) 808}
  [\href{https://arxiv.org/abs/1002.0840}{{\ttfamily 1002.0840}}].

\bibitem{Staub:2012pb}
F.~Staub, \emph{{SARAH 3.2: Dirac Gauginos, UFO output, and more}},
  \href{https://doi.org/10.1016/j.cpc.2013.02.019}{\emph{Computer Physics
  Communications} {\bfseries 184} (2013) pp. 1792}
  [\href{https://arxiv.org/abs/1207.0906}{{\ttfamily 1207.0906}}].

\bibitem{Staub:2013tta}
F.~Staub, \emph{{SARAH 4: A tool for (not only SUSY) model builders}},
  \href{https://doi.org/10.1016/j.cpc.2014.02.018}{\emph{Comput.Phys.Commun.}
  {\bfseries 185} (2014) 1773}
  [\href{https://arxiv.org/abs/1309.7223}{{\ttfamily 1309.7223}}].

\bibitem{Porod:2003um}
W.~Porod, \emph{{SPheno, a program for calculating supersymmetric spectra, SUSY
  particle decays and SUSY particle production at e+ e- colliders}},
  \href{https://doi.org/10.1016/S0010-4655(03)00222-4}{\emph{Comput. Phys.
  Commun.} {\bfseries 153} (2003) 275}
  [\href{https://arxiv.org/abs/hep-ph/0301101}{{\ttfamily hep-ph/0301101}}].

\bibitem{Porod:2011nf}
W.~Porod and F.~Staub, \emph{{SPheno 3.1: Extensions including flavour,
  CP-phases and models beyond the MSSM}},
  \href{https://doi.org/10.1016/j.cpc.2012.05.021}{\emph{Comput. Phys. Commun.}
  {\bfseries 183} (2012) 2458}
  [\href{https://arxiv.org/abs/1104.1573}{{\ttfamily 1104.1573}}].

\bibitem{Allanach:2001kg}
B.C.~Allanach, \emph{{SOFTSUSY: a program for calculating supersymmetric
  spectra}}, \href{https://doi.org/10.1016/S0010-4655(01)00460-X}{\emph{Comput.
  Phys. Commun.} {\bfseries 143} (2002) 305}
  [\href{https://arxiv.org/abs/hep-ph/0104145}{{\ttfamily hep-ph/0104145}}].

\bibitem{Allanach:2017hcf}
B.C.~Allanach and T.~Cridge, \emph{{The Calculation of Sparticle and Higgs
  Decays in the Minimal and Next-to-Minimal Supersymmetric Standard Models:
  SOFTSUSY4.0}}, \href{https://doi.org/10.1016/j.cpc.2017.07.021}{\emph{Comput.
  Phys. Commun.} {\bfseries 220} (2017) 417}
  [\href{https://arxiv.org/abs/1703.09717}{{\ttfamily 1703.09717}}].

\bibitem{Athron:2014yba}
P.~Athron, J.-h.~Park, D.~St\"ockinger and A.~Voigt,
  \emph{{FlexibleSUSY\textemdash{}A spectrum generator generator for
  supersymmetric models}},
  \href{https://doi.org/10.1016/j.cpc.2014.12.020}{\emph{Comput. Phys. Commun.}
  {\bfseries 190} (2015) 139}
  [\href{https://arxiv.org/abs/1406.2319}{{\ttfamily 1406.2319}}].

\bibitem{Athron:2017fvs}
P.~Athron, M.~Bach, D.~Harries, T.~Kwasnitza, J.-h.~Park, D.~St\"ockinger
  et~al., \emph{{FlexibleSUSY 2.0: Extensions to investigate the phenomenology
  of SUSY and non-SUSY models}},
  \href{https://doi.org/10.1016/j.cpc.2018.04.016}{\emph{Comput. Phys. Commun.}
  {\bfseries 230} (2018) 145}
  [\href{https://arxiv.org/abs/1710.03760}{{\ttfamily 1710.03760}}].

\bibitem{Athron:2016fuq}
P.~Athron, J.-h.~Park, T.~Steudtner, D.~St\"ockinger and A.~Voigt,
  \emph{{Precise Higgs mass calculations in (non-)minimal supersymmetry at both
  high and low scales}},
  \href{https://doi.org/10.1007/JHEP01(2017)079}{\emph{JHEP} {\bfseries 01}
  (2017) 079} [\href{https://arxiv.org/abs/1609.00371}{{\ttfamily
  1609.00371}}].

\bibitem{Staub:2017jnp}
F.~Staub and W.~Porod, \emph{{Improved predictions for intermediate and heavy
  Supersymmetry in the MSSM and beyond}},
  \href{https://doi.org/10.1140/epjc/s10052-017-4893-7}{\emph{Eur. Phys. J. C}
  {\bfseries 77} (2017) 338}
  [\href{https://arxiv.org/abs/1703.03267}{{\ttfamily 1703.03267}}].

\bibitem{Bagnaschi:2022zvd}
E.~Bagnaschi, M.~Goodsell and P.~Slavich, \emph{{Higgs-mass prediction in the
  NMSSM with heavy BSM particles}},
  \href{https://doi.org/10.1140/epjc/s10052-022-10810-2}{\emph{Eur. Phys. J. C}
  {\bfseries 82} (2022) 853}
  [\href{https://arxiv.org/abs/2206.04618}{{\ttfamily 2206.04618}}].

\bibitem{Baglio:2019nlc}
J.~Baglio, T.N.~Dao and M.~M\"uhlleitner, \emph{{One-Loop Corrections to the
  Two-Body Decays of the Neutral Higgs Bosons in the Complex NMSSM}},
  \href{https://doi.org/10.1140/epjc/s10052-020-08520-8}{\emph{Eur. Phys. J. C}
  {\bfseries 80} (2020) 960}
  [\href{https://arxiv.org/abs/1907.12060}{{\ttfamily 1907.12060}}].

\bibitem{Bagnaschi:2014rsa}
E.~Bagnaschi, G.F.~Giudice, P.~Slavich and A.~Strumia, \emph{{Higgs Mass and
  Unnatural Supersymmetry}},
  \href{https://doi.org/10.1007/JHEP09(2014)092}{\emph{JHEP} {\bfseries 09}
  (2014) 092} [\href{https://arxiv.org/abs/1407.4081}{{\ttfamily 1407.4081}}].

\bibitem{Bahl:2020jaq}
H.~Bahl and I.~Sobolev, \emph{{Two-loop matching of renormalizable operators:
  general considerations and applications}},
  \href{https://doi.org/10.1007/JHEP03(2021)286}{\emph{JHEP} {\bfseries 03}
  (2021) 286} [\href{https://arxiv.org/abs/2010.01989}{{\ttfamily
  2010.01989}}].

\bibitem{Buttazzo:2013uya}
D.~Buttazzo, G.~Degrassi, P.P.~Giardino, G.F.~Giudice, F.~Sala, A.~Salvio
  et~al., \emph{{Investigating the near-criticality of the Higgs boson}},
  \href{https://doi.org/10.1007/JHEP12(2013)089}{\emph{JHEP} {\bfseries 12}
  (2013) 089} [\href{https://arxiv.org/abs/1307.3536}{{\ttfamily 1307.3536}}].

\bibitem{Schienbein:2018fsw}
I.~Schienbein, F.~Staub, T.~Steudtner and K.~Svirina, \emph{{Revisiting RGEs
  for general gauge theories}},
  \href{https://doi.org/10.1016/j.nuclphysb.2018.12.001}{\emph{Nucl. Phys. B}
  {\bfseries 939} (2019) 1} [\href{https://arxiv.org/abs/1809.06797}{{\ttfamily
  1809.06797}}].

\bibitem{Chetyrkin:2012rz}
K.G.~Chetyrkin and M.F.~Zoller, \emph{{Three-loop
  \textbackslash{}beta-functions for top-Yukawa and the Higgs self-interaction
  in the Standard Model}},
  \href{https://doi.org/10.1007/JHEP06(2012)033}{\emph{JHEP} {\bfseries 06}
  (2012) 033} [\href{https://arxiv.org/abs/1205.2892}{{\ttfamily 1205.2892}}].

\bibitem{Chetyrkin:2016ruf}
K.G.~Chetyrkin and M.F.~Zoller, \emph{{Leading QCD-induced four-loop
  contributions to the \ensuremath{\beta}-function of the Higgs self-coupling
  in the SM and vacuum stability}},
  \href{https://doi.org/10.1007/JHEP06(2016)175}{\emph{JHEP} {\bfseries 06}
  (2016) 175} [\href{https://arxiv.org/abs/1604.00853}{{\ttfamily
  1604.00853}}].

\bibitem{Martin:1993yx}
S.P.~Martin and M.T.~Vaughn, \emph{{Regularization dependence of running
  couplings in softly broken supersymmetry}},
  \href{https://doi.org/10.1016/0370-2693(93)90136-6}{\emph{Phys. Lett. B}
  {\bfseries 318} (1993) 331}
  [\href{https://arxiv.org/abs/hep-ph/9308222}{{\ttfamily hep-ph/9308222}}].

\bibitem{Braathen:2018htl}
J.~Braathen, M.D.~Goodsell and P.~Slavich, \emph{{Matching renormalisable
  couplings: simple schemes and a plot}},
  \href{https://doi.org/10.1140/epjc/s10052-019-7093-9}{\emph{Eur. Phys. J. C}
  {\bfseries 79} (2019) 669}
  [\href{https://arxiv.org/abs/1810.09388}{{\ttfamily 1810.09388}}].

\bibitem{Machacek:1983tz}
M.E.~Machacek and M.T.~Vaughn, \emph{{Two Loop Renormalization Group Equations
  in a General Quantum Field Theory. 1. Wave Function Renormalization}},
  \href{https://doi.org/10.1016/0550-3213(83)90610-7}{\emph{Nucl. Phys. B}
  {\bfseries 222} (1983) 83}.

\bibitem{Machacek:1983fi}
M.E.~Machacek and M.T.~Vaughn, \emph{{Two Loop Renormalization Group Equations
  in a General Quantum Field Theory. 2. Yukawa Couplings}},
  \href{https://doi.org/10.1016/0550-3213(84)90533-9}{\emph{Nucl. Phys. B}
  {\bfseries 236} (1984) 221}.

\bibitem{Machacek:1984zw}
M.E.~Machacek and M.T.~Vaughn, \emph{{Two Loop Renormalization Group Equations
  in a General Quantum Field Theory. 3. Scalar Quartic Couplings}},
  \href{https://doi.org/10.1016/0550-3213(85)90040-9}{\emph{Nucl. Phys. B}
  {\bfseries 249} (1985) 70}.

\bibitem{Sperling:2013eva}
M.~Sperling, D.~St\"ockinger and A.~Voigt, \emph{{Renormalization of vacuum
  expectation values in spontaneously broken gauge theories}},
  \href{https://doi.org/10.1007/JHEP07(2013)132}{\emph{JHEP} {\bfseries 07}
  (2013) 132} [\href{https://arxiv.org/abs/1305.1548}{{\ttfamily 1305.1548}}].

\bibitem{Sperling:2013xqa}
M.~Sperling, D.~St\"ockinger and A.~Voigt, \emph{{Renormalization of vacuum
  expectation values in spontaneously broken gauge theories: Two-loop
  results}}, \href{https://doi.org/10.1007/JHEP01(2014)068}{\emph{JHEP}
  {\bfseries 01} (2014) 068} [\href{https://arxiv.org/abs/1310.7629}{{\ttfamily
  1310.7629}}].

\bibitem{PardoVega:2015eno}
J.~Pardo~Vega and G.~Villadoro, \emph{{SusyHD: Higgs mass Determination in
  Supersymmetry}}, \href{https://doi.org/10.1007/JHEP07(2015)159}{\emph{JHEP}
  {\bfseries 07} (2015) 159}
  [\href{https://arxiv.org/abs/1504.05200}{{\ttfamily 1504.05200}}].

\bibitem{Bagnaschi:2019esc}
E.~Bagnaschi, G.~Degrassi, S.~Pa\ss{}ehr and P.~Slavich, \emph{{Full two-loop
  QCD corrections to the Higgs mass in the MSSM with heavy superpartners}},
  \href{https://doi.org/10.1140/epjc/s10052-019-7417-9}{\emph{Eur. Phys. J. C}
  {\bfseries 79} (2019) 910}
  [\href{https://arxiv.org/abs/1908.01670}{{\ttfamily 1908.01670}}].

\bibitem{Gabelmann:2018axh}
M.~Gabelmann, M.~M\"uhlleitner and F.~Staub, \emph{{Automatised matching
  between two scalar sectors at the one-loop level}},
  \href{https://doi.org/10.1140/epjc/s10052-019-6570-5}{\emph{Eur. Phys. J. C}
  {\bfseries 79} (2019) 163}
  [\href{https://arxiv.org/abs/1810.12326}{{\ttfamily 1810.12326}}].

\bibitem{Kublbeck:1990xc}
J.~Kublbeck, M.~Bohm and A.~Denner, \emph{{Feyn Arts: Computer Algebraic
  Generation of Feynman Graphs and Amplitudes}},
  \href{https://doi.org/10.1016/0010-4655(90)90001-H}{\emph{Comput.Phys.Commun.}
  {\bfseries 60} (1990) 165}.

\bibitem{Hahn:2000kx}
T.~Hahn, \emph{{Generating Feynman diagrams and amplitudes with FeynArts 3}},
  \href{https://doi.org/10.1016/S0010-4655(01)00290-9}{\emph{Comput.Phys.Commun.}
  {\bfseries 140} (2001) 418}
  [\href{https://arxiv.org/abs/hep-ph/0012260}{{\ttfamily hep-ph/0012260}}].

\bibitem{FeynCalc}
R.~Mertig, M.~B{\"o}hm and A.~Denner, \emph{Feyn calc - computer-algebraic
  calculation of feynman amplitudes},
  \href{https://doi.org/http://dx.doi.org/10.1016/0010-4655(91)90130-D}{\emph{Computer
  Physics Communications} {\bfseries 64} (1991) 345 }.

\bibitem{Shtabovenko:2016sxi}
V.~Shtabovenko, R.~Mertig and F.~Orellana, \emph{{New Developments in FeynCalc
  9.0}}, \href{https://doi.org/10.1016/j.cpc.2016.06.008}{\emph{Comput. Phys.
  Commun.} {\bfseries 207} (2016) 432}
  [\href{https://arxiv.org/abs/1601.01167}{{\ttfamily 1601.01167}}].

\bibitem{Shtabovenko:2020gxv}
V.~Shtabovenko, R.~Mertig and F.~Orellana, \emph{{FeynCalc 9.3: New features
  and improvements}},
  \href{https://doi.org/10.1016/j.cpc.2020.107478}{\emph{Comput. Phys. Commun.}
  {\bfseries 256} (2020) 107478}
  [\href{https://arxiv.org/abs/2001.04407}{{\ttfamily 2001.04407}}].

\bibitem{Dao:2019nxi}
T.N.~Dao, L.~Fritz, M.~Krause, M.~M\"uhlleitner and S.~Patel, \emph{{Gauge
  dependences of higher-order corrections to NMSSM Higgs boson masses and the
  charged Higgs Decay ${H^{\pm } \rightarrow W^\pm h_{i}}$}},
  \href{https://doi.org/10.1140/epjc/s10052-020-7837-6}{\emph{Eur. Phys. J. C}
  {\bfseries 80} (2020) 292}
  [\href{https://arxiv.org/abs/1911.07197}{{\ttfamily 1911.07197}}].

\bibitem{Domingo:2020wiy}
F.~Domingo and S.~Pa\ss{}ehr, \emph{{Towards Higgs masses and decay widths
  satisfying the symmetries in the (N)MSSM}},
  \href{https://doi.org/10.1140/epjc/s10052-020-08655-8}{\emph{Eur. Phys. J. C}
  {\bfseries 80} (2020) 1124}
  [\href{https://arxiv.org/abs/2007.11010}{{\ttfamily 2007.11010}}].

\bibitem{ParticleDataGroup:2022pth}
{\scshape Particle Data Group} collaboration, \emph{{Review of Particle
  Physics}}, \href{https://doi.org/10.1093/ptep/ptac097}{\emph{PTEP} {\bfseries
  2022} (2022) 083C01}.

\bibitem{Bahl:2022igd}
H.~Bahl, T.~Biek\"otter, S.~Heinemeyer, C.~Li, S.~Paasch, G.~Weiglein et~al.,
  \emph{{HiggsTools: BSM scalar phenomenology with new versions of HiggsBounds
  and HiggsSignals}},  \href{https://arxiv.org/abs/2210.09332}{{\ttfamily
  2210.09332}}.

\bibitem{Bechtle:2020pkv}
P.~Bechtle, D.~Dercks, S.~Heinemeyer, T.~Klingl, T.~Stefaniak, G.~Weiglein
  et~al., \emph{{HiggsBounds-5: Testing Higgs Sectors in the LHC 13 TeV Era}},
  \href{https://arxiv.org/abs/2006.06007}{{\ttfamily 2006.06007}}.

\bibitem{Bechtle:2020uwn}
P.~Bechtle, S.~Heinemeyer, T.~Klingl, T.~Stefaniak, G.~Weiglein and
  J.~Wittbrodt, \emph{{HiggsSignals-2: Probing new physics with precision Higgs
  measurements in the LHC 13 TeV era}},
  \href{https://doi.org/10.1140/epjc/s10052-021-08942-y}{\emph{Eur. Phys. J. C}
  {\bfseries 81} (2021) 145}
  [\href{https://arxiv.org/abs/2012.09197}{{\ttfamily 2012.09197}}].

\bibitem{Masip:1998jc}
M.~Masip, R.~Munoz-Tapia and A.~Pomarol, \emph{{Limits on the mass of the
  lightest Higgs in supersymmetric models}},
  \href{https://doi.org/10.1103/PhysRevD.57.R5340}{\emph{Phys. Rev. D}
  {\bfseries 57} (1998) R5340}
  [\href{https://arxiv.org/abs/hep-ph/9801437}{{\ttfamily hep-ph/9801437}}].

\bibitem{Kniehl:2016enc}
B.A.~Kniehl, A.F.~Pikelner and O.L.~Veretin, \emph{{mr: a C++ library for the
  matching and running of the Standard Model parameters}},
  \href{https://doi.org/10.1016/j.cpc.2016.04.017}{\emph{Comput. Phys. Commun.}
  {\bfseries 206} (2016) 84}
  [\href{https://arxiv.org/abs/1601.08143}{{\ttfamily 1601.08143}}].

\bibitem{Bednyakov:2012rb}
A.V.~Bednyakov, A.F.~Pikelner and V.N.~Velizhanin, \emph{{Anomalous dimensions
  of gauge fields and gauge coupling beta-functions in the Standard Model at
  three loops}}, \href{https://doi.org/10.1007/JHEP01(2013)017}{\emph{JHEP}
  {\bfseries 01} (2013) 017} [\href{https://arxiv.org/abs/1210.6873}{{\ttfamily
  1210.6873}}].

\bibitem{Bednyakov:2012en}
A.V.~Bednyakov, A.F.~Pikelner and V.N.~Velizhanin, \emph{{Yukawa coupling
  beta-functions in the Standard Model at three loops}},
  \href{https://doi.org/10.1016/j.physletb.2013.04.038}{\emph{Phys. Lett. B}
  {\bfseries 722} (2013) 336}
  [\href{https://arxiv.org/abs/1212.6829}{{\ttfamily 1212.6829}}].

\bibitem{Bednyakov:2013eba}
A.V.~Bednyakov, A.F.~Pikelner and V.N.~Velizhanin, \emph{{Higgs self-coupling
  beta-function in the Standard Model at three loops}},
  \href{https://doi.org/10.1016/j.nuclphysb.2013.07.015}{\emph{Nucl. Phys. B}
  {\bfseries 875} (2013) 552}
  [\href{https://arxiv.org/abs/1303.4364}{{\ttfamily 1303.4364}}].

\bibitem{Kniehl:2015nwa}
B.A.~Kniehl, A.F.~Pikelner and O.L.~Veretin, \emph{{Two-loop electroweak
  threshold corrections in the Standard Model}},
  \href{https://doi.org/10.1016/j.nuclphysb.2015.04.010}{\emph{Nucl. Phys. B}
  {\bfseries 896} (2015) 19}
  [\href{https://arxiv.org/abs/1503.02138}{{\ttfamily 1503.02138}}].

\bibitem{vanRitbergen:1997va}
T.~van Ritbergen, J.A.M.~Vermaseren and S.A.~Larin, \emph{{The Four loop beta
  function in quantum chromodynamics}},
  \href{https://doi.org/10.1016/S0370-2693(97)00370-5}{\emph{Phys. Lett. B}
  {\bfseries 400} (1997) 379}
  [\href{https://arxiv.org/abs/hep-ph/9701390}{{\ttfamily hep-ph/9701390}}].

\bibitem{Vermaseren:1997fq}
J.A.M.~Vermaseren, S.A.~Larin and T.~van Ritbergen, \emph{{The four loop quark
  mass anomalous dimension and the invariant quark mass}},
  \href{https://doi.org/10.1016/S0370-2693(97)00660-6}{\emph{Phys. Lett. B}
  {\bfseries 405} (1997) 327}
  [\href{https://arxiv.org/abs/hep-ph/9703284}{{\ttfamily hep-ph/9703284}}].

\bibitem{Bezrukov:2012sa}
F.~Bezrukov, M.Y.~Kalmykov, B.A.~Kniehl and M.~Shaposhnikov, \emph{{Higgs Boson
  Mass and New Physics}},
  \href{https://doi.org/10.1007/JHEP10(2012)140}{\emph{JHEP} {\bfseries 10}
  (2012) 140} [\href{https://arxiv.org/abs/1205.2893}{{\ttfamily 1205.2893}}].

\bibitem{Jegerlehner:2001fb}
F.~Jegerlehner, M.Y.~Kalmykov and O.~Veretin, \emph{{MS versus pole masses of
  gauge bosons: Electroweak bosonic two loop corrections}},
  \href{https://doi.org/10.1016/S0550-3213(02)00613-2}{\emph{Nucl. Phys. B}
  {\bfseries 641} (2002) 285}
  [\href{https://arxiv.org/abs/hep-ph/0105304}{{\ttfamily hep-ph/0105304}}].

\bibitem{Jegerlehner:2002em}
F.~Jegerlehner, M.Y.~Kalmykov and O.~Veretin, \emph{{MS-bar versus pole masses
  of gauge bosons. 2. Two loop electroweak fermion corrections}},
  \href{https://doi.org/10.1016/S0550-3213(03)00177-9}{\emph{Nucl. Phys. B}
  {\bfseries 658} (2003) 49}
  [\href{https://arxiv.org/abs/hep-ph/0212319}{{\ttfamily hep-ph/0212319}}].

\bibitem{Jegerlehner:2003py}
F.~Jegerlehner and M.Y.~Kalmykov, \emph{{O(alpha alpha(s)) correction to the
  pole mass of the t quark within the standard model}},
  \href{https://doi.org/10.1016/j.nuclphysb.2003.10.012}{\emph{Nucl. Phys. B}
  {\bfseries 676} (2004) 365}
  [\href{https://arxiv.org/abs/hep-ph/0308216}{{\ttfamily hep-ph/0308216}}].

\bibitem{Martin:2019lqd}
S.P.~Martin and D.G.~Robertson, \emph{{Standard model parameters in the
  tadpole-free pure $\overline{\rm{MS}}$ scheme}},
  \href{https://doi.org/10.1103/PhysRevD.100.073004}{\emph{Phys. Rev. D}
  {\bfseries 100} (2019) 073004}
  [\href{https://arxiv.org/abs/1907.02500}{{\ttfamily 1907.02500}}].

\bibitem{Martin:2005qm}
S.P.~Martin and D.G.~Robertson, \emph{{TSIL: A Program for the calculation of
  two-loop self-energy integrals}},
  \href{https://doi.org/10.1016/j.cpc.2005.08.005}{\emph{Comput. Phys. Commun.}
  {\bfseries 174} (2006) 133}
  [\href{https://arxiv.org/abs/hep-ph/0501132}{{\ttfamily hep-ph/0501132}}].

\bibitem{Martin:2014cxa}
S.P.~Martin and D.G.~Robertson, \emph{{Higgs boson mass in the Standard Model
  at two-loop order and beyond}},
  \href{https://doi.org/10.1103/PhysRevD.90.073010}{\emph{Phys. Rev. D}
  {\bfseries 90} (2014) 073010}
  [\href{https://arxiv.org/abs/1407.4336}{{\ttfamily 1407.4336}}].

\bibitem{Martin:2015lxa}
S.P.~Martin, \emph{{Pole Mass of the W Boson at Two-Loop Order in the Pure
  $\overline {MS}$ Scheme}},
  \href{https://doi.org/10.1103/PhysRevD.91.114003}{\emph{Phys. Rev. D}
  {\bfseries 91} (2015) 114003}
  [\href{https://arxiv.org/abs/1503.03782}{{\ttfamily 1503.03782}}].

\bibitem{Martin:2015rea}
S.P.~Martin, \emph{{$Z$-Boson Pole Mass at Two-Loop Order in the Pure
  $\overline{MS}$ Scheme}},
  \href{https://doi.org/10.1103/PhysRevD.92.014026}{\emph{Phys. Rev. D}
  {\bfseries 92} (2015) 014026}
  [\href{https://arxiv.org/abs/1505.04833}{{\ttfamily 1505.04833}}].

\bibitem{Martin:2016xsp}
S.P.~Martin, \emph{{Top-quark pole mass in the tadpole-free $\overline {MS}$
  scheme}}, \href{https://doi.org/10.1103/PhysRevD.93.094017}{\emph{Phys. Rev.
  D} {\bfseries 93} (2016) 094017}
  [\href{https://arxiv.org/abs/1604.01134}{{\ttfamily 1604.01134}}].

\bibitem{Martin:2016bgz}
S.P.~Martin and D.G.~Robertson, \emph{{Evaluation of the general 3-loop vacuum
  Feynman integral}},
  \href{https://doi.org/10.1103/PhysRevD.95.016008}{\emph{Phys. Rev. D}
  {\bfseries 95} (2017) 016008}
  [\href{https://arxiv.org/abs/1610.07720}{{\ttfamily 1610.07720}}].

\bibitem{Martin:2017lqn}
S.P.~Martin, \emph{{Effective potential at three loops}},
  \href{https://doi.org/10.1103/PhysRevD.96.096005}{\emph{Phys. Rev. D}
  {\bfseries 96} (2017) 096005}
  [\href{https://arxiv.org/abs/1709.02397}{{\ttfamily 1709.02397}}].

\bibitem{Martin:2022qiv}
S.P.~Martin, \emph{{Three-loop QCD corrections to the electroweak boson
  masses}}, \href{https://doi.org/10.1103/PhysRevD.106.013007}{\emph{Phys. Rev.
  D} {\bfseries 106} (2022) 013007}
  [\href{https://arxiv.org/abs/2203.05042}{{\ttfamily 2203.05042}}].

\bibitem{Alam:2022cdv}
Z.~Alam and S.P.~Martin, \emph{{Standard model at 200~GeV}},
  \href{https://doi.org/10.1103/PhysRevD.107.013010}{\emph{Phys. Rev. D}
  {\bfseries 107} (2023) 013010}
  [\href{https://arxiv.org/abs/2211.08576}{{\ttfamily 2211.08576}}].

\bibitem{Kwasnitza:2020wli}
T.~Kwasnitza, D.~St\"ockinger and A.~Voigt, \emph{{Improved MSSM Higgs mass
  calculation using the 3-loop FlexibleEFTHiggs approach including
  $x_{t}$-resummation}},
  \href{https://doi.org/10.1007/JHEP06(2023)201}{\emph{JHEP} {\bfseries 07}
  (2020) 197} [\href{https://arxiv.org/abs/2003.04639}{{\ttfamily
  2003.04639}}].

\bibitem{Bahl:2016brp}
H.~Bahl and W.~Hollik, \emph{{Precise prediction for the light MSSM Higgs boson
  mass combining effective field theory and fixed-order calculations}},
  \href{https://doi.org/10.1140/epjc/s10052-016-4354-8}{\emph{Eur. Phys. J. C}
  {\bfseries 76} (2016) 499}
  [\href{https://arxiv.org/abs/1608.01880}{{\ttfamily 1608.01880}}].

\bibitem{Bahl:2018ykj}
H.~Bahl, \emph{{Pole mass determination in presence of heavy particles}},
  \href{https://doi.org/10.1007/JHEP02(2019)121}{\emph{JHEP} {\bfseries 02}
  (2019) 121} [\href{https://arxiv.org/abs/1812.06452}{{\ttfamily
  1812.06452}}].

\bibitem{Bahl:2019hmm}
H.~Bahl, S.~Heinemeyer, W.~Hollik and G.~Weiglein, \emph{{Theoretical
  uncertainties in the MSSM Higgs boson mass calculation}},
  \href{https://doi.org/10.1140/epjc/s10052-020-8079-3}{\emph{Eur. Phys. J. C}
  {\bfseries 80} (2020) 497}
  [\href{https://arxiv.org/abs/1912.04199}{{\ttfamily 1912.04199}}].

\bibitem{Bahl:2017aev}
H.~Bahl, S.~Heinemeyer, W.~Hollik and G.~Weiglein, \emph{{Reconciling EFT and
  hybrid calculations of the light MSSM Higgs-boson mass}},
  \href{https://doi.org/10.1140/epjc/s10052-018-5544-3}{\emph{Eur. Phys. J. C}
  {\bfseries 78} (2018) 57} [\href{https://arxiv.org/abs/1706.00346}{{\ttfamily
  1706.00346}}].

\bibitem{ACME:2018yjb}
{\scshape ACME} collaboration, \emph{{Improved limit on the electric dipole
  moment of the electron}},
  \href{https://doi.org/10.1038/s41586-018-0599-8}{\emph{Nature} {\bfseries
  562} (2018) 355}.

\end{thebibliography}\endgroup
%%%%%%%%%%%%%%%%%%%%%%%%%%%%%%%%%%%%%%%%%%%%%%%%%%%%%%%%%%

\end{document}